\documentclass[%
 reprint,
superscriptaddress,
 amsmath,amssymb,
 aps,
]{revtex4-2}

\usepackage[colorlinks = true,
            linkcolor = magenta,
            urlcolor  = magenta,
            citecolor = magenta,
            anchorcolor = magenta]{hyperref}

\usepackage{graphicx}
\usepackage{dsfont}
\usepackage{dcolumn}
\usepackage{bm}
\usepackage[mathlines]{lineno}
\newcommand*\chem[1]{\ensuremath{\mathrm{#1}}}
\usepackage{amsmath,amssymb}
\usepackage{gensymb}
\usepackage{float}
\usepackage{xcolor}
\usepackage{comment}

\begin{document}

\title{Classical spin models of the windmill lattice and their relevance for PbCuTe$_2$O$_6$}

\author{Anna Fancelli}
\affiliation{Helmholtz-Zentrum Berlin f\"ur Materialien und Energie, Hahn-Meitner-Platz 1, 14109 Berlin, Germany}
\affiliation{Dahlem Center for Complex Quantum Systems and Fachbereich Physik, Freie Universit\"at Berlin, Arnimallee 14, 14195 Berlin, Germany}
\author{Johannes Reuther}
\affiliation{Helmholtz-Zentrum Berlin f\"ur Materialien und Energie, Hahn-Meitner-Platz 1, 14109 Berlin, Germany}
\affiliation{Dahlem Center for Complex Quantum Systems and Fachbereich Physik, Freie Universit\"at Berlin, Arnimallee 14, 14195 Berlin, Germany}
\affiliation{Department of Physics and Quantum Center for Diamond and Emergent Materials (QuCenDiEM), Indian Institute of Technology Madras, Chennai 600036, India}
\author{Bella Lake}
\affiliation{Helmholtz-Zentrum Berlin f\"ur Materialien und Energie, Hahn-Meitner-Platz 1, 14109 Berlin, Germany}
\affiliation{Institut f\"ur Festk\"orperforschung, Technische Universit\"at Berlin, 10623 Berlin, Germany}

\date{\today}

\begin{abstract}
We investigate classical Heisenberg models on the distorted windmill lattice and discuss their applicability to the spin-$1/2$ spin liquid candidate PbCuTe$_2$O$_6$. We first consider a general Heisenberg model on this lattice with antiferromagnetic interactions $J_n$ ($n=1,2,3,4$) up to fourth neighbors. Setting $J_1=J_2$ (as approximately realized in PbCuTe$_2$O$_6$) we map out the classical ground state phase diagram in the remaining parameter space and identify a competition between $J_3$ and $J_4$ that opens up interesting magnetic scenarios. Particularly, these couplings tune the ground states from coplanar commensurate or non-coplanar incommensurate magnetically ordered states to highly degenerate ground state manifolds with subextensive or extensive degeneracies. In the latter case, we uncover an unusual classical spin liquid defined on a lattice of corner sharing octahedra. We then focus on the particular set of interaction parameters $J_n$ that has previously been proposed for PbCuTe$_2$O$_6$ and investigate the system's incommensurate magnetic ground state order and finite temperature multistage ordering mechanism. We perform extensive finite temperature simulations of the system's dynamical spin structure factor and compare it with published neutron scattering data for PbCuTe$_2$O$_6$ at low temperatures. Our results demonstrate that thermal fluctuations in the classical model can largely explain the signal distribution in the measured spin structure factor but we also identify distinct differences. Our investigations make use of a variety of different analytical and numerical approaches for classical spin systems, such as Luttinger-Tisza, classical Monte Carlo, iterative minimization, and molecular dynamics simulations.
\end{abstract}

\maketitle

\section{Introduction}
\label{sec:intro}

Magnetic frustration plays a central role in determining the collective behavior of interacting spins at low temperatures. In particular, it can prevent conventional magnetic long-range order and give rise to exotic phases, with spin liquids perhaps representing the most notable ones~\cite{balents2010spin}. From a classical point of view, frustrated spin systems are often characterized by a large ground state degeneracy at zero temperature~\cite{ramirez1999zero,Chalker2011ChapterLacroix}. A typical situation where this is the case arises when the lattice consists of clusters of sites, in such a way that neighboring clusters only have one site in common~\cite{Chalker1992Hiddenorder,Reimers1992Absence,Petrenko2000garnet,Hopkinson2007ClassicalHyp,Henley2010Coulombphase,Rehn2016classicalSLHoney}. This is, for example, realized in the two-dimensional kagome~\cite{Chalker1992Hiddenorder} and in the three-dimensional pyrochlore~\cite{Reimers1992Absence} lattices, where the corner-sharing units are given by triangles and tetrahedra, respectively. If, additionally, antiferromagnetic Heisenberg interactions couple all spins within a cluster, the classical ground states are determined by the condition that the spins within each cluster have to sum up to zero. Depending on details of the spin degrees of freedom (number of spin components) and the precise lattice geometry (number of sites in a cluster and number of clusters per unit cell) these constraints typically leave sufficient freedom for the relative spin orientations within a cluster, such that the constraints can be satisfied by an infinite number of spin configurations~\cite{Reimers1991meafieldmagnorder,Moessner1998propertiesclassSL}. This creates a situation where, on the one hand, the classical spins are free to fluctuate, but on the other hand are constrained in their collective behaviors. A system with such a ground state was called by Villain a \textit{cooperative paramagnet}~\cite{villain1979insulating} and is commonly identified as a classical spin liquid.

Adding quantum fluctuations enables tunneling between the classical ground states, such that the new (and now possibly unique) ground state is a macroscopic superposition of the formerly degenerate classical states, a situation that is particularly promising for producing a quantum spin liquid~\cite{Hermele2004pyrophotons,Benton2012light,Shannon2012quantumice}. Because of the rich variety of phenomena arising from this construction, the lattices hosting corner-sharing units have been of central interest in the field of frustrated magnetism. The most celebrated examples are the Heisenberg models on the kagome~\cite{Huse1992Classicalkag,Garanin1999infinitecomp,Zhitomirsky2008Octupolar} and pyrochlore~\cite{Reimers1991meafieldmagnorder,Reimers1992Absence,Moessner1998propertiesclassSL,Moessner1998lowtempproperties} lattices which, classically, both have an extensive ground state degeneracy, that means the number of ground states scales exponentially with the number of lattice sites. When turning on quantum fluctuations the kagome Heisenberg model is widely believed to realize a quantum spin liquid~\cite{Sachdev1992Kagome,Yan2011KagomeDMRG,han2012fractionalized}. Quantum spin systems on the pyrochlore lattice are, generally, also good candidates for quantum spin liquids, however, this is well established only in the case when the Ising model is perturbed by small transverse interactions~\cite{Hermele2004pyrophotons,Benton2012light,Taillefumier2017CompetingSL,Benton2018QuantumSL}, while latest results for the pyrochlore Heisenberg antiferromagnet rather indicate a symmetry broken ground state~\cite{Hagymasi2021possibleinv,schafer2022abundancehardhex,Astrakhantsev2021brokensymm,Hering2022}.

Besides the well-known kagome and pyrochlore networks, the distorted windmill lattice~\cite{Nakamura1997distwind,Canals200bMnmeanfield,Isakov2008FateTrillium,Bergholtz2010SymmBreakHyp} is an alternative and less explored corner-sharing lattice geometry that, likewise, appears very promising in the context of possible spin liquid behavior. In fact, the distorted windmill lattice represents a family of lattices that can be constructed from the hyperkagome lattice. The latter is a three-dimensional arrangement of corner-sharing triangles, where each site participates in two triangles and which can be thought of as a three-dimensional generalization of the kagome lattice. Interestingly, there exists a particular way of deforming the hyperkagome lattice -- described by one real parameter -- such that its point group remains unchanged. By tuning this parameter, one can realize different corner-sharing geometries~\cite{Canals200bMnmeanfield,Hopkinson2007ClassicalHyp}. Moreover, several material realizations are known~\cite{Schiffer1995frustrGa,Nakamura1997BetaMn,Okamoto2007NaIr,Paddison2013emergfrustrbMn,Chillal2020Evidence,Koteswararao2014Magprop,Khuntia2016muonspin}, providing concrete possibilities for obtaining quantum spin liquids. Among them, \chem{Pb Cu Te_2 O_6} with spin-1/2 Cu$^{2+}$ ions has recently attracted the greatest attention. This material realizes a distorted windmill lattice in which the nearest neighbors form isolated triangles, while the second neighbors form a hyperkagome network. Since the theoretical predictions for the Heisenberg interactions indicate that first and second-neighbor couplings are antiferromagnetic and almost equally strong~\cite{Chillal2020Evidence}, the dominant couplings form a network of corner-sharing triangles, where each spin is shared by {\it three} triangles. The classical Heisenberg model for this lattice (which has recently been dubbed the hyper-hyperkagome model~\cite{Chillal2020Evidence}) admits an infinitely large ground state degeneracy, which, however, is only {\it subextensive}, i.e., the number of ground states scales exponentially only in the {\it linear} system size~\cite{Isakov2008FateTrillium,Canals200bMnmeanfield}. Couplings beyond second neighbors which are also present in the Heisenberg Hamiltonian for \chem{Pb Cu Te_2 O_6} lift this degeneracy~\cite{Chillal2020Evidence}.

The experimental findings are consistent with a spin liquid phase at low temperatures. Particularly, thermodynamic probes do not find any signature of symmetry breaking via a magnetic phase transition~\cite{Koteswararao2014Magprop} and muon spin relaxation experiments show no signs of static magnetism~\cite{Khuntia2016muonspin}. Moreover, inelastic neutron scattering on single crystals shows a broad dispersionless continuum of magnetic excitations~\cite{Chillal2020Evidence} that can be interpreted as resulting from fractional spinon quasiparticles which are characteristic for quantum spin liquids. The diffuse spin structure factor measured by neutron scattering is compatible with either a $U(1)$ gapless or a  ${\mathbb{Z}}_{2}$ gapped quantum spin liquid according to a fermionic parton mean field theory~\cite{Chern2021PSGJ1J2}.

Inspired by the experimental results on \chem{Pb Cu Te_2 O_6}, this work adds a different perspective on this material by performing theoretical investigations of the {\it classical} version of the model Hamiltonian in Ref.~\cite{Chillal2020Evidence}. A first focus in Sec.~\ref{sec:latticeH} is the reexamination of the lattice structure and the unusual change of connectivity of lattice bonds upon varying the tuning parameter for the site positions. Section~\ref{sect:T0PD} is dedicated to understanding the precise role of the interactions $J_n$ ($n=1,2,3,4$) up to fourth neighbors in determining the magnetic ground state. The classical ground state phase diagram, which has so far only been investigated up to third neighbor interactions~\cite{Jin2020Classicalquantum}, is mapped out including all four interactions and setting $J_1=J_2$ (as is approximately realized in  \chem{Pb Cu Te_2 O_6}). Surprisingly, the seemingly inconspicuous fourth neighbor coupling is found to have significant impact on the system's magnetic properties and can tune the network of interacting spins towards an interesting and previously unexplored lattice of corner-sharing octahedra. We identify an extensive ground state degeneracy in this system giving rise to an unusual type of classical spin liquid. Moreover, the phase diagram contains regions with subextensive ground-state degeneracies as well as commensurate and incommensurate ground-state magnetic orders. 

In Sec.~\ref{Sec:pbcuteo}, we focus on the particular set of couplings that have previously been proposed to be realized in \chem{Pb Cu Te_2 O_6}~\cite{Chillal2020Evidence}. First, we investigate in detail the magnetic properties of this system such as the nature of its incommensurate ground state and the sequence of two finite temperature phase transitions at which this order builds up. In the second part of Sec.~\ref{Sec:pbcuteo}, inspired by recent work~\cite{Samarakoon2017ClassicalKitaev,Samarakoon2018quantclass,Hosoi2022Uncovering,Smith2022Case,Bhardwaj2022-sf,Zhang2019Dynamicalstruct,Pohle2021ca10,Bai2019,Franke2022} according to which quantum fluctuations in a variety of systems show a surprising resemblance to disordered thermal fluctuations, we aim to identify a similar effect in \chem{Pb Cu Te_2 O_6}. Specifically, we investigate the classical model Hamiltonian for \chem{Pb Cu Te_2 O_6} in the paramagnetic regime, i.e., above the ordering transitions. Calculating the system's dynamical spin structure factor and comparing it with measured neutron scattering data, we investigate whether thermal fluctuations can mimic the effects of quantum fluctuations in \chem{Pb Cu Te_2 O_6}. We indeed find that the overall shape of the simulated spin structure factor agrees well with the measured data, except for a feature of strong intensity in our simulations that can be associated with the magnetic long-range order below the critical temperature. These results shed light on the nature of the observed spin fluctuations in \chem{Pb Cu Te_2 O_6} and reveal a partial quantum-to-classical correspondence. 

\section{Lattice and Hamiltonian}
\label{sec:latticeH}
\begin{table}[b]
\caption{\label{tab:pos}Positions of the twelve atoms in the cubic unit cell (with the lattice constant set to unity) of the distorted windmill lattice, parameterized by the real parameter $y\in\mathds{R}$. For \chem{Pb Cu Te_2 O_6} this value is given by $y =-0.2258$~\cite{Chillal2020Evidence}. All other sites of the lattice are obtained by adding integer multiples of the lattice vectors $\bm{\hat{x}}=(1,0,0)$, $\bm{\hat{y}}=(0,1,0)$, $\bm{\hat{z}}=(0,0,1)$.}
\begin{ruledtabular}
\begin{tabular}{lcdr}
\textrm{Sublattice}&
\textrm{Position}\\
\colrule
1 &  (3/4 + $y$, 3/8, 1 - $y$)\\
2 &  (1/2 + $y$, 1/4 - $y$, 7/8)\\
3 &  (5/8, 1/2 - $y$, 3/4 - $y$)\\
4 &  (1/2 - $y$, 3/4 - $y$, 5/8)\\
5 &  (3/4 - $y$, 5/8, 1/2 - $y$)\\
6 &  (7/8, 1/2 + $y$, 1/4 - $y$)\\
7 &  (1 - $y$, 3/4 + $y$, 3/8)\\
8 &  (1/4 - $y$, 7/8, 1/2 + $y$)\\
9 &  (3/8,  1 - $y$, 3/4 + $y$)\\
10 & (1/4 + $y$, 1/8, $y$)\\
11 & (1/8, $y$, 1/4 + $y$)\\
12 & ($y$, 1/4 + $y$, 1/8)\\
\end{tabular}
\end{ruledtabular}
\end{table}
The magnetic behavior of \chem{Pb Cu Te_2 O_6} is determined by \chem{Cu^{2+}} ions with spin $S=1/2$, and these ions form a distorted windmill lattice~\cite{Koteswararao2014Magprop}. This lattice has twelve atoms in its cubic unit cell, located at the $12d$ Wyckoff positions of the $P4_132$ space group~\cite{Nakamura1997distwind}. These positions depend on a real parameter $y \in\mathds{R}$, as reported in Table~\ref{tab:pos}. By comparing the crystallography data on \chem{Pb Cu Te_2 O_6} with the $12d$ Wyckoff positions of the $P4_132$ space group, one obtains $y=-0.2258$ for this compound~\cite{Chillal2020Evidence,Chern2021PSGJ1J2}.

The Hamiltonian for \chem{Pb Cu Te_2 O_6} is given by 
\begin{equation}
H=\sum_{n=1}^4J_n\sum_{{\langle i<j \rangle}_n} \bm{S}_{i}\cdot \bm{S}_{j},
\label{eq:Ham}
\end{equation}
where $\langle\ldots \rangle_n$ indicates the sum over the $n^{th}$-nearest neighbours~\cite{Chillal2020Evidence}. The estimates for the interaction strengths $J_n$ are reported in Table~\ref{tab:J}. Since we focus on the classical model, the spins are treated as three-component vectors ${\bm S}_{i}=(S^x_{i},S^y_{i},S^z_{i})$ with unitary norm $\lvert {\bm S}_{i} \rvert = 1$.

\begin{table}[b]
\caption{\label{tab:J}
Exchange parameters of the Heisenberg Hamiltonian \eqref{eq:Ham} for \chem{Pb Cu Te_2 O_6} from DFT calculations~\cite{Chillal2020Evidence}.
}
\begin{ruledtabular}
\begin{tabular}{cccc}
$J_1$ & $J_2$ & $J_3$ & $J_4$\\
\colrule
1.13 meV & 1.07 meV & 0.59 meV & 0.12 meV\\
\end{tabular}
\end{ruledtabular}
\end{table}

\begin{figure}[h!]
\includegraphics[width=0.49\textwidth]{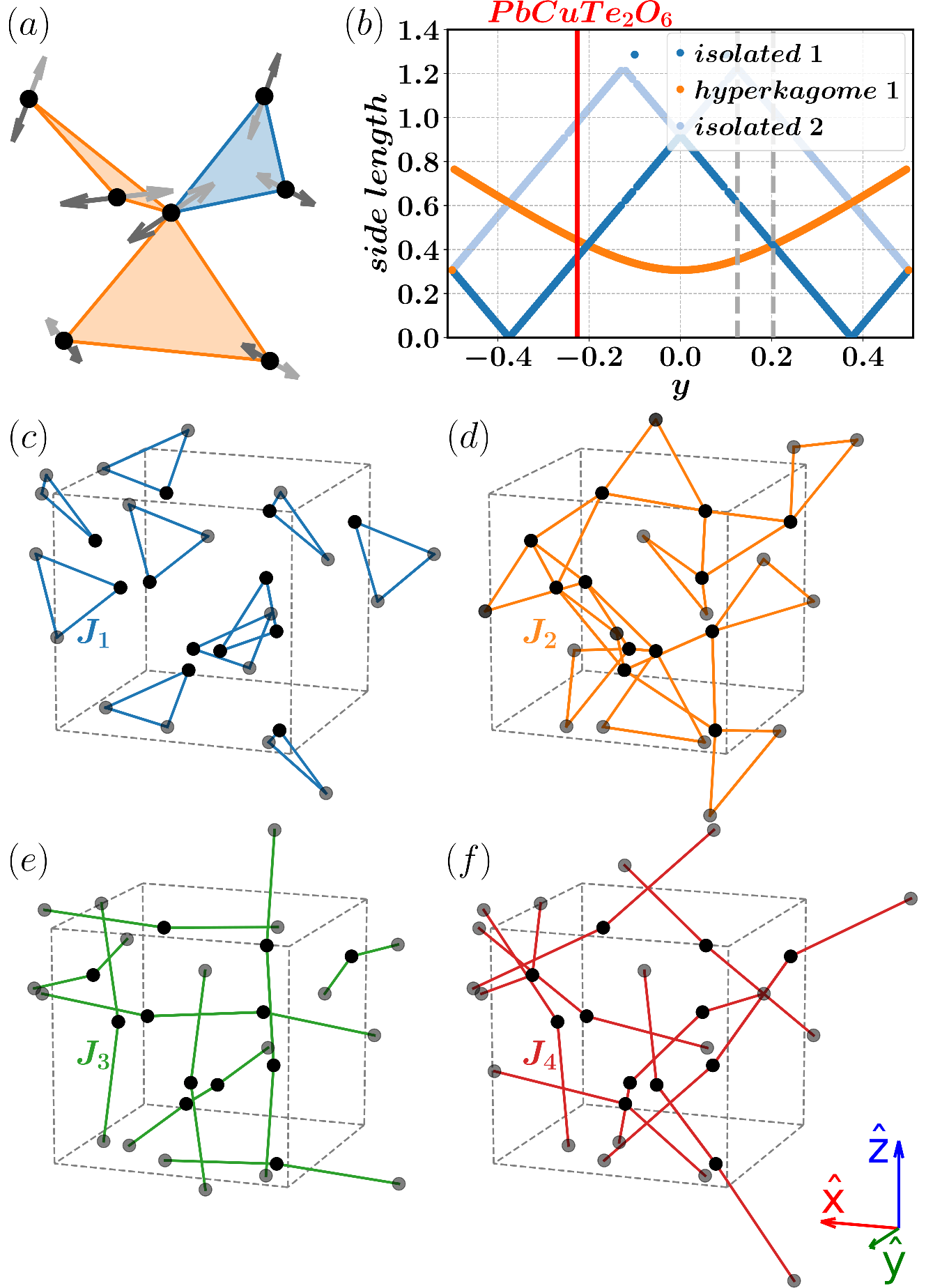}
\caption{\label{fig:UnitCell}(a) Unit of the distorted windmill lattice: each site is shared by one isolated triangle (in blue) with coordination number two [as shown in (c)] and two hyperkagome triangles (in orange) with coordination number four [as shown in (d)]. The light (dark) gray arrows indicate the directions along which the sites move when decreasing (increasing) $y$. (b) Side lengths of the smallest triangles in the distorted windmill lattice, as a function of $y$. Only the two smallest types of isolated triangles (labeled `isolated 1' and `isolated 2') are shown. At the special points $y= \pm y^*$ with $y^*=(9-\sqrt{33})/16\simeq0.2035$ the hyperkagome and isolated triangles have equal size and the system has six nearest neighbors. At $y= \pm 3/8$ the smallest isolated triangles collapse into one point and the positions in the unit cell become three times degenerate, resulting in a lattice with four sites per unit cell. The classical nearest neighbor Heisenberg models with $y=0.125$ and $y=y^*$ (dotted gray lines) have already been studied in Ref.~\cite{Hopkinson2007ClassicalHyp} and Refs.~\cite{Canals200bMnmeanfield,Isakov2008FateTrillium}, respectively. (c)-(f) First to fourth nearest neighbor bonds of the distorted windmill lattice with $y=-0.2258$ as realized in \chem{Pb Cu Te_2 O_6}. The black and grey dots correspond to sites inside and outside the cubic unit cell, respectively. (b) First nearest neighbors form isolated triangles. (d) Second nearest neighbors create a network of corner-sharing (hyperkagome) triangles. (e) Third nearest neighbours form chains parallel to the $\bm{\hat{x}}$, $\bm{\hat{y}}$, $\bm{\hat{z}}$ Cartesian directions. (f) Fourth nearest neighbors form chains along the body diagonals.}
\end{figure}

In the distorted windmill lattice, bonds of the same length often form connected or disconnected arrangements of equilateral triangles. As an example, in Fig.~\ref{fig:UnitCell}(a) we plot the triangle configuration obtained by connecting one site to its first and second nearest neighbors, for fixed value of the position parameter $y=-0.2258$. Each site is shared by three triangles that constitute the unit of the distorted windmill lattice: two corner-sharing (hyperkagome) triangles (in orange) and one isolated triangle (in blue). By increasing or decreasing the value of $y$, the site positions move as indicated by the gray arrows in Fig.~\ref{fig:UnitCell}(a) and the size of the triangles changes. To visualize this, in Fig.~\ref{fig:UnitCell}(b) we plot the side lengths of the smallest equilateral triangles created by connecting $n^{th}$-nearest neighbors, as a function of the position parameter $y$. Other bond types, which do not form equilateral triangles are omitted in Fig.~\ref{fig:UnitCell}(b). Since the change $y\rightarrow y+1$ leaves the lattice invariant, it is sufficient to consider $y$ in the interval $y\in[-1/2,1/2]$ only. The triangles with the smallest side length for a fixed value of $y$ also correspond to the first nearest neighbours among all bonds. For $-1/2<y< -y^*$ and $y^*<y<1/2$, with $y^*=(9-\sqrt{33})/16\simeq0.2035$, the first nearest neighbours create isolated triangles, see Fig.~\ref{fig:UnitCell}(c). For $-y^* < y < y^*$ and $y=\pm 1/2$, the first nearest neighbours create the hyperkagome triangles, where each spin is shared by two triangles, see Fig.~\ref{fig:UnitCell}(d). This lattice has been referred to as the hyperkagome lattice~\cite{Hopkinson2007ClassicalHyp}. At the special points  $y= \pm y^*$ the corner-sharing and isolated triangles have equal size and each spin belongs to three triangles~\cite{Isakov2008FateTrillium}. At $y=\pm 3/8$ the isolated triangles formed by the first nearest neighbors collapse into one point and the positions in the unit cell become three times degenerate. This can be understood by looking at the direction along which the sites belonging to the isolated triangle move when decreasing $y$ [light gray arrows in Fig.~\ref{fig:UnitCell}(a)]. The resulting lattice for $y= \pm 3/8$ with four sites per unit cell corresponds to the so-called trillium lattice~\cite{Hopkinson2006Trillium}. Various discontinuities in the side lengths as a function of $y$ can be observed in Fig.~\ref{fig:UnitCell}(b). For example, at $y=0.5$ a hyperkagome network is created with much smaller bond lengths than at slightly smaller or larger values of $y$. This is explained by the observation that at this point, the two smallest isolated triangles have equal size and, when combining them, become a hyperkagome network. The smallest triangles that are truly isolated then occur with larger bond lengths outside the plotted region. Other discontinuities in Fig.~\ref{fig:UnitCell}(b) have a similar origin.

For $y=-0.2258$ realized in \chem{Pb Cu Te_2 O_6}, the first and second nearest neighbours create, respectively, a network of isolated and corner-sharing triangles [Fig.~\ref{fig:UnitCell}(c), (d)]. This value [red line in Fig.~\ref{fig:UnitCell}(b)] lies close to the special point $1-y^*$. In fact, it was found that $J_1 \simeq J_2$ (Table~\ref{tab:J}), reflecting the similar geometrical distance between the first and second neighbor sites. As a result, each spin interacts almost equally with six other spins, forming a network of corner-sharing triangles where each site contributes to three triangles. This network has the same connectivity as the one at $y=\pm y^*$ formed by the first nearest neighbors. Because of the higher connectivity with respect to the hyperkagome lattice, the present case has been referred to as a hyper-hyperkagome lattice~\cite{Chillal2020Evidence}.

The magnetic Hamiltonian for \chem{Pb Cu Te_2 O_6} also contains antiferromagnetic interactions between the third and fourth nearest neighbors (Table~\ref{tab:J}). These form chains along the $\bm{\hat{x}}$, $\bm{\hat{y}}$, $\bm{\hat{z}}$ Cartesian directions [Fig.~\ref{fig:UnitCell}(e)]
and the body diagonals [Fig.~\ref{fig:UnitCell}(f)], respectively. In the next section, we will focus on the role of all these interactions in determining the magnetic ground state.

\section{Hyper-hyperkagome model and magnetic phase diagram} \label{sect:T0PD}
In this section, we present the $T=0$ classical phase diagram for the $J_1$-$J_2$-$J_3$-$J_4$ model at $J_1=J_2$, obtained by combining the Luttinger-Tisza (LT)~\cite{Luttinger1946LT} and the iterative minimization (IM)~\cite{Sklan2013Noncop} techniques. Particularly, we also discuss the role of the $J_4$ coupling which has not been considered in previous works~\cite{Jin2020Classicalquantum}. Details of the LT and IM methods and their comparison are given, respectively, in the Appendices~\ref{appendix:LT} and~\ref{appendix:IM}. The LT method is an analytic approach where the classical energy is minimized under the approximation of a partially relaxed spin length constraint where only the total spin but not necessarily the individual spins are normalized. The IM approach, on the other hand, consists of minimizing the classical energy numerically, by aligning the spins to the effective magnetic field created by the surrounding spins, starting from a random configuration. As an iterative technique, it is prone to detecting local energy minima instead of global ones. Our procedure to identify the ground state magnetic order consists of comparing results from IM and LT. A first insight into the ground state spin configuration is obtained by the magnetic ordering wave vector $\bm{Q}$ that minimizes the classical energy within LT. This wave vector can also be calculated with IM, as it corresponds to the maxima of the equal-time spin structure factor for spins belonging to the same sublattice, summed over the sublattices~\cite{Jin2020Classicalquantum}
\begin{equation}
\mathcal{S}_\text{sub}(\bm{q}) = \frac{12}{N}\sum_{I,J}^{L\times L \times L} \sum_{\alpha = 1}^{12}\bm{S}_{\alpha,I} \cdot \bm{S}_{\alpha,J} e^{i \bm{q} \cdot (\bm{R}_I-\bm{R}_J)},
\label{eq:Ssub}
\end{equation}
where $I,J$ index the unit cell, $\alpha$ the sublattice and $\bm{R}_I$ denotes the real space position of the unit cell $I$. Furthermore, $N$ is the total number of spins in the simulated system, which is a cube of $L\times L\times L$ crystallographic unit cells ($N=12L^3$). Since $\mathcal{S}_\text{sub}(\bm{q})$ only contains spatial Fourier transforms of spin correlations on the same sublattice, it describes how spins rotate between different unit cells but neglects information on how spins are correlated within a unit cell. Using $\mathcal{S}_\text{sub}(\bm{q})$ for characterizing magnetic order can be useful when the lattice has a large unit cell. Besides that, it can be directly compared to the LT results, by checking if $\bm{Q}$ is one of the wave vectors that minimizes the energy within LT. The quantity $\mathcal{S}_\text{sub}(\bm{q})$ should be distinguished from the equal-time spin structure factor
\begin{equation}
\mathcal{S}(\bm{q})=\frac{1}{N}\sum_{i,j}\bm{S}_i  \cdot \bm{S}_j e^{i\bm{q} \cdot (\bm{r}_i-\bm{r}_j)},\label{eq:S}
\end{equation}
which uses the actual site positions $\bm{r}_i$ and considers correlations between all sublattices and which will also be used below. 

We consider the set of interactions described in Fig.~\ref{fig:UnitCell}, fixing $J_1=J_2=J$ (hyper-hyperkagome case) and letting $J_3$ and $J_4$ vary in the range $0\leq J_3,J_4 \leq J$. Setting $J_1=J_2$ is motivated by the physical situation in \chem{Pb Cu Te_2 O_6} where $y\simeq y^*$, i.e., the isolated nearest neighbor triangles and the hyperkagome triangles are approximately of the same size, and the same is true for the corresponding interactions.
\begin{figure}[t]
\includegraphics[width=0.35\textwidth]{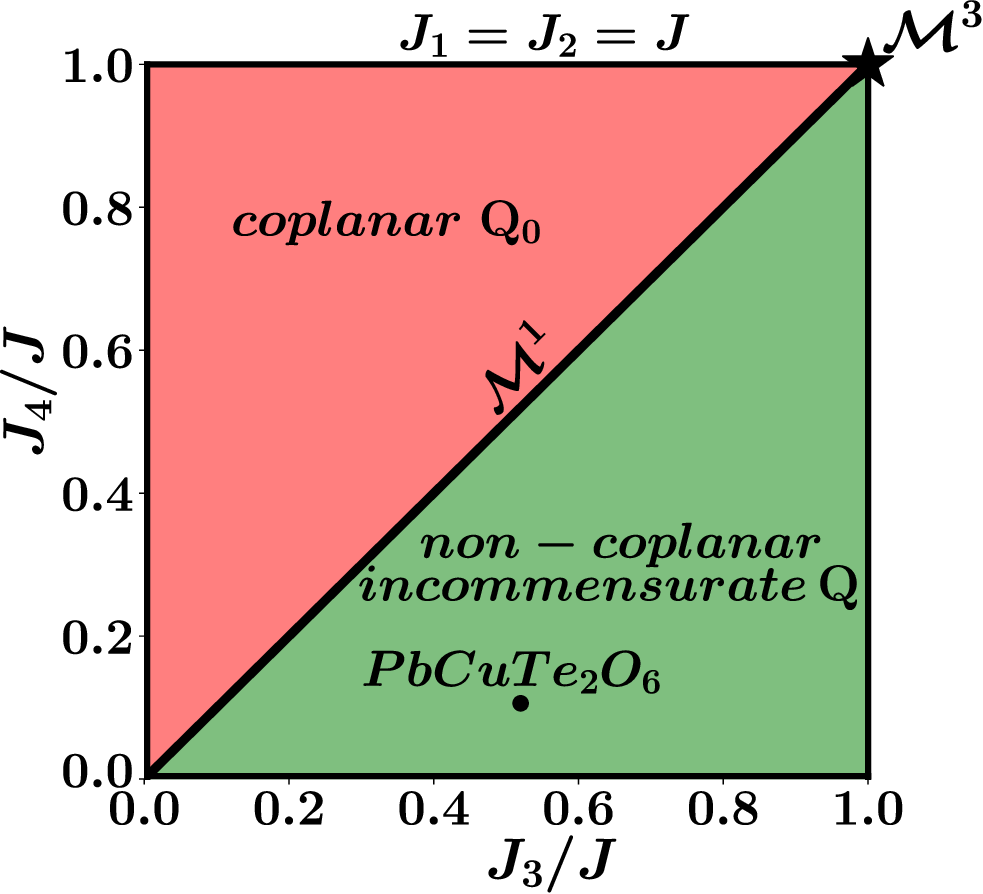}
\caption{Classical zero temperature phase diagram for the Heisenberg model on the distorted windmill lattice with fixed $J_1=J_2=J$, as a function of $J_3$ and $J_4$. Four different cases can be distinguished. (1) In the green region where $J_4 < J_3$ the ground state is magnetically ordered with an incommensurate wave vector $\bm{Q}$. (2) In the red region where $J_3 < J_4$ the ground state has coplanar magnetic order with $\bm{Q}_0=0$. (3) Along the $J_3=J_4$ line with the point $J_3=J_4=0$ included and the point $J_3=J_4=J$ excluded, the degenerate ground states form a sub-extensive (1D) manifold ($\mathcal{M}^1$), equal to the one found for the $J_3=J_4=0$ case~\cite{Isakov2008FateTrillium}. (4) At the point $J_3=J_4=J$ the degenerate ground states form an extensive (3D) manifold ($\mathcal{M}^3$), as further explained in the main text. The point associated with PbCuTe$_2$O$_6$ is drawn by approximating $J_2$ by $J_1$. It corresponds to the ratios $J_3/J_1$ and $J_4/J_1$, with the values of $J_i$ reported in Table~\ref{tab:J}.} 
\label{fig:T0phase}
\end{figure}
\begin{figure}[t!]
\includegraphics[width=0.42\textwidth]{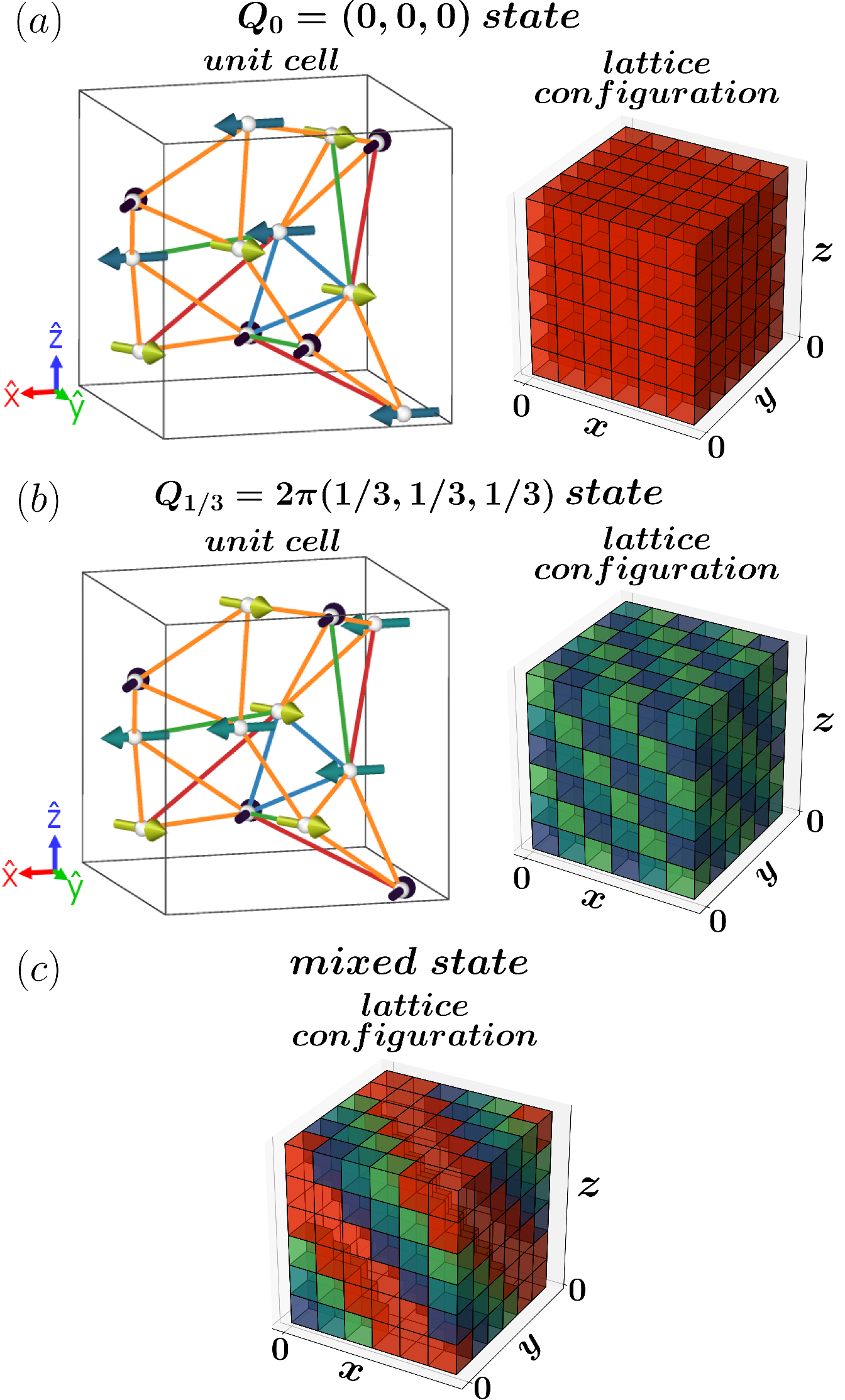}
\caption{(a), (b) Left: Spin configurations in one unit cell corresponding to (a) the $\bm{Q}_0=(0,0,0)$ coplanar order and (b) the $\bm{Q}_{1/3}=2\pi(1/3,1/3,1/3)$ coplanar order. The coloring of the bonds with interactions $J_1$ (blue), $J_2$ (orange), $J_3$ (green) and $J_4$ (red) matches Fig.~\ref{fig:UnitCell}(b)-(e). Only the bonds within the unit cell are shown. Spins with the same orientation have the same color. Spins with different colors enclose an angle of 120$\degree$ around each triangle. Right: Corresponding construction of spin states over the full lattice. The $\bm{Q}_0=(0,0,0)$ unit cell configuration illustrated by red boxes in (a) is repeated in each unit cell. The spins in the $\bm{Q}_{1/3}=2\pi(1/3,1/3,1/3)$ unit cell undergo a rotation about the axis perpendicular to the plane they span, by an angle $\bm{Q}_{1/3}\cdot \bm{R}$ when proceeding to the other unit cells, where $\bm{R}$ denotes the position of the unit cell. Consequently, there are three distinct spin configurations in one unit cell, indicated by different colors, that are arranged over the full lattice as depicted in (b). (c) Possible ground state configuration for the $J_3=J_4$ case obtained by mixing the $\bm{Q}_0=(0,0,0)$ (red) and the $\bm{Q}_{1/3}=2\pi(1/3,1/3,1/3)$ unit cells (from blue to green) with periodic boundary conditions. The obtained configuration consists of an alternating stacking of $\bm{Q}_{1/3}$ and $\bm{Q}_{0}$ unit cells along the $(1,1,1)$ direction.}
\label{fig:states}
\end{figure}

As shown in Fig.~\ref{fig:T0phase}, the phase diagram contains two different extended phases and different types of ground state degeneracies at the boundaries, described in more detail below.\\
\newline
\textit{(1) $J_4<J_3<J$: Incommensurate $\bm{Q}$ order}\\
In the region where $J_4<J_3<J$ the system orders magnetically in a non-coplanar spin state with incommensurate wave vector $\bm{Q}$, which varies as $J_3$ and $J_4$ vary. A similar result was found respectively for the $J_4=0$ case~\cite{Jin2020Classicalquantum} and for the hyperkagome model in presence of Dzyaloshinskii-Moriya interaction~\cite{Chen2008SpinOrbit}. The procedure used to identify the incommensurate order is explained in the next section, where we discuss the system's ground state and finite temperature properties for the set of interactions of \chem{Pb Cu Te_2 O_6} (black dot in the phase diagram in Fig.~\ref{fig:T0phase}).\\
\newline
\textit{(2) $J_3<J_4<J$: Coplanar $\bm{Q}=0$ order}\\
In the region where $J_3<J_4<J$ the system orders magnetically in a coplanar state with wave vector $\bm{Q}_0=0$. This means that the spin arrangement, as depicted in Fig.~\ref{fig:states}(a), is the same in each unit cell. Pairs of spins coupled by $J_1$, $J_2$, and $J_4$ form an angle of 120$\degree$ with each other. On the other hand, the chains formed by $J_3$ have a ferromagnetic spin alignment. As a consequence, increasing the value of $J_3$ increases the level of frustration, as this bond gives a positive contribution to the energy, see Table \ref{tab:copJ} for $\bm{Q}_0$. This spin configuration is unique up to global spin rotations. A special subgroup of such rotations consists of permutations of the three spin directions which form this state.\\ 
\newline
\textit{(3) $J_3=J_4 \neq J$: One-dimensional degenerate ground state manifold}\\
Along the $J_3=J_4 \neq J$ line, the system is characterized by a sub-extensive ground state manifold, equal to the one found for the $J_3=J_4=0$ case~\cite{Isakov2008FateTrillium}, where the number of degenerate ground states scales exponentially in the linear system size $L$. We will first briefly review the $J_3=J_4=0$ case and then show how the results can be generalized to the $J_3=J_4  \neq J$ case. When $J_3=J_4=0$, there are two different types of coplanar ground states, characterized by $\bm{Q}_{0}=0$ and $\bm{Q}_{1/3}=2\pi(\pm1/3,\pm1/3,\pm1/3)$, where the signs $\pm$ in the three components of $\bm{Q}_{1/3}$ can be chosen independently. The $\bm{Q}_{0}=0$ ground state corresponds to the one found in the region (2). An example for a $\bm{Q}_{1/3}$ state with the choice of signs given by $\bm{Q}_{1/3}=2\pi(1/3,1/3,1/3)$ is depicted in Fig. \ref{fig:states}(b) for one unit cell. (Note that here and in the following, the term `unit cell' refers to the crystallographic unit cell, not to the magnetic one.) Other combinations of signs correspond to other spin configurations not shown. The configurations in the neighboring unit cells are found by rotating the spins about the axis perpendicular to the plane in which the spins lie, by an angle given by $\bm{Q}_{1/3} \cdot \bm{R}$, where $\bm{R}$ denotes the position of the unit cell. 

\begin{table}[b]
\caption{\label{tab:copJ}
Bond contributions $c_n = \frac{1}{N}\sum_{{\langle i<j \rangle}_n} \bm{S}_{i}\cdot \bm{S}_{j}$ for the coplanar $\bm{Q}_{0}$, $\bm{Q}_{1/3}$ and mixed states, defined such that $H=N\sum_{n=1}^4 J_n c_n$. The mixed state is obtained by mixing $a$ unit cells with $\bm{Q}_{0}$ order and $b$ unit cells with $\bm{Q}_{1/3}$ order in such a way that $J_1$ and $J_2$ triangles remain in energetically optimal $120\degree$ spin configurations, see main text for details. As indicated in the last row, when $J_3=J_4$, the energy contribution from $J_3$ and $J_4$ bonds given by $\sim J_3 c_3+J_4 c_4=J_3(c_3+c_4)$ is independent of the parameters $a$ and $b$ and equal to the $\bm{Q}_{0}$ and $\bm{Q}_{1/3}$ states.}
\begin{ruledtabular}
\begin{tabular}{cccc}
\textrm{ }&
\textrm{$\bm{Q}_{0}$ state} &
\textrm{$\bm{Q}_{1/3}$ state}&
\textrm{mixed state}\\\colrule
$c_1$ & -0.5 & -0.5 & -0.5\\
$c_2$ & -1 & -1 & -1\\
$c_3$ & 1 & -0.125 & $\frac{a-0.125b}{a+b}$ \\
$c_4$ & -0.5 & 0.625& $\frac{-0.5a+0.625b}{a+b}$\\
$c_3 + c_4$ & 0.5 & 0.5& 0.5\\
\end{tabular}
\end{ruledtabular}
\end{table}
These rotations always involve an angle of 120$\degree$. Therefore, for each combination of signs, there are three possible arrangements of spins in the unit cell denoted $A$, $B$, $C$ that are cyclically or anti-cyclically repeated along each Cartesian direction. Particularly, for the signs as given in $\bm{Q}_{1/3}=2\pi(1/3,1/3,1/3)$, a cyclic progression $A\rightarrow B\rightarrow C\rightarrow A$ occurs along all three Cartesian directions which results in stacked layers of different unit cells along the $(111)$ direction [see Fig.~\ref{fig:states}(b)], while other sign choices lead to stacks along other body diagonals.

Both the $\bm{Q}_{0}$ and the $\bm{Q}_{1/3}$ coplanar ground states feature spins in the isolated and hyperkagome triangles that have angles of 120$\degree$ between each other, see Fig.~\ref{fig:states}(b) and Fig.~\ref{fig:states}(c), respectively. Having the same mutual arrangement of spins in the triangles allows the construction of an infinite set of ground states, by mixing stacked layers of $A$, $B$, $C$ unit cells (corresponding to a $\bm{Q}_{1/3}$ order) with stacked layers of $\bm{Q}_{0}$ order along the respective body diagonal. Such a composition of unit cells along the body diagonal has to be done under a precise set of rules, explained in Ref.~\cite{Isakov2008FateTrillium}, to ensure that the triangles shared by neighboring unit cells maintain the 120$\degree$ angles. An example of a possible ground state configuration is depicted in Fig.~\ref{fig:states}(c). The existence of mixed ground states gives rise to an exponentially large ground state degeneracy $\sim 2^{L}/6$, where the exponent scales linearly with the system size, see Ref.~\cite{Isakov2008FateTrillium}.

Remarkably, all the degenerate ground states in the case $J_3=J_4=0$ remain ground states when $J_3=J_4 \neq 0$. This can be understood by examining the energy contributions from the individual $J_1$, $J_2$, $J_3$ and $J_4$ bonds in the $\bm{Q}_0$, $\bm{Q}_{1/3}$ and mixed states, respectively, reported in Table~\ref{tab:copJ}. Particularly, when $J_3=J_4$ the energy contributions from the chain interactions are equal in all three states and also independent of the precise mixing of $\bm{Q}_0$ and $\bm{Q}_{1/3}$ orders.
Furthermore, the corresponding energies agree with the ones found within IM and LT, indicating that these degenerate states are also ground states. Thus, the ground state manifold of the $J_3=J_4<J$ case is identical to the one in the $J_3=J_4=0$ case.

\begin{figure}[t]
\includegraphics[width=0.48\textwidth]{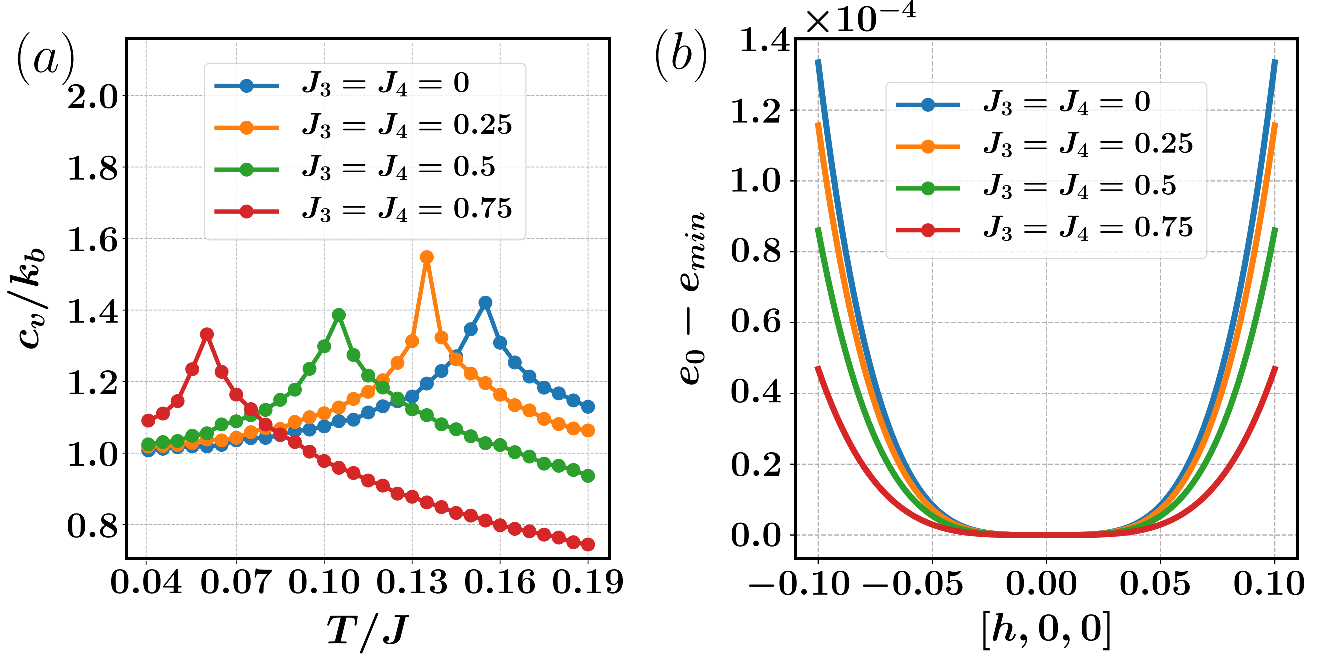}
 \caption{\label{fig:J3J4eq}(a) Specific heat as a function of temperature obtained by MC simulations with $L=8$, for $J_1=J_2=J$ and four different values of $J_3=J_4\in\{0,0.25,0.5,0.75\}$. For every value of $J_3=J_4$, the system undergoes an order-by-disorder transition, as indicated by the peak. By increasing the value of $J_3=J_4$, the temperature of the transition decreases. (b) Energy ${\bm e}_0$ of the lowest band from the LT method, along the momentum cut $[h,0,0]$ that contains the energy minimum ${\bm e}_{min}$ at $\bm{Q}_0=(0,0,0)$. For better comparison of the behaviors at low energies, the bands are shifted by the respective minimum $\bm{e}_{min}$ for each value of $J_3=J_4$. The flattening of the band with increasing $J_3=J_4$ causes a downward shift of the order-by-disorder transition temperature.}
\end{figure}

\begin{figure}[t]
\includegraphics[width=0.48\textwidth]{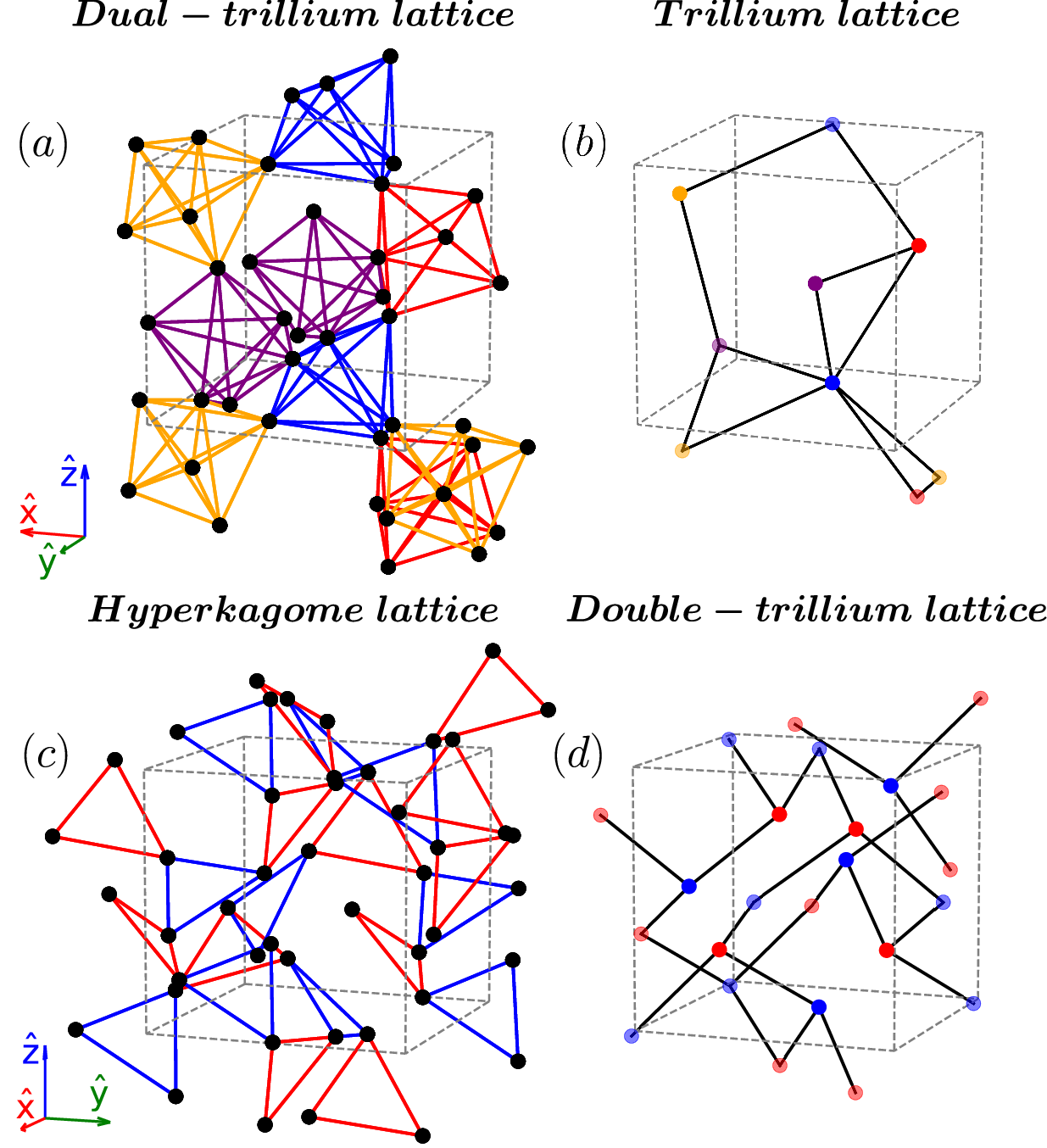}
 \caption{(a) Distorted windmill lattice with $y=-0.2258$. The colored bonds represent the $J_1$, $J_2$, $J_3$, and $J_4$ couplings which are all of equal strength $J$. We refer to this network as the dual-trillium lattice. All bonds within an irregular octahedron are illustrated with the same color. (b) Centers of the irregular octahedra plotted in (a). The centers of the octahedra that share a site are connected by a black line, creating a nearest neighbor trillium network. The trillium lattice has four sites per unit cell which are represented by four different colors of the sites. These colors match the ones used for the octahedral bonds in (a). (c) Hyperkagome lattice where the colored bonds correspond to the hyperkagome couplings $J_2$ shown in Fig.~\ref{fig:UnitCell}(c). The same color is used for the hyperkagome triangles corresponding to the same trillium sublattice of the double-trillium lattice formed by the centers of the triangles plotted in (d). The centers of the hyperkagome triangles that share a site are connected by a black line in (d), creating a double-trillium network. The double-trillium lattice is bipartite and composed of two trillium sublattices highlighted here by the two colors that match the colors of the corresponding hyperkagome triangles in (c).}
\label{fig:unitcellocta}
\end{figure}

A finite-temperature order-by-disorder transition towards the $\bm{Q}_0$ state was previously observed in the $J_3=J_4=0$ case~\cite{Isakov2008FateTrillium}. To investigate whether an analogous behavior also occurs for $J_3=J_4\neq 0$ we perform finite-temperature classical Monte Carlo (MC) simulations of $N=12\times L\times L \times L$ spins with periodic boundary conditions, interacting according to the Hamiltonian in Eq.~\eqref{eq:Ham}. Details of the simulations are given in Appendix~\ref{appendix:CMC}. We fix $J_1=J_2$ and change the value of $J_3=J_4$. The specific heat as a function of temperature shows a peak for each value of $J_3=J_4$, indicating the presence of an order-by-disorder transition [Fig.~\ref{fig:J3J4eq}(a)]. Below the transition the system is always found to be in the $\bm{Q}_0$ state, in analogy with the $J_3=J_4=0$ case. The temperature of the order-by-disorder transition decreases as $J_3=J_4$ increases. This can be understood by investigating the energy bands from LT which are obtained by diagonalizing the coupling matrix $J_{ij}$. As shown in Fig.~\ref{fig:J3J4eq}(b) increasing the value of $J_3=J_4$ causes a flattening of the lowest band around the minimum (which forms a line in momentum space), making the system more prone to explore different configurations driven by thermal fluctuations. As a consequence, the order-by-disorder transition shifts towards lower temperatures.
\\
\newline
\textit{(4) $J_3=J_4=J$: Three-dimensional degenerate ground state manifold}\\
At the point $J_3=J_4=J$ the system features an extensive ground state manifold where the number of degenerate ground states scales exponentially in the number of sites $N$. This is due to the fact that when $J_1=J_2=J_3=J_4=J$, the Hamiltonian can be rewritten as a sum over spin clusters that have the shape of irregular octahedra,
\begin{equation}
H=\frac{J}{2}\sum_\text{octa} (\bm{S}_\text{octa})^2 + \text{const.},
\label{eq:octa}
\end{equation}
where $\bm{S}_\text{octa}$ is the sum of the spins in each irregular octahedral cluster.
More precisely, in this specific case, each spin is coupled to ten other spins with the same interaction strength. These ten spins can be divided into two groups of five spins, where each group together with the reference spin forms an irregular octahedra. This property holds for each spin in the lattice, creating thus a network of corner-sharing clusters of six spins. Within each cluster, all pairs of spins are coupled with the same interaction $J$. We call these clusters irregular octahedra because they have six vertices and eight faces, however, the triangles composing the faces are not equilateral (as they would be in a regular octahedron) even though all interaction strengths are the same [see Fig.~\ref{fig:unitcellocta}(a)]. The underlying lattice created by the center points of the octahedral clusters is a trillium lattice with four sites in the cubic unit cell where each site is shared by three triangles~\cite{Hopkinson2006Trillium}, see Fig.~\ref{fig:unitcellocta}(b). Due to this property, we will refer to this network as the dual-trillium lattice.

The present case is similar to the well-known classical Heisenberg model on the pyrochlore lattice, in which the spins create a network of corner-sharing tetrahedra. As for the pyrochlore lattice, the origin of the extensive degeneracy on the dual trillium lattice can be understood from a Maxwellian counting argument~\cite{Reimers1991meafieldmagnorder,Moessner1998propertiesclassSL}. The number of degrees of freedom can be estimated as $F=\frac{N_\text{c}}{2}(n-1)q$, where $n$ is the number of spin components, $q$ is the number of spins in a cluster (for the dual trillium lattice $q=6$), and $N_\text{c}$ is the number of clusters. The factor $n-1$ comes from the fact that the spin's length constraint fixes one of its components, while the factor $1/2$ is due to the fact that each spin belongs to two clusters. The number of constraints $\bm{S}_\text{octa}=0$ that determine the ground state manifold is given by $K=nN_\text{c}$. If the constraints are independent the dimension of the ground state manifold is given by $D=F-K=\frac{N_\text{c}}{2}[n(q-2)-q]$, which for $q=6$ is extensive for Heisenberg spins with $n=3$ (where $D=3N_\text{c}$) and for XY spins with $n=2$ (where $D=N_\text{c}$). Interestingly, these estimates for the dimension of the ground state manifolds are even larger than for the pyrochlore lattice where $q=4$ and, consequently, $D=N_\text{c}$ for Heisenberg spins and $D=0$ for XY spins.  

It is also instructive to compare the magnetic properties of the dual trillium Heisenberg model and the hyperkagome Heisenberg model, since the latter, likewise, features an extensive ground state degeneracy~\cite{Hopkinson2007ClassicalHyp}. As mentioned earlier the hyperkagome network is a special type of distorted windmill lattice where only $J_2$ bonds but no other $J_1$, $J_3$, or $J_4$ bonds contribute. In the hyperkagome lattice, the spins create a network of corner-sharing triangles and the dual lattice that is formed by connecting the centers of these triangles corresponds to a lattice with two interconnected trillium lattices, that we call the double-trillium lattice, see Fig.~\ref{fig:unitcellocta}(c) and (d) and Ref.~\cite{zivkovic21}. This lattice has a doubled unit cell with respect to the trillium lattice, and each site is connected to three sites of the respective other trillium sublattice (consequently the double-trillium lattice is bipartite). The classical hyperkagome Heisenberg model is characterized by dipolar spin correlations, which manifest as pinch points in the equal time spin structure factor~\cite{Hopkinson2007ClassicalHyp}. On the other hand, however, the system shows partial order at low temperatures $T/J \lesssim 0.001$~\cite{Hopkinson2007ClassicalHyp}, where an order-by-disorder transition drives the system into a phase with coplanar spin configurations. 

\begin{figure}[t]
\includegraphics[width=0.48\textwidth]{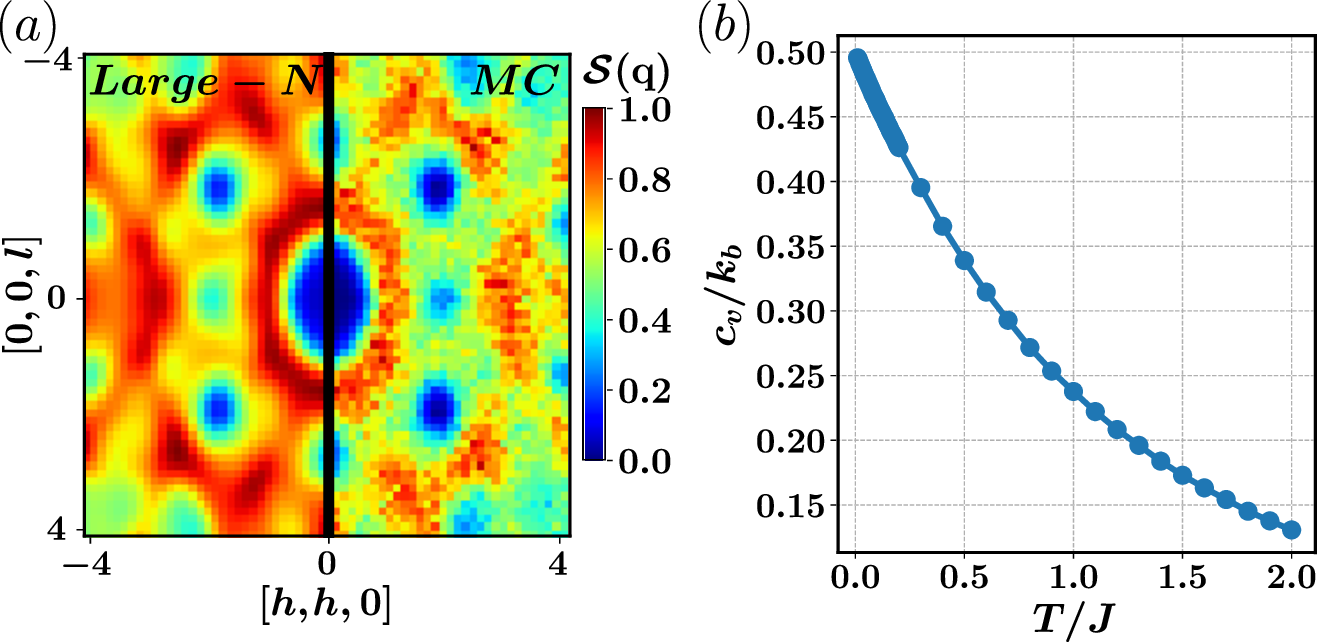}
 \caption{(a) Equal time spin structure factor [Eq.~(\ref{eq:S})] of the classical dual trillium Heisenberg model with $J_1=J_2=J_3=J_4=J$ in the $[h,h,l]$ plane at $T/J=0.01$ obtained by means of MC simulations (right), compared with the one obtained by means of the large-$N$ approach~\cite{Garanin1999infinitecomp} (left). (b) Specific heat per spin for the same model as a function of temperature obtained by MC simulations. No peak is observed down to $T/J=0.01$.}
\label{fig:Sqcv}
\end{figure}

The classical dual trillium Heisenberg model obtained at $J_1=J_2=J_3=J_4=J$ shows significant differences compared to the hyperkagome Heisenberg model. First, the equal time spin structure factor [see Eq.~(\ref{eq:S})] does not show any pinch points, neither within the large-$N$ approximation~\cite{Garanin1999infinitecomp} nor within classical MC, see Fig.~\ref{fig:Sqcv}(a) for a comparison of the equal time spin structure factors in the $[h,h,l]$ plane obtained by both methods. As explained in Ref.~\cite{Rehn2017classicalZ2}, the absence of pinch points can be traced back to the fact that the parent trillium lattice (that is created by connecting the centers of the irregular octahedra) is {\it non-bipartite}. In contrast, the double trillium lattice (that is created by connecting the centers of the hyperkagome triangles) is bipartite. A non-bipartite nature of a lattice of corner-sharing clusters gives rise to fascinating physical phenomena, most strikingly, these systems have been proposed to host ``classical $\mathbb{Z}_2$ spin liquids''~\cite{Rehn2017classicalZ2}, which can be understood as the classical counterparts of quantum $\mathbb{Z}_2$ spin liquids (see also the recent works in Refs.~\cite{Han2305.00155,Han2305.19189,Davier2304.10906} for general classification schemes for classical spin liquids). In the classical case, these phases are characterized by exponentially decaying spin correlations and an absence of pinch points in the spin structure factor.

Furthermore, the specific heat from MC simulations of the classical dual trillium Heisenberg model does not show any sign of an order-by-disorder transition [see Fig.~\ref{fig:Sqcv}(b)].
This is in agreement with the argument given in Ref.~\cite{Moessner1998lowtempproperties}, according to which a system is expected to show an order-by-disorder transition if $n<(q+2)/(q-2)$. In the present case with six-site clusters ($q=6$) and Heisenberg spins ($n=3$) this condition is not satisfied. Our MC calculations predict a specific heat per spin that approaches the value $c_v=k_B/2$ at low temperatures. 
This value can again be explained within the aforementioned Maxwellian counting argument and is also consistent with the absence of an order-by-disorder transition in the following way. The $D$ modes in the ground state manifold (which correspond to flat bands in LT) do not contribute to the specific heat. The other $F-D=K=nN_c$ modes contribute with $k_B/2$ to the specific heat, provided that these are all harmonic modes. Hence the specific heat per spin is expected to be $c_v=n N_c k_B/(2 N)$. Since the number of sites $N$ and the number of octahedral clusters $N_c$ are related by $N=3N_c$ and assuming Heisenberg spins ($n=3$) one indeed reproduces the value $c_v=k_B/2$ from MC calculations. For this agreement, it is essential that all $K$ non-flat modes outside the ground-state manifold are harmonic. If that was not the case, particularly, if a subset of the $K$ modes were quartic~\cite{Chalker1992Hiddenorder,Moessner1998propertiesclassSL} (which would then contribute $k_B/4$ to the specific heat), these modes would be selected within an order-by-disorder mechanism, in contradiction with our observations.

The dual trillium lattice constitutes an interesting platform for investigating novel types of classical spin liquids. We defer a more detailed investigation of this system to future work and now focus on the relevance of the distorted windmill lattice for \chem{Pb Cu Te_2 O_6}.

\section{Classical study of $\mathbf{PbCuTe_2O_6}$}\label{Sec:pbcuteo}

In this section, we focus on the classical version of the Heisenberg Hamiltonian for \chem{Pb Cu Te_2 O_6} in Eq.~\eqref{eq:Ham} with the couplings given in Table~\ref{tab:J} and the site positions from Table~\ref{tab:pos} with $y=-0.2258$. Besides an investigation of the magnetic properties of this model, we discuss to which extent previous experimental results on \chem{Pb Cu Te_2 O_6} can already be explained on a classical level, despite the $S=1/2$ quantum nature of this compound, following a similar strategy as in Refs.~\cite{Samarakoon2017ClassicalKitaev,Samarakoon2018quantclass,Hosoi2022Uncovering,Smith2022Case,Bhardwaj2022-sf,Zhang2019Dynamicalstruct,Pohle2021ca10,Bai2019,Franke2022}. We simulate $N=12\times L\times L \times L$ classical spins with periodic boundary conditions, using different analytical and numerical techniques. To characterize the magnetic order at $T=0$, we proceed similarly to the previous section by applying IM and LT. We then study the finite temperature behavior by means of MC simulations. Finally, we investigate the magnetic excitations, i.e., the dynamical spin structure factor in the paramagnetic regime using molecular dynamics (MD) simulations.

\subsection{Static magnetic order at zero and finite temperatures}

\begin{figure}[b]
\includegraphics[width=0.499\textwidth]{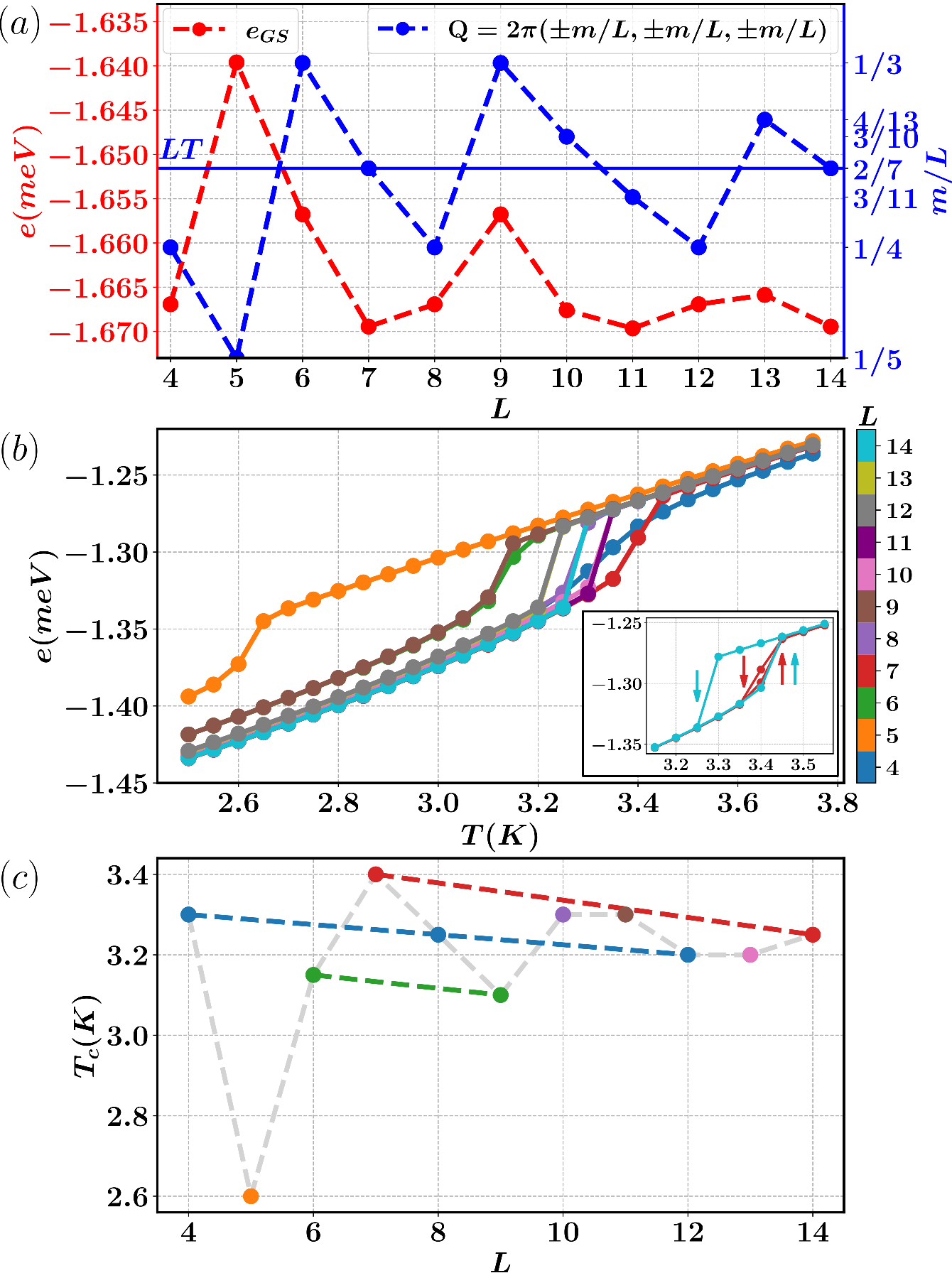}
 \caption{(a) Ground state energy per spin $e_{\text{GS}}$ (left) and ordering wave vector $\bm{Q} =2\pi (\pm m/L,\pm m/L,\pm m/L)$ (right) of the classical Heisenberg Hamiltonian for \chem{Pb Cu Te_2 O_6} obtained within IM as a function of the linear system size $L$. The horizontal blue line marks the optimal wave vector $\bm{Q} = 2\pi (\pm \xi,\pm \xi,\pm \xi)$ with $\xi=0.285756(1)$ found by means of LT. (b) Average energy per spin $e$ for different linear system sizes $L$ as a function of the temperature $T$, obtained on cooling by means of MC simulations. The energy drop indicates a phase transition towards a magnetically ordered state. The curve for $L=13$ is almost completely covered by the curve for $L=12$. Inset: Hysteresis loop for $L=7,14$. The arrow pointing down (up) indicates that the curve is obtained by decreasing (increasing) the temperature during a MC simulation. (c) Critical temperature from the MC simulations on cooling as a function of the linear size. Linear sizes with the same ordering vector [see (a)] are marked with the same color and connected by a dashed line. The light gray dashed line serves as a guide to the eye. }
\label{fig:eL}
\end{figure}

As indicated in the phase diagram in Fig.~\ref{fig:T0phase}, the classical Hamiltonian for \chem{Pb Cu Te_2 O_6} has a magnetically ordered ground state with an incommensurate wave vector $\bm{Q}$. Although this type of magnetic long-range order does not seem realized in \chem{Pb Cu Te_2 O_6}, we investigate its properties here, to complete our study of the classical model Hamiltonian for this compound. To identify the incommensurate order by means of IM we proceeded similarly to Ref.~\cite{Jin2020Classicalquantum}, by investigating different linear system sizes $L$. As reported in Fig.~\ref{fig:eL}(a), the ground state energy and the corresponding ordering wave vector $\bm{Q} = 2\pi(\pm m/L,\pm m/L,\pm m/L)$  with $m$ integer, depend on the simulated linear size $L$. The lowest energy is found for $L=11$ and the ordering wave vector for this system size is given by $\bm{Q} = 2\pi(\pm 3/11,\pm 3/11,\pm 3/11)$, where all combinations of signs occur and yield identical energies. For $L=7$ and $L=14$ we find only slightly larger ground state energies which differ from the one at $L=11$ only after the fourth decimal digit. The corresponding ordering wave vector for $L=7$ and $L=14$ is given by $\bm{Q} = 2\pi(\pm 2/7,\pm 2/7,\pm 2/7)$. Similar energies for these two wave vectors are expected, since $2/7$ and $3/11$ only differ by approximately $5\%$. These results are also consistent with LT, where the ordering wave vector $\bm{Q}$ is found to be $\bm{Q} = 2\pi (\pm \xi,\pm \xi,\pm \xi)$ where $\xi=0.285756(1)$, i.e., very close to $2/7\simeq0.285714$. The dependence of these results on the linear size and the presence of two almost degenerate estimates of the ground state energy indicate that the exact ground state is not reached by finite-size simulations. This is the case when $\bm{Q}$ approaches an irrational number, which corresponds to a ground state characterized by incommensurate magnetic order.

Next, we study the finite temperature behavior by means of MC simulations. By decreasing the temperature, the energy per spin as a function of temperature exhibits a drop that becomes steeper with increasing linear size $L$ [Fig.~\ref{fig:eL}(b)], which is a clear signature of a phase transition. The position of the energy drop, i.e., the critical temperature $T_c$ of the transition again strongly depends on the linear system size $L$, see Fig.~\ref{fig:eL}(c). This size dependence of $T_c$ is expected since different linear sizes realize different magnetic orders, corresponding to finite size approximations of the actual incommensurate order. For small values of $L$ the available $m/L$ values of the ordering vector can significantly differ from the actual incommensurate number, see for example the $\bm{Q}$ ordering vector for $L=5$ and the LT estimate of $\bm{Q}$ in Fig.~\ref{fig:eL}(a). The inability of the systems with these linear sizes to approximate the ground state results in a lower critical temperature and a higher energy below the transition, see Fig.~\ref{fig:eL}(b). By increasing the linear size, the $\bm{Q}$ ordering vector is discretized in smaller steps and the difference between $m/L$ and the actual incommensurate value decreases. Consequently, this effect is suppressed and the fluctuations of the critical temperature as a function of the linear size decrease, see Fig.~\ref{fig:eL}(c). By comparing the critical temperatures of the linear sizes characterized by the same ordering wave vector $\bm{Q}$, one recognizes another type of dependence on the linear size: The critical temperature $T_c$ decreases with increasing system sizes, see Fig.~\ref{fig:eL}(c). Such a behavior could be due to thermal hysteresis effects~\cite{Hopkinson2007ClassicalHyp,Isakov2008FateTrillium}, that are more pronounced with increasing the linear size~\cite{Masuda2012Hyst}. To verify this possibility, we perform simulations starting from a low-temperature configuration and gradually heating up the system for the two linear sizes $L=7,14$ with the same ordering vector. The curves obtained, respectively, by heating and cooling the system do not match around the critical temperature and the mismatch is larger for the largest size, confirming the presence of a hysteresis effect [Fig.~\ref{fig:eL}(b) inset]. The energy jump at the critical temperature and the thermal hysteresis effect are indications that the phase transition is of first order~\cite{Zheng1998ThermalHysteresis}.

Because of the strong dependence of the results on the linear system size, it is difficult to give a value of the critical temperature in the thermodynamic limit. Our best estimates are $T_c=3.25K$ ($L=14$) and $T_c=3.3K$ ($L=11$), obtained by taking into account the largest linear sizes with the lowest ground state energy.  

\begin{figure}[t!]
\includegraphics[width=0.48\textwidth]{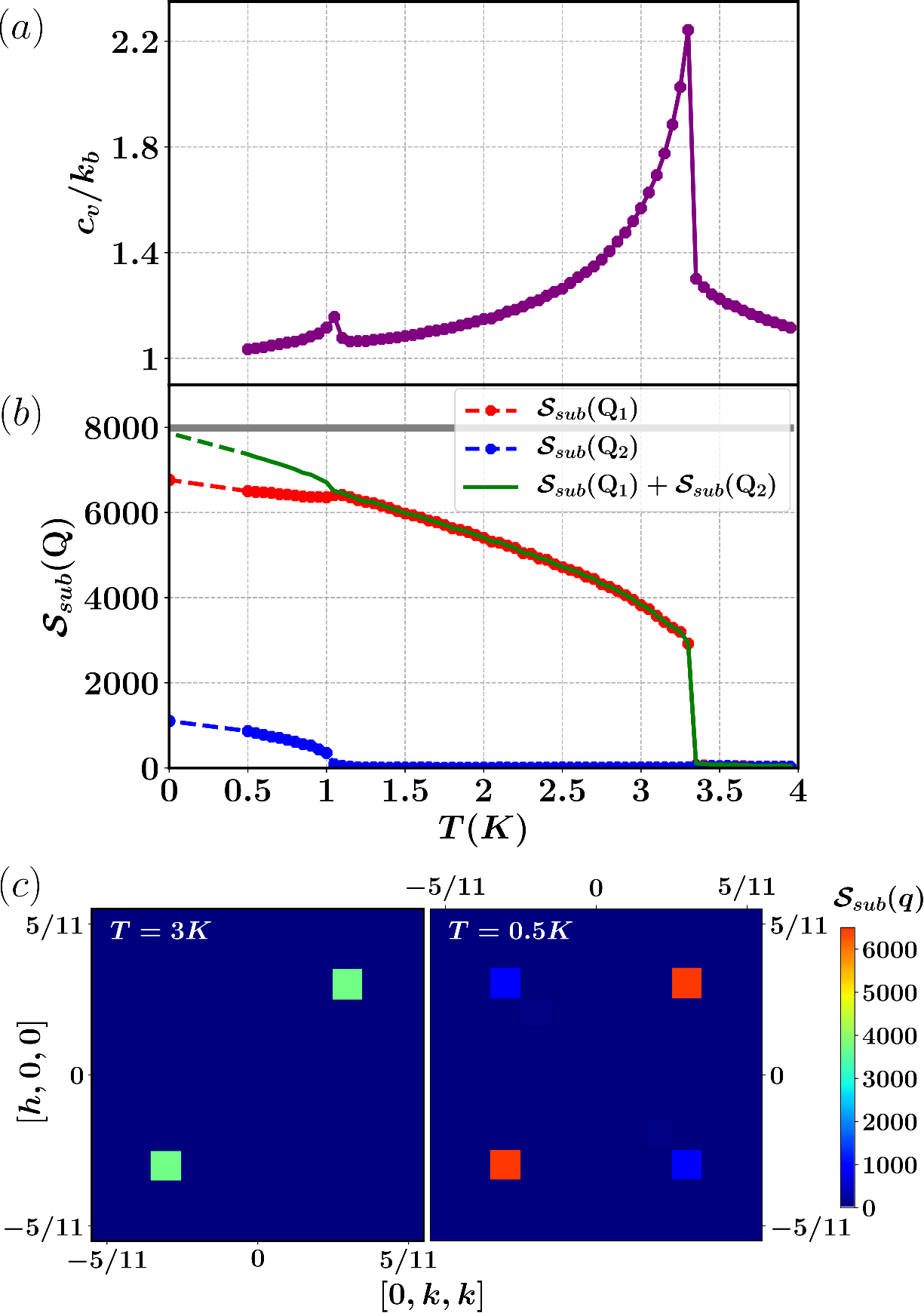}
\caption{\label{fig:smallpeak}(a) Specific heat for $L=11$ as a function of temperature for the classical Heisenberg Hamiltonian for \chem{Pb Cu Te_2 O_6}, obtained by means of MC simulations. The peak at $T_{c}=3.3K$ corresponds to the onset of the magnetic long-range order. The peak at $T_{c_2}=1.05K$ is associated with additional discrete lattice symmetry breaking, as discussed in the main text. (b) $\mathcal{S}_\text{sub}(\bm{q})$ at $\bm{q}=\bm{Q}_1$ and $\bm{q}=\bm{Q}_2$ as a function of the temperature, obtained by means of MC simulations at finite temperature and with IM at $T=0$. The vectors $\bm{Q}_1$ and $\bm{Q}_2$ are two distinct combinations of signs in the ordering wave vector $\bm{Q}= 2\pi (\pm 3/11,\pm3/11,\pm3/11)$. The gray line marks the total intensity $\sum_{\bm{q}\in\text{HBZ}}\mathcal{S}_\text{sub}(\bm{q})=N/2$ where the summation is only done over half of the first Brillouin zone (HBZ) since the other half is identical due to $\mathcal{S}_\text{sub}(\bm{q})=\mathcal{S}_\text{sub}(-\bm{q})$.
This value $N/2$ results from a sum rule and is independent of the temperature. 
(c) $\mathcal{S}_\text{sub}(\bm{q})$ in the $[h,k,k]$ plane, obtained from one MC run at $T=3K$ (left) and $T=0.5K$ (right). At $T_{c_2}<T=3K<T_{c}$ a peak emerges in $\mathcal{S}_\text{sub}(\bm{q})$ at $ \bm{Q}_1=\pm  2\pi (-3/11,3/11,3/11)$. At $T=0.5K<T_{c_2}$ a second less intense peak emerges in $\mathcal{S}_\text{sub}(\bm{q})$ at $ \bm{Q}_2=\pm  2\pi (3/11,3/11,3/11) $.}
\end{figure}

To study the magnetically ordered low-temperature regime in more detail, we focus on $L=11$ for the rest of this subsection. According to our previous results, this value provides the lowest ground state energy among all simulated system sizes. The specific heat as a function of temperature displays two peaks, one at $T_{c}=3.3K$ and another, less pronounced peak at $T_{c_2}=1.05K$ [Fig.~\ref{fig:smallpeak}(a)]. The peak at $T_c$ corresponds to the position of the energy drop in Fig.~\ref{fig:eL}(b) and marks the onset of magnetic long-range order. To investigate the origin of the second peak in the specific heat at $T_{c_2}$ we consider the Fourier transform of the spin-spin correlations $\mathcal{S}_\text{sub}(\bm{q})$ as defined in Eq.~\eqref{eq:Ssub}. (Since $\mathcal{S}_\text{sub}(\bm{q})$ only takes into account the spin-spin correlations within the same sublattice, it is periodic in momentum space with respect to the first Brillouin zone.) Particularly insightful is $\mathcal{S}_\text{sub}(\bm{q}=\bm{Q})$ as a function of temperature at the ordering wave vector $\bm{Q}$, which is $\bm{Q} =2\pi(\pm3/11,\pm3/11,\pm3/11)$ for $L=11$, where we now explicitly distinguish between different combinations of signs in this vector [see Fig.~\ref{fig:smallpeak}(b)].
The results at $T>0$ ($T=0$) in Fig.~\ref{fig:smallpeak}(b) are obtained by means of MC calculations (IM calculations). We find that the onset of magnetic order at $T_c$ is associated with the selection of a particular sign combination in $\bm{Q} =2\pi(\pm3/11,\pm3/11,\pm3/11)$ (due to the random initialization and spin updates within MC, this selection is also random in each MC run). Denoting this chosen wave vector as $\bm{Q}_1$, below $T_c$ we observe the onset of a finite weight in $\mathcal{S}_\text{sub}(\bm{q}=\bm{Q}_1)$. Note that, since Eq.~(\ref{eq:Ssub}) is invariant with respect to momentum inversion $\bm{q}\rightarrow-\bm{q}$ the same signal is also found at $-\bm{Q}_1$, i.e., $\mathcal{S}_\text{sub}(\bm{q}=\bm{Q}_1)=\mathcal{S}_\text{sub}(\bm{q}=-\bm{Q}_1)$. 
Below $T_{c_2}$, we find that, additionally, $\mathcal{S}_\text{sub}(\bm{q}=\bm{Q}_2)$ acquires a finite weight, where $\bm{Q}_2$ corresponds to {\it another} selected combination of signs in $\bm{Q} =2\pi(\pm3/11,\pm3/11,\pm3/11)$. The sum $\mathcal{S}_\text{sub}(\bm{Q}_1)+\mathcal{S}_\text{sub}(\bm{Q}_2)$ exhibits a small kink at $T_{c_2}$ and accounts for almost the total intensity at $T=0$. This indicates that the changes which the system undergoes at $T_{c_2}$ do not just correspond to a redistribution of signal in $\mathcal{S}_\text{sub}(\bm{q})$ but rather the generation of additional signal, pointing towards another magnetic ordering transition associated with the formation of enhanced static magnetic moments. To further illustrate the presence of the additional magnetic Bragg peak, in Fig.~\ref{fig:smallpeak}(c) we show $\mathcal{S}_\text{sub}(\bm{q})$ in a plane that contains both $\bm{Q}_1$ and $\bm{Q}_2$, at two different temperatures $T_{c_2}<T<T_{c}$ and $T<T_{c_2}$, from a single MC run. The fact that we observe this transition for all simulated system sizes, indicates that this is an intrinsic feature of the system, rather than a finite size effect.

The physical picture that emerges is the following: The spins undergo a two-stage ordering mechanism, in which at $T_c$ they first develop magnetic order characterized by a single ordering wave vector $\bm{Q}_1$. At the lower temperature $T_{c_2}$ additional long-range correlations described by another (symmetry related) wave vector $\bm{Q}_2$ set in. This corresponds to the loss of certain types of fluctuations that are still allowed for $T_{c_2} < T < T_{c}$. Since the presence of two peaks of unequal height at $\bm{Q}_1$ and $\bm{Q}_2$ represents a weight distribution in $\mathcal{S}_\text{sub}(\bm{q})$ with lower momentum space symmetries compared to the presence of just one peak at $\bm{Q}_1$, we expect that the magnetic order that develops at $T_{c_2}$ is associated with additional lattice symmetry breaking. A possible explanation for the occurrence of two consecutive phase transitions is provided by the LT approach. Within this method a single-$\bm{Q}$ magnetic ordering in the ground state is allowed, but with the caveat that the length constraint of individual spins is not fulfilled. At sufficiently large temperatures (i.e., in the temperature range $T_{c_2} < T < T_{c}$) a single-$\bm{Q}$ state can nevertheless be realized since thermal fluctuations reduce the static magnetic moment such that the spin's length constraint is effectively relaxed. With decreasing temperature and the loss of thermal fluctuations, however, the fulfillment of the individual length constraints eventually requires an additional ordering wave vector $\bm{Q}_2$ which leads to the observed second phase transition at $T_{c_2}$. Such a multi-$\bm{Q}$ ordering in the ground state is not common but can occur when the interactions are particularly complex and generate a high level of frustration~\cite{Okubo2011multiple-q,Shimokawa2019multiqhex}.

In summary, the classical Heisenberg model corresponding to \chem{Pb Cu Te_2 O_6} orders magnetically in a state with incommensurate $\bm{Q}$. This behavior is very different from the one obtained experimentally for \chem{Pb Cu Te_2 O_6}. In fact, both thermodynamic probes and inelastic neutron scattering (INS) experiments show no sign of magnetic order down to $T=0.01K$~\cite{Koteswararao2014Magprop,Chillal2020Evidence}. This indicates that quantum effects, completely neglected in the classical model, play an important role in determining the behavior of the real material at low temperatures. 

\subsection{Dynamics}
As pointed out in the previous section, the low-temperature magnetic behavior of the classical Heisenberg model in Eq.~(\ref{eq:Ham}) differs strongly from the one of the real material \chem{Pb Cu Te_2 O_6}: While the model Hamiltonian exhibits magnetic order below $T_{c}\simeq 3.3K$, the material shows no sign of a magnetic phase transition down to $T=0.01K$. These differences are also reflected in the low-temperature properties of the dynamical spin structure factor. A magnetically ordered classical spin system gives rise to well-defined dispersive spin wave features in the spin structure factor, in contrast to the dispersionless and broad features found in INS experiments on \chem{Pb Cu Te_2 O_6}~\cite{Chillal2020Evidence}. Thus, classical simulations performed at the same temperature as INS experiments (i.e., below the critical temperature $T_c$ of the classical model), do not reproduce the dynamical spin structure factors from experiments. On the other hand, in the paramagnetic regime at $T>T_c$ the classical model also shows broad and dispersionless features, due to the absence of magnetic order caused by thermal fluctuations. Consequently, in this regime, it is possible to attempt a comparison between the simulated classical model at suitably chosen temperatures $T>T_c$ and the experimental data. In the following, we will investigate whether thermal fluctuations at $T>T_c$ can indeed mimic the strong quantum fluctuations observed in \chem{Pb Cu Te_2 O_6} at low temperatures.

We compute the dynamical spin structure factor at $T>T_{c}$ by means of molecular dynamics (MD) simulations. Due to the absence of magnetic order in this temperature regime, possible finite size effects resulting from the incompatibility of the finite system and incommensurate ordering wave vectors are not present. As a consequence, we can fix the linear system size to a moderate (i.e., not too large) value $L=8$. For this system size, the critical temperature is given by $T_{c}=3.25 K$. MD simulations consist of numerically solving the classical equations of motion for the spins, where the starting spin configuration is taken from a snapshot of a MC run at a given temperature, when the system has reached thermal equilibrium. Further numerical details are given in Appendix~\ref{appendix:MD}. Within MD simulations the dynamical spin structure factor $\mathcal{S}(\bm{q},\omega)$ is accessible via the Fourier transform of the time-dependent spin-spin correlation function
\begin{equation}
\mathcal{S}(\bm{q},\omega) =\frac{\beta\omega\lvert{\mathcal{F}(\bm{q})}\rvert^2}{N\sqrt{N_t}}\sum_{n_t = 0}^{N_t}  e^{-i\omega n_t \delta t}  \sum_{i,j}^{N}  e^{i \bm{q} \cdot (\bm{r}_i-\bm{r}_j)} \bm{S}_{j}(t) \cdot \bm{S}_{i}(0),
\label{eq:Smunu}
\end{equation}
where $\bm{r}_i$ is the position of site $i$. Furthermore, $N_t$ is the number of time steps and $ \delta t$ is the time step size. The factors $\mathcal{F}(\bm{q})$ and $\beta \omega$ correspond respectively to the form factor for the \chem{Cu^{2+}} ions~\cite{brown2006magneticform} and a commonly used factor to take into account the differences between quantum and classical statistics~\cite{Zhang2019Dynamicalstruct}.

In the specific case of an isotropic Heisenberg model as the one under exam, $\mathcal{S}(\bm{q},\omega)$ in Eq.\eqref{eq:Smunu} corresponds to the \textit{perpendicular} structure factor, used to compare theoretical calculations with INS data~\cite{pohle2017spinCa,Zhang2019Dynamicalstruct}, because it takes into account the impossibility of neutrons to probe excitations with momentum parallel to the scattering direction. Also note that, since $\mathcal{S}(\bm{q},\omega)$ contains the actual site positions $\bm{r}_i$ which are non-integer multiples of the lattice vectors in direct space, this quantity is not periodic in the first Brillouin zone. 
To take into account the spin length $S=1/2$ of the \chem{Cu^{2+}} ions in our comparisons between theory and experiment, we rescale the frequency $\omega$ by a factor of $2$. This can be justified by linear spin wave theory where the spin wave energies are proportional to $S$ and match the classical ones from MD simulations (where spins are normalized as $|\bm{S}_i|=1$) when $S=1$. Consequently, to adjust the energy scales of a classical MD simulation with the energy scales of a spin-wave approach at $S=1/2$, we need to rescale our MD results according to $\omega\rightarrow\omega/2$. 

We compute the dynamical spin structure factor by means of MD simulations for four different temperatures $T=4K, 5K, 6K, 7K$ in the paramagnetic regime and compare it with INS data from Ref.~\cite{Chillal2020Evidence}. We emphasize that these temperatures are significantly smaller than the temperature corresponding to the typical energy scale of exchange couplings ($\sim 13$ K). Therefore, spins are already strongly correlated at such temperatures, defining the {\it correlated paramagnetic} regime.
\begin{figure*}
\includegraphics[width=0.98\textwidth]{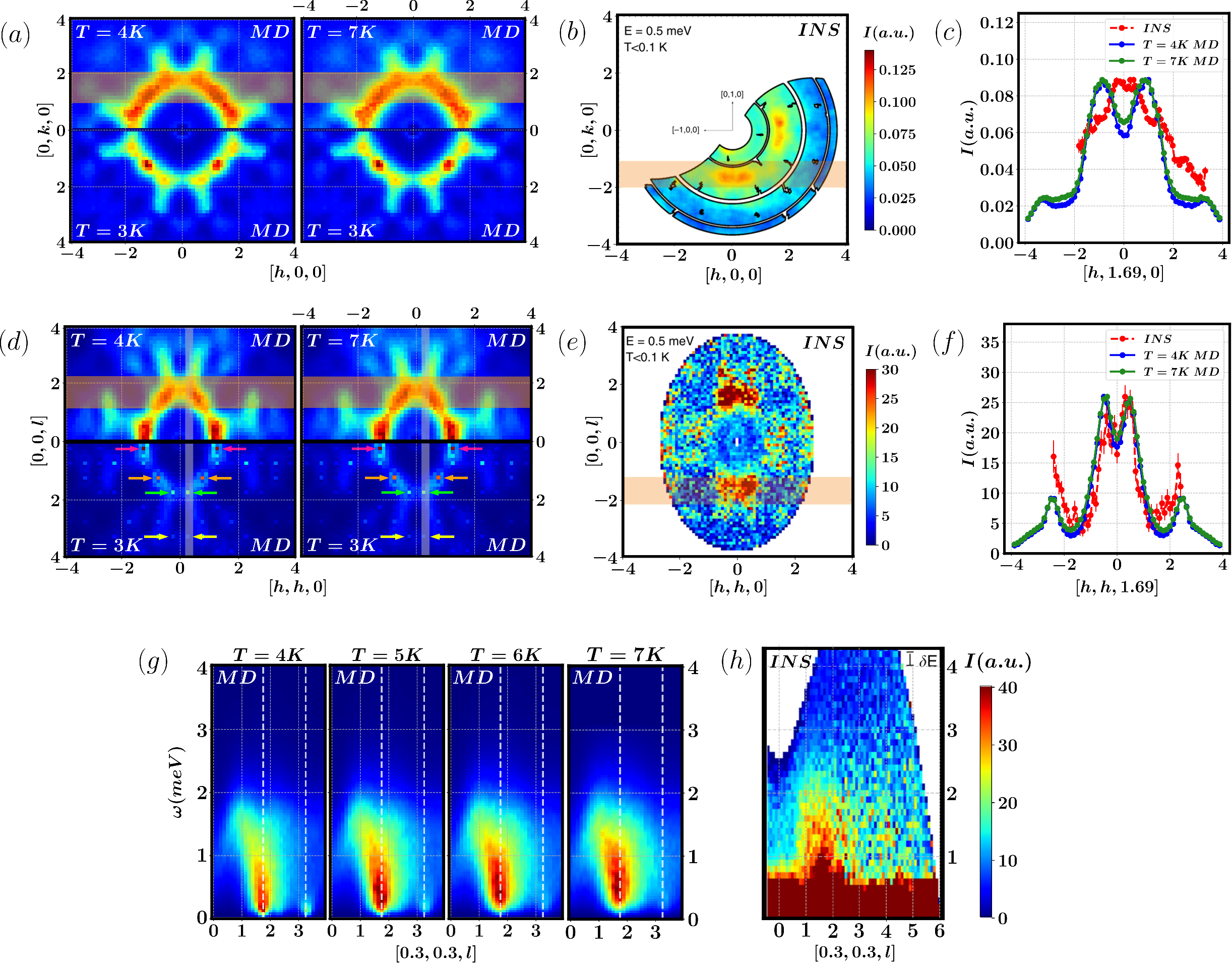}
\caption{\label{fig:sqw}(a) Dynamical spin structure factor from MD in the $[h,k,0]$ plane, integrated in the energy range $0.415\text{meV}\leq \omega \leq 0.604\text{meV}$, at $T=4K$ and $T=7K$ (top) compared with the one at $T=3K$ (bottom). (b) Dynamical structure factor at $T=0.1K$ from INS experiments in the $[h,k,0]$ plane, integrated in the energy range $0.4 \text{meV}\leq \omega \leq 0.6\text{meV}$~\cite{Chillal2020Evidence}. (c) Dynamical spin structure factor from INS and MD as a function of $h$ along the line cut $[h,1.69,0]$ where the signal is integrated in the perpendicular direction $[0,k,0]$ respectively over $1.125\leq k\leq 2$ (MD) and $1.1 \lesssim k \lesssim 2$ (INS), as indicated by the orange shaded region in (a) and (b). Subfigures (d) and (e) show the same as (a) and (b) but now in the $[h,h,l]$ plane. The colored arrows in the simulated data at $T=3K$ mark the Bragg peaks discussed in the main text. The experimental data are additionally integrated along the perpendicular direction $[h,-h,0]$ in the range $ -0.1 \leq h \leq 0.1$. (f) Dynamical spin structure factor from INS and MD as a function of $h$ along the line cut $[h,h,1.69]$ where the MD signal is integrated in the perpendicular direction $[0,0,l]$ respectively over $1.25\leq l\leq2.125$ (MD) and $1.2 \lesssim k \lesssim 2.1$ (INS), as indicated by the orange shaded region in (d) and (e). (g) From left to right: dynamical spin structure factor along the $[0.3,0.3,l]$ direction and as a function of frequency $\omega$, computed by means of MD simulations at $T=4K,5K,6K,7K$. The white dashed lines mark the positions of the Bragg peaks below $T_c$ at $(\pm 0.25,\pm 0.25, \pm 1.75)$ [green arrow in (d)] and $(\pm 0.25,\pm 0.25, \pm 3.25)$ [yellow arrow in (d)]. The simulated data are integrated along the perpendicular direction $[h,h,0]$ in the range $ 0.25\leq h \leq 0.375$ [white shaded area in (d)]. The finite resolution in energy of the INS data ($\delta E = 0.18 meV$) is added to the simulated data by means of a Gaussian broadening. (h) Dynamical spin structure factor along the $[0.3,0.3,l]$ direction from INS experiments at $T<0.1K$~\cite{Chillal2020Evidence}. The intensity at low frequencies is due to non-magnetic incoherent background and is not reproduced in MD simulations. In all plots, the simulated intensity is rescaled such that the maximum matches the one from the INS experiments.}
\end{figure*}
In Fig.~\ref{fig:sqw}(a) and (b) the comparison is shown for the $[h,k,0]$ plane while Fig.~\ref{fig:sqw}(d) and (e) present the comparison in the $[h,h,l]$ plane. The MD results in these plots are obtained at $T=4K$ and $T=7K$, while the INS data has been measured at $T < 0.1 K$. The simulated inelastic spin structure factor is also shown at $T=3K < T_c$ to highlight the changes which the system undergoes when entering the magnetically ordered phase.

In the $[h,k,0]$ plane [see Fig.~\ref{fig:sqw}(a), (b)], the MD simulations are able to reproduce the ring-shape features found in experiments. However, the regions of highest intensities on this ring do not occur around $(h,k,0)=(\pm2,0,0)$ and $(0,\pm2,0)$, as seen in the experimental data, but are shifted towards $(\pm1,\pm1,0)$. At $T=3K$, i.e., below the critical temperature, the inelastic structure factor continues to display similar broad and ring-shaped features. This is because the $[h,k,0]$ plane does not contain magnetic Bragg peaks and, consequently, the signal at $T=3K$ stems from the fluctuating part of the spins.

In the $[h,h,l]$ plane [see Fig.~\ref{fig:sqw}(d), (e)], the simulated inelastic spin structure factors at $T=4K$ and $T=7K$ again reproduce the overall ring-like distribution of signal observed in INS experiments, except for the region around $(\pm1,\pm1,0)$ which appears overemphasized in our simulations, as already mentioned above. Interestingly, even weaker features outside the ring are contained in our MD results in the $[h,h,l]$ plane such as two radial streaks at $l>2$ and a feature of enhanced intensity at $h \simeq 2$. Below $T_c$ the simulated magnetic structure factor shows multiple magnetic Bragg peaks in the $[h,h,l]$ plane, the one with the highest intensity is located at $(h,h,l)=(\pm 1.25,\pm 1.25, \pm 0.25)$ [pink arrow in Fig.~\ref{fig:sqw}(d)]. This peak position is consistent with our findings for the system's ground state order from IM in Fig.~\ref{fig:eL}(a) where an ordering wave vector $\bm{Q}=2\pi(\pm m/L,\pm m/L,\pm m/L)$ with $m/L=1/4$ was found for $L=8$. More precisely, the dominant magnetic Bragg peak at $(\pm 1.25,\pm 1.25, \pm 0.25)$ can be written as $(\pm (1+m/L),\pm (1+m/L), \pm m/L)$ where the integer parts correspond to reciprocal lattices vectors. This dominant peak also explains the strong simulated signal at $T>T_c$ in the nearby region around $(h,h,l)=(\pm1,\pm1,0)$ which can be interpreted as a molten remnant of the magnetic long-range order. In contrast, the INS data does not show a particularly strong signal around $(h,h,l)=(\pm 1.25,\pm 1.25, \pm 0.25)$. This indicates that despite the obvious similarities between the calculated and the measured data, the spin structure factor from INS cannot be simply interpreted as displaying smeared remnants of classical magnetic Bragg peaks. 

The plots in Fig.~\ref{fig:sqw}(c) and (f) show the same comparison between MD results and INS data but now along line cuts in the $[h,k,0]$ and $[h,h,l]$ planes as indicated by the shaded regions in Fig.~\ref{fig:sqw}(a), (b) and Fig.~\ref{fig:sqw}(d), (e), respectively. Furthermore, to match the processing of the INS data, the calculated spin structure factor has been integrated in a direction perpendicular to the line cut, as indicated by the width of the shaded regions. These plots demonstrate that in the temperature range $4K \leq T \leq 7K$ the width of simulated features in the spin structure factor approximately agrees with the typical broadening from quantum fluctuations in the INS data (overall, however, the changes of our results within this temperature range are rather small). They also confirm our observation that the best agreement between MD and INS data occurs in the $[h,h,l]$ plane away from the $(\pm1,\pm1,0)$ region, where the simulated results along the investigated line cut reproduce the experimental data even within error bars [Fig.~\ref{fig:sqw}(f)]. On the other hand, for the plotted line cut in the $[h,k,0]$ plane which contains contributions from the strong signal at $(\pm1,\pm1,0)$, larger deviations between theory and experiment are observed [Fig.~\ref{fig:sqw}(c)].

Finally, in Fig.~\ref{fig:sqw}(g) we report the simulated dynamical spin structure factor in a plane spanned by the frequency $\omega$ (vertical axis) and the $[0.3,0.3,l]$ momentum direction (horizontal axis). At all simulated temperatures $T=4K, 5K, 6K, 7K$, the MD results exhibit a rather featureless region of strong signal. With decreasing temperature, the features in the simulated data become slightly more distinct and the intensity shifts towards lower frequencies, reminiscent of spin waves. The wave vectors where the intensities are largest at low frequencies correspond to the magnetic Bragg peaks at $(\pm m/L,\pm m/L, \pm(2-m/L))=(\pm 0.25,\pm 0.25, \pm 1.75)$ and $(\pm m/L,\pm m/L, \pm(3+m/L))=(\pm 0.25,\pm 0.25, \pm 3.25)$ (where $m/L=1/4$ for the simulated system size $L=8$), see, respectively, green and yellow arrows in Fig.~\ref{fig:sqw}(d). The second peak is clearly less intense than the first peak, also due to the form factor in $\mathcal{S}(\bm{q},\omega)$ that reduces the intensity at large wave vectors. Comparing these observations with experimental results, INS data [Fig.~\ref{fig:sqw}(h)] show a similar dispersionless and smeared feature, whose intensity reaches up to maximal energies $1.5 \text{meV} \lesssim \omega \lesssim 2 \text{meV}$, in agreement with our calculations. The MD simulations are also able to reproduce this maximum intensity region in the correct momentum space position. More precisely, the INS data show large intensities at $(\pm 0.3,\pm 0.3,-1.67)$ which is compatible with the Bragg peak position at $(\pm 0.25,\pm 0.25, \pm 1.75)$ found in the classical simulations below $T_c$.

Overall, our classical simulations at finite temperatures correctly reproduce the main features of the dynamical spin structure factor of \chem{Pb Cu Te_2 O_6} measured at much lower temperatures and subject to strong quantum fluctuations. An exception is the region around $(h,k,l)=(\pm1,\pm1,0)$ where our calculated signal is comparably large. This region in momentum space is close to the dominant Bragg peak position observed in the magnetically ordered low-temperature regime of the classical model. This indicates that despite the demonstrated similar effects of thermal fluctuations in the classical model and quantum fluctuations in \chem{Pb Cu Te_2 O_6}, the spin structure factor of \chem{Pb Cu Te_2 O_6} cannot be purely explained by molten remnants of classical magnetic long-range order. On the other hand, the signal distribution in the measured spin structure factor of \chem{Pb Cu Te_2 O_6} {\it can be} consistently interpreted in terms of spin liquid behavior, as already demonstrated in previous works~\cite{Chillal2020Evidence,Chern2021PSGJ1J2}.

\section{Conclusions} 
In the first part of this work, we investigated the classical $T=0$ phase diagram of the hyper-hyperkagome $J_1=J_2$ Heisenberg model with varying chain interactions $J_3$ and $J_4$. We identified a rich variety of ground states in this phase diagram, with a subextensive degenerate manifold along the $J_3=J_4$ line, and an extensive manifold for the point $J_1=J_2=J_3=J_4$. We expect that, when extending the phase diagram for $J_3,J_4>J_1=J_2$, further subextensively degenerate lines can be found, which connect to the $J_1=J_2=J_3=J_4$ point. In fact, highly degenerate points are often conjunctions of subextensively degenerate lines in parameter space~\cite{Sklan2013Noncop,Balla2020phasediagfcc}. Our preliminary LT calculations, indeed, show that for the line cuts $J_1=J_2=J_3$, $J_4>J_1$ and $J_1=J_2=J_4$, $J_3>J_1$ the minima of the lowest energy band form a surface in momentum space. 

It remains an open question how the presented phase diagram changes when quantum fluctuations are taken into account.
The latter usually lift classical ground state degeneracies and stabilize new phases that do not have a classical counterpart~\cite{,Mambrini2006plaq,Yamamoto2014quantumphasediagTri,Iqbal2019quantclass}. 
The experimental results on \chem{Pb Cu Te_2 O_6} can be considered as a first indication of how quantum fluctuations affect the classical system. In fact, the results of the classical study presented in this work strongly differ from the quantum picture that has emerged from experiments. First, the classical model has a phase transition at around $T_c=3.3K$ towards a magnetically ordered state, that is absent in the real material. It is worth mentioning that further experimental works on \chem{Pb Cu Te_2 O_6} identified a ferroelectric transition at $T\simeq1K$~\cite{Hanna2021Crystalgrowth,Thurn2021ferroel}. For a rough estimate of whether this ferroelectric transition could be related to the magnetic transition we found at $T_c=3.3K$, one needs to adjust the temperature (energy) scales of our classical simulation with spins normalized to one ($|\bm{S}|=1$) and the $S=1/2$ degrees of freedom of the \chem{Cu^{2+}} magnetic ions. Since critical temperatures of magnetic transitions are expected to scale as $\sim S^2$, this means that the ferroelectric transition temperature has to be divided by $S^2=1/4$ to compare it with our simulated transition temperature (assuming that the ferroelectric transition was accompanied by magnetic order). 
This would give a critical temperature of $T=4 K$, that is higher than $T_{c}=3.3 K$ found for the classical model. If the observed ferroelectric transition was accompanied by magnetic order occurring at the same transition temperature, we would expect the classical model to predict a {\it higher} critical temperature than observed in experiments. This is because quantum fluctuations are expected to lower magnetic transition temperatures. Indeed, there is no evidence for magnetic long-range order at any temperature in \chem{Pb Cu Te_2 O_6} and quantum fluctuations have been experimentally demonstrated to play an important role in suppressing order.

Finally, our simulated dynamical spin structure factors in the paramagnetic regime show an overall good agreement with experimental data. This indicates that thermal fluctuations are able to reproduce the broad features in the experimental dynamical spin structure factor caused by quantum fluctuations. There are already existing examples for materials that show a good agreement between the dynamical structure factors from classical simulations and INS experiments at different temperatures~\cite{Zhang2019Dynamicalstruct,Pohle2021ca10}. However, in these cases, the behavior of the system does not change drastically when quantum fluctuations are taken into account. More specifically, no phase transition occurs between the temperature used in the simulations and the one at which the INS experiments are carried out. The temperature for the classical simulations is tuned to get the best match between simulated and measured dynamical structure factors. This usually results in a higher temperature with respect to the experimental one, in agreement with the fact that the materials are also subject to quantum fluctuations. In the present case, at the temperature of the INS experiments, the classical model is magnetically ordered, while the real material does not show any magnetic phase transition. Nevertheless, the dynamical spin structure factor in the paramagnetic regime of the classical model is able to reproduce the main features found by means of INS experiments. Although it is not possible to draw specific conclusions on the nature of quantum fluctuations based on this agreement, identifying and explaining such quantum-to-classical correspondences~\cite{Wang2020quantclasscorr} will remain an interesting research direction that may give new insights into the quantum ground states of strongly frustrated spin systems.

\begin{acknowledgments}
We thank S. Chillal, A. R. N. Hanna, and R. Moessner for their useful insights. The simulations were performed on the Tron cluster at the Physics Department of Freie Universit{\"a}t Berlin. B.L. acknowledges the support of Deutsche Forschungsgemeinschaft (DFG) through project B06 of SFB 1143: Correlated Magnetism: From Frustration To Topology (ID 247310070). 
\end{acknowledgments}

\appendix
\section{Luttinger-Tsiza (LT)}\label{appendix:LT}
To apply the LT method we proceed as follows. First we rewrite the Hamiltonian in Eq.~\eqref{eq:Ham} as
\begin{equation}
H = \frac{1}{2}  \sum_{\alpha,\beta}\sum_{i\in \alpha,j\in \beta} J_{i,j} (\Delta \bm{r}_{i,j})  \bm{S}_{i} \cdot \bm{S}_{j},
\label{eq:Hsub}
\end{equation}
where $i,j$ index the sites and $\alpha,\beta=1\ldots12$ index the sublattices. Furthermore, $\bm{r}_i$ is the position of the site $i$ and $\Delta \bm{r}_{i,j}=\bm{r}_i-\bm{r}_j$.

Then we define the Fourier transform of the spins on the individual sublattices
\begin{equation}
\tilde{ \bm{S}}_{\alpha} ( \bm{q}) = \frac{1}{\sqrt{N/12}} \sum_{i\in \alpha} e^{-i \bm{q} \cdot \bm{r}_i} \bm{S}_{i},
\label{eq:FTS}
\end{equation}
and its inverse
\begin{equation}
\bm{S}_{i\in \alpha} = \frac{1}{\sqrt{N/12}} \sum_{\bm{q}} e^{i \bm{q} \cdot \bm{r}_{i}} \tilde{ \bm{S}}( \bm{q}).
\label{eq:invFTS}
\end{equation}
Inserting Eq.~\eqref{eq:invFTS} into Eq.~\eqref{eq:Hsub} we obtain
\begin{equation}
H = \sum_{\bm{q}} \sum_{\alpha,\beta} \tilde J_{\alpha,\beta}( \bm{q})  \tilde{ \bm{S}}_{\alpha}( \bm{q}) \cdot \tilde{ \bm{S}}_{\beta}( \bm{q}),
\label{eq:HFT}
\end{equation}
where
\begin{equation}
\tilde J_{\alpha,\beta} ( \bm{q}) = \frac{1}{2}\sum_{i\in \alpha,j\in \beta}  J_{i,j} (\Delta \bm{r}_{i,j}) e^{i \bm{q} \cdot (\bm{r}_i-\bm{r}_j)},
\label{eq:Jmat}
\end{equation}
is the Fourier transform of the interaction matrix.

The ground state energy and the magnetic ordering vector correspond respectively to the minimum in $\bm{q}$ space of the lowest eigenvalue of the matrix $\tilde J_{\alpha,\beta} ( \bm{q})$ in Eq.~\eqref{eq:Jmat} and the wave vector $\bm{q}$ where this minimum is located. In the case of a non-Bravais lattice (such as our distorted windmill lattice), the LT method only provides a lower bound to the ground state energy and the predicted $\bm{Q}$ ordering may not correspond to the exact ground state order~\cite{Sklan2013noncoplanaroctahedra}. In fact, this procedure ensures that the spin length constraint $\lvert {\bm S}_{i} \rvert=1$ (the so-called strong constraint) is satisfied only globally, $\sum_i|\bm{S}_i|^2= N$ (which is referred to as the weak constraint). As a consequence, the obtained ground state could be ruled out when the strong constraint is also imposed.

In this work, we performed the diagonalization of $\tilde J_{\alpha,\beta} ( \bm{q})$ numerically: each component of $\bm{q} = (q_x,q_y,q_z)$ in Eq.~\eqref{eq:Jmat} is discretised in finite steps between $-\pi$ and $+\pi$, and for each value of $\bm{q}$ the diagonalization of the $12 \times 12$ matrix in Eq.~\eqref{eq:Jmat} is carried out.

\section{Iterative Minimisation (IM)}
\label{appendix:IM}
The IM method consists of finding the classical ground state, based on the property that in a ground state, all spins are aligned to the effective magnetic field created by the surrounding spins with which they interact. More precisely, we start by rewriting the Hamiltonian in Eq.~\eqref{eq:Ham} as
\begin{equation}
H = - \sum_i \bm{B}_{i}^{\text{eff}} \cdot \bm{S}_{i},
\label{eq:HB}
\end{equation}
where 
\begin{equation}
\bm{B}_{i}^{\text{eff}} = - \frac{1}{2} \sum_j J_{i,j} \bm{S}_{j}
\label{eq:Beff}
\end{equation}
is the effective local magnetic field to which the spin $\bm{S}_i$ is subject, due to the Heisenberg interactions with other spins. The procedure is simple, we first initialize a finite number of spins, in our case $N= 12\times L \times L \times L$ spins, in a random configuration. We then select a random site and align it to its effective magnetic field
\begin{equation}
\bm{S}_{i} \rightarrow \frac{\bm{B}_{i}^{\text{eff}}}{ \lvert \bm{B}_{i}^{\text{eff}}  \rvert},
\label{eq:IMmove}
\end{equation}
where the magnetic field $\bm{B}_{i}^{\text{eff}}$ on the right-hand side is divided by its norm to keep the spin length equal to one.
We iterate this procedure until the energy per spin does not change up to the $10^{th}$ decimal digit. The $\bm{Q}$ ordering wave vector is obtained by the wave vector $\bm{q}$ at which $\mathcal{S}_\text{sub}(\bm{q})$ in Eq.~\eqref{eq:Ssub} calculated from the resulting spin configuration has its maximum.

Coplanar configurations are characterized by the nematic order parameter~\cite{Zhitomirsky2008Octupolar}
\begin{equation}
Q^{\nu,\mu} = \frac{1}{N} \sum_i \left( \bm{S}^{\nu}_{i} \bm{S}^{\mu}_{i} - \frac{1}{3} \delta^{\nu,\mu} \right),
\label{eq:cop}    
\end{equation}
where $\nu,\mu={x,y,z}$.
In particular, the trace of the second moment of \eqref{eq:cop}
\begin{equation}
\sum_{\mu,\nu}  Q^{\mu,\nu} Q^{\nu,\mu} = \frac{1}{N^2} \sum_{i,j} \left( ( \bm{S}_{i} \cdot \bm{S}_{j})^2 - \frac{1}{3} \right),
\label{eq:order_cop}    
\end{equation}
is equal to $1/6$ in a coplanar configuration~\cite{Zhitomirsky2008Octupolar}. This quantity has been used to determine whether a spin configuration is coplanar or not.

We remark that IM does not have any control over whether it reaches the real ground state or a metastable state. Thus, one has to perform different attempts with the same linear system size $L$, and then check for convergence to the same result when varying the linear size $L$.

\section{Classical Monte Carlo (MC)}
\label{appendix:CMC}
We simulated the Heisenberg model given in Eq. \eqref{eq:Ham} on the distorted windmill lattice with periodic boundary conditions using different linear sizes $L$, which correspond to a total number of spins $N=12\times L \times L \times L$. A single MC step includes $N$ local heat-bath moves~\cite{Miyatake1986heatbath}, followed by $3$ overrelaxation moves~\cite{Creutz1987overrelax}. Each run is initialized with a random spin configuration and gradually cooled down from $T \simeq 2J$ to $T \simeq 0.02 J$, where $J$ corresponds to the largest coupling. Physical quantities are averaged over $2 \cdot 10^5$ MC steps, after $3 \cdot 10^5$ MC steps for thermalization.

\section{Molecular Dynamics (MD)}
\label{appendix:MD}
We performed molecular dynamics simulations to compute the Fourier transform of the time-dependent spin-spin correlation function $\mathcal{S}(\bm{q},\omega)$ defined in Eq. \eqref{eq:Smunu} at a fixed temperature. 
The starting configuration is taken from the MC simulations. The spins are then evolved in time according to the equation of motion
\begin{equation}
\frac{d\bm{S}_i}{dt}=\bm{S}_i \times \bm{B}_{i}^{\text{eff}},
\label{eq:motion}
\end{equation}
where $\bm{B}_{i}^{\text{eff}}$ is defined in Eq.~\eqref{eq:Beff}. This equation describes the precession of the spins around their local effective field. We numerically integrate Eq.~\eqref{eq:motion} with a fourth-order Runge-Kutta method. This procedure is repeated for different starting configurations at a fixed temperature and then averaged over 50 independent runs. To compute $\mathcal{S}(\bm{q},\omega)$, we employed the approach described in Ref.~\cite{Zhang2019Dynamicalstruct}. This consists of accumulating the time Fourier transform of each spin component during the simulation, rather than computing the time-dependent correlations and Fourier transform afterward.

\bibliography{apssamp}

\begin{thebibliography}{75}%
\makeatletter
\providecommand \@ifxundefined [1]{%
 \@ifx{#1\undefined}
}%
\providecommand \@ifnum [1]{%
 \ifnum #1\expandafter \@firstoftwo
 \else \expandafter \@secondoftwo
 \fi
}%
\providecommand \@ifx [1]{%
 \ifx #1\expandafter \@firstoftwo
 \else \expandafter \@secondoftwo
 \fi
}%
\providecommand \natexlab [1]{#1}%
\providecommand \enquote  [1]{``#1''}%
\providecommand \bibnamefont  [1]{#1}%
\providecommand \bibfnamefont [1]{#1}%
\providecommand \citenamefont [1]{#1}%
\providecommand \href@noop [0]{\@secondoftwo}%
\providecommand \href [0]{\begingroup \@sanitize@url \@href}%
\providecommand \@href[1]{\@@startlink{#1}\@@href}%
\providecommand \@@href[1]{\endgroup#1\@@endlink}%
\providecommand \@sanitize@url [0]{\catcode `\\12\catcode `\$12\catcode
  `\&12\catcode `\#12\catcode `\^12\catcode `\_12\catcode `\%12\relax}%
\providecommand \@@startlink[1]{}%
\providecommand \@@endlink[0]{}%
\providecommand \url  [0]{\begingroup\@sanitize@url \@url }%
\providecommand \@url [1]{\endgroup\@href {#1}{\urlprefix }}%
\providecommand \urlprefix  [0]{URL }%
\providecommand \Eprint [0]{\href }%
\providecommand \doibase [0]{https://doi.org/}%
\providecommand \selectlanguage [0]{\@gobble}%
\providecommand \bibinfo  [0]{\@secondoftwo}%
\providecommand \bibfield  [0]{\@secondoftwo}%
\providecommand \translation [1]{[#1]}%
\providecommand \BibitemOpen [0]{}%
\providecommand \bibitemStop [0]{}%
\providecommand \bibitemNoStop [0]{.\EOS\space}%
\providecommand \EOS [0]{\spacefactor3000\relax}%
\providecommand \BibitemShut  [1]{\csname bibitem#1\endcsname}%
\let\auto@bib@innerbib\@empty
\bibitem [{\citenamefont {Balents}(2010)}]{balents2010spin}%
  \BibitemOpen
  \bibfield  {author} {\bibinfo {author} {\bibfnamefont {L.}~\bibnamefont
  {Balents}},\ }\bibfield  {title} {\bibinfo {title} {{Spin liquids in
  frustrated magnets}},\ }\href {https://doi.org/10.1038/nature08917}
  {\bibfield  {journal} {\bibinfo  {journal} {Nature}\ }\textbf {\bibinfo
  {volume} {464}},\ \bibinfo {pages} {199} (\bibinfo {year}
  {2010})}\BibitemShut {NoStop}%
\bibitem [{\citenamefont {Ramirez}\ \emph {et~al.}(1999)\citenamefont
  {Ramirez}, \citenamefont {Hayashi}, \citenamefont {Cava}, \citenamefont
  {Siddharthan},\ and\ \citenamefont {Shastry}}]{ramirez1999zero}%
  \BibitemOpen
  \bibfield  {author} {\bibinfo {author} {\bibfnamefont {A.~P.}\ \bibnamefont
  {Ramirez}}, \bibinfo {author} {\bibfnamefont {A.}~\bibnamefont {Hayashi}},
  \bibinfo {author} {\bibfnamefont {R.~J.}\ \bibnamefont {Cava}}, \bibinfo
  {author} {\bibfnamefont {R.}~\bibnamefont {Siddharthan}},\ and\ \bibinfo
  {author} {\bibfnamefont {B.}~\bibnamefont {Shastry}},\ }\bibfield  {title}
  {\bibinfo {title} {{Zero-point entropy in ‘spin ice’}},\ }\href
  {https://doi.org/https://doi.org/10.1038/20619} {\bibfield  {journal}
  {\bibinfo  {journal} {Nature}\ }\textbf {\bibinfo {volume} {399}},\ \bibinfo
  {pages} {333} (\bibinfo {year} {1999})}\BibitemShut {NoStop}%
\bibitem [{\citenamefont {Chalker}(2011)}]{Chalker2011ChapterLacroix}%
  \BibitemOpen
  \bibfield  {author} {\bibinfo {author} {\bibfnamefont {J.~T.}\ \bibnamefont
  {Chalker}},\ }\bibinfo {title} {{Geometrically Frustrated Antiferromagnets:
  Statistical Mechanics and Dynamics}},\ in\ \href@noop {} {\emph {\bibinfo
  {booktitle} {{Introduction to Frustrated Magnetism: Materials, Experiments,
  Theory}}}},\ \bibinfo {editor} {edited by\ \bibinfo {editor} {\bibfnamefont
  {C.}~\bibnamefont {Lacroix}}, \bibinfo {editor} {\bibfnamefont
  {P.}~\bibnamefont {Mendels}},\ and\ \bibinfo {editor} {\bibfnamefont
  {F.}~\bibnamefont {Mila}}}\ (\bibinfo  {publisher} {Springer Berlin
  Heidelberg},\ \bibinfo {year} {2011})\ pp.\ \bibinfo {pages}
  {3--22}\BibitemShut {NoStop}%
\bibitem [{\citenamefont {Chalker}\ \emph {et~al.}(1992)\citenamefont
  {Chalker}, \citenamefont {Holdsworth},\ and\ \citenamefont
  {Shender}}]{Chalker1992Hiddenorder}%
  \BibitemOpen
  \bibfield  {author} {\bibinfo {author} {\bibfnamefont {J.~T.}\ \bibnamefont
  {Chalker}}, \bibinfo {author} {\bibfnamefont {P.~C.~W.}\ \bibnamefont
  {Holdsworth}},\ and\ \bibinfo {author} {\bibfnamefont {E.~F.}\ \bibnamefont
  {Shender}},\ }\bibfield  {title} {\bibinfo {title} {{Hidden order in a
  frustrated system: Properties of the {H}eisenberg Kagom\'e
  antiferromagnet}},\ }\href {https://doi.org/10.1103/PhysRevLett.68.855}
  {\bibfield  {journal} {\bibinfo  {journal} {Phys. Rev. Lett.}\ }\textbf
  {\bibinfo {volume} {68}},\ \bibinfo {pages} {855} (\bibinfo {year}
  {1992})}\BibitemShut {NoStop}%
\bibitem [{\citenamefont {Reimers}(1992)}]{Reimers1992Absence}%
  \BibitemOpen
  \bibfield  {author} {\bibinfo {author} {\bibfnamefont {J.~N.}\ \bibnamefont
  {Reimers}},\ }\bibfield  {title} {\bibinfo {title} {{Absence of long-range
  order in a three-dimensional geometrically frustrated antiferromagnet}},\
  }\href {https://doi.org/10.1103/PhysRevB.45.7287} {\bibfield  {journal}
  {\bibinfo  {journal} {Phys. Rev. B}\ }\textbf {\bibinfo {volume} {45}},\
  \bibinfo {pages} {7287} (\bibinfo {year} {1992})}\BibitemShut {NoStop}%
\bibitem [{\citenamefont {Petrenko}\ and\ \citenamefont
  {McK.~Paul}(2000)}]{Petrenko2000garnet}%
  \BibitemOpen
  \bibfield  {author} {\bibinfo {author} {\bibfnamefont {O.~A.}\ \bibnamefont
  {Petrenko}}\ and\ \bibinfo {author} {\bibfnamefont {D.}~\bibnamefont
  {McK.~Paul}},\ }\bibfield  {title} {\bibinfo {title} {{Classical {H}eisenberg
  antiferromagnet on a garnet lattice: A {M}onte {C}arlo simulation}},\ }\href
  {https://doi.org/10.1103/PhysRevB.63.024409} {\bibfield  {journal} {\bibinfo
  {journal} {Phys. Rev. B}\ }\textbf {\bibinfo {volume} {63}},\ \bibinfo
  {pages} {024409} (\bibinfo {year} {2000})}\BibitemShut {NoStop}%
\bibitem [{\citenamefont {Hopkinson}\ \emph {et~al.}(2007)\citenamefont
  {Hopkinson}, \citenamefont {Isakov}, \citenamefont {Kee},\ and\ \citenamefont
  {Kim}}]{Hopkinson2007ClassicalHyp}%
  \BibitemOpen
  \bibfield  {author} {\bibinfo {author} {\bibfnamefont {J.~M.}\ \bibnamefont
  {Hopkinson}}, \bibinfo {author} {\bibfnamefont {S.~V.}\ \bibnamefont
  {Isakov}}, \bibinfo {author} {\bibfnamefont {H.-Y.}\ \bibnamefont {Kee}},\
  and\ \bibinfo {author} {\bibfnamefont {Y.~B.}\ \bibnamefont {Kim}},\
  }\bibfield  {title} {\bibinfo {title} {{Classical Antiferromagnet on a
  Hyperkagome Lattice}},\ }\href
  {https://doi.org/10.1103/PhysRevLett.99.037201} {\bibfield  {journal}
  {\bibinfo  {journal} {Phys. Rev. Lett.}\ }\textbf {\bibinfo {volume} {99}},\
  \bibinfo {pages} {037201} (\bibinfo {year} {2007})}\BibitemShut {NoStop}%
\bibitem [{\citenamefont {Henley}(2010)}]{Henley2010Coulombphase}%
  \BibitemOpen
  \bibfield  {author} {\bibinfo {author} {\bibfnamefont {C.~L.}\ \bibnamefont
  {Henley}},\ }\bibfield  {title} {\bibinfo {title} {{The “{C}oulomb Phase”
  in Frustrated Systems}},\ }\href
  {https://doi.org/10.1146/annurev-conmatphys-070909-104138} {\bibfield
  {journal} {\bibinfo  {journal} {Annual Review of Condensed Matter Physics}\
  }\textbf {\bibinfo {volume} {1}},\ \bibinfo {pages} {179} (\bibinfo {year}
  {2010})}\BibitemShut {NoStop}%
\bibitem [{\citenamefont {Rehn}\ \emph {et~al.}(2016)\citenamefont {Rehn},
  \citenamefont {Sen}, \citenamefont {Damle},\ and\ \citenamefont
  {Moessner}}]{Rehn2016classicalSLHoney}%
  \BibitemOpen
  \bibfield  {author} {\bibinfo {author} {\bibfnamefont {J.}~\bibnamefont
  {Rehn}}, \bibinfo {author} {\bibfnamefont {A.}~\bibnamefont {Sen}}, \bibinfo
  {author} {\bibfnamefont {K.}~\bibnamefont {Damle}},\ and\ \bibinfo {author}
  {\bibfnamefont {R.}~\bibnamefont {Moessner}},\ }\bibfield  {title} {\bibinfo
  {title} {{Classical Spin Liquid on the Maximally Frustrated Honeycomb
  Lattice}},\ }\href {https://doi.org/10.1103/PhysRevLett.117.167201}
  {\bibfield  {journal} {\bibinfo  {journal} {Phys. Rev. Lett.}\ }\textbf
  {\bibinfo {volume} {117}},\ \bibinfo {pages} {167201} (\bibinfo {year}
  {2016})}\BibitemShut {NoStop}%
\bibitem [{\citenamefont {Reimers}\ \emph {et~al.}(1991)\citenamefont
  {Reimers}, \citenamefont {Berlinsky},\ and\ \citenamefont
  {Shi}}]{Reimers1991meafieldmagnorder}%
  \BibitemOpen
  \bibfield  {author} {\bibinfo {author} {\bibfnamefont {J.~N.}\ \bibnamefont
  {Reimers}}, \bibinfo {author} {\bibfnamefont {A.~J.}\ \bibnamefont
  {Berlinsky}},\ and\ \bibinfo {author} {\bibfnamefont {A.-C.}\ \bibnamefont
  {Shi}},\ }\bibfield  {title} {\bibinfo {title} {{Mean-field approach to
  magnetic ordering in highly frustrated pyrochlores}},\ }\href
  {https://doi.org/10.1103/PhysRevB.43.865} {\bibfield  {journal} {\bibinfo
  {journal} {Phys. Rev. B}\ }\textbf {\bibinfo {volume} {43}},\ \bibinfo
  {pages} {865} (\bibinfo {year} {1991})}\BibitemShut {NoStop}%
\bibitem [{\citenamefont {Moessner}\ and\ \citenamefont
  {Chalker}(1998{\natexlab{a}})}]{Moessner1998propertiesclassSL}%
  \BibitemOpen
  \bibfield  {author} {\bibinfo {author} {\bibfnamefont {R.}~\bibnamefont
  {Moessner}}\ and\ \bibinfo {author} {\bibfnamefont {J.~T.}\ \bibnamefont
  {Chalker}},\ }\bibfield  {title} {\bibinfo {title} {{Properties of a
  Classical Spin Liquid: The {H}eisenberg Pyrochlore Antiferromagnet}},\ }\href
  {https://doi.org/10.1103/PhysRevLett.80.2929} {\bibfield  {journal} {\bibinfo
   {journal} {Phys. Rev. Lett.}\ }\textbf {\bibinfo {volume} {80}},\ \bibinfo
  {pages} {2929} (\bibinfo {year} {1998}{\natexlab{a}})}\BibitemShut {NoStop}%
\bibitem [{\citenamefont {Villain}(1979)}]{villain1979insulating}%
  \BibitemOpen
  \bibfield  {author} {\bibinfo {author} {\bibfnamefont {J.}~\bibnamefont
  {Villain}},\ }\bibfield  {title} {\bibinfo {title} {{Insulating spin
  glasses}},\ }\href@noop {} {\bibfield  {journal} {\bibinfo  {journal}
  {Zeitschrift f{\"u}r Physik B Condensed Matter}\ }\textbf {\bibinfo {volume}
  {33}},\ \bibinfo {pages} {31} (\bibinfo {year} {1979})}\BibitemShut {NoStop}%
\bibitem [{\citenamefont {Hermele}\ \emph {et~al.}(2004)\citenamefont
  {Hermele}, \citenamefont {Fisher},\ and\ \citenamefont
  {Balents}}]{Hermele2004pyrophotons}%
  \BibitemOpen
  \bibfield  {author} {\bibinfo {author} {\bibfnamefont {M.}~\bibnamefont
  {Hermele}}, \bibinfo {author} {\bibfnamefont {M.~P.~A.}\ \bibnamefont
  {Fisher}},\ and\ \bibinfo {author} {\bibfnamefont {L.}~\bibnamefont
  {Balents}},\ }\bibfield  {title} {\bibinfo {title} {{Pyrochlore photons: The
  {$U(1)$} spin liquid in a {$S=\frac{1}{2}$} three-dimensional frustrated
  magnet}},\ }\href {https://doi.org/10.1103/PhysRevB.69.064404} {\bibfield
  {journal} {\bibinfo  {journal} {Phys. Rev. B}\ }\textbf {\bibinfo {volume}
  {69}},\ \bibinfo {pages} {064404} (\bibinfo {year} {2004})}\BibitemShut
  {NoStop}%
\bibitem [{\citenamefont {Benton}\ \emph {et~al.}(2012)\citenamefont {Benton},
  \citenamefont {Sikora},\ and\ \citenamefont {Shannon}}]{Benton2012light}%
  \BibitemOpen
  \bibfield  {author} {\bibinfo {author} {\bibfnamefont {O.}~\bibnamefont
  {Benton}}, \bibinfo {author} {\bibfnamefont {O.}~\bibnamefont {Sikora}},\
  and\ \bibinfo {author} {\bibfnamefont {N.}~\bibnamefont {Shannon}},\
  }\bibfield  {title} {\bibinfo {title} {{Seeing the light: Experimental
  signatures of emergent electromagnetism in a quantum spin ice}},\ }\href
  {https://doi.org/10.1103/PhysRevB.86.075154} {\bibfield  {journal} {\bibinfo
  {journal} {Phys. Rev. B}\ }\textbf {\bibinfo {volume} {86}},\ \bibinfo
  {pages} {075154} (\bibinfo {year} {2012})}\BibitemShut {NoStop}%
\bibitem [{\citenamefont {Shannon}\ \emph {et~al.}(2012)\citenamefont
  {Shannon}, \citenamefont {Sikora}, \citenamefont {Pollmann}, \citenamefont
  {Penc},\ and\ \citenamefont {Fulde}}]{Shannon2012quantumice}%
  \BibitemOpen
  \bibfield  {author} {\bibinfo {author} {\bibfnamefont {N.}~\bibnamefont
  {Shannon}}, \bibinfo {author} {\bibfnamefont {O.}~\bibnamefont {Sikora}},
  \bibinfo {author} {\bibfnamefont {F.}~\bibnamefont {Pollmann}}, \bibinfo
  {author} {\bibfnamefont {K.}~\bibnamefont {Penc}},\ and\ \bibinfo {author}
  {\bibfnamefont {P.}~\bibnamefont {Fulde}},\ }\bibfield  {title} {\bibinfo
  {title} {{Quantum Ice: A Quantum {M}onte {C}arlo Study}},\ }\href
  {https://doi.org/10.1103/PhysRevLett.108.067204} {\bibfield  {journal}
  {\bibinfo  {journal} {Phys. Rev. Lett.}\ }\textbf {\bibinfo {volume} {108}},\
  \bibinfo {pages} {067204} (\bibinfo {year} {2012})}\BibitemShut {NoStop}%
\bibitem [{\citenamefont {Huse}\ and\ \citenamefont
  {Rutenberg}(1992)}]{Huse1992Classicalkag}%
  \BibitemOpen
  \bibfield  {author} {\bibinfo {author} {\bibfnamefont {D.~A.}\ \bibnamefont
  {Huse}}\ and\ \bibinfo {author} {\bibfnamefont {A.~D.}\ \bibnamefont
  {Rutenberg}},\ }\bibfield  {title} {\bibinfo {title} {{Classical
  antiferromagnets on the Kagom\'e lattice}},\ }\href
  {https://doi.org/10.1103/PhysRevB.45.7536} {\bibfield  {journal} {\bibinfo
  {journal} {Phys. Rev. B}\ }\textbf {\bibinfo {volume} {45}},\ \bibinfo
  {pages} {7536} (\bibinfo {year} {1992})}\BibitemShut {NoStop}%
\bibitem [{\citenamefont {Garanin}\ and\ \citenamefont
  {Canals}(1999)}]{Garanin1999infinitecomp}%
  \BibitemOpen
  \bibfield  {author} {\bibinfo {author} {\bibfnamefont {D.~A.}\ \bibnamefont
  {Garanin}}\ and\ \bibinfo {author} {\bibfnamefont {B.}~\bibnamefont
  {Canals}},\ }\bibfield  {title} {\bibinfo {title} {{Classical spin liquid:
  Exact solution for the infinite-component antiferromagnetic model on the
  kagom\'e lattice}},\ }\href {https://doi.org/10.1103/PhysRevB.59.443}
  {\bibfield  {journal} {\bibinfo  {journal} {Phys. Rev. B}\ }\textbf {\bibinfo
  {volume} {59}},\ \bibinfo {pages} {443} (\bibinfo {year} {1999})}\BibitemShut
  {NoStop}%
\bibitem [{\citenamefont {Zhitomirsky}(2008)}]{Zhitomirsky2008Octupolar}%
  \BibitemOpen
  \bibfield  {author} {\bibinfo {author} {\bibfnamefont {M.~E.}\ \bibnamefont
  {Zhitomirsky}},\ }\bibfield  {title} {\bibinfo {title} {{Octupolar ordering
  of classical kagome antiferromagnets in two and three dimensions}},\ }\href
  {https://doi.org/10.1103/PhysRevB.78.094423} {\bibfield  {journal} {\bibinfo
  {journal} {Phys. Rev. B}\ }\textbf {\bibinfo {volume} {78}},\ \bibinfo
  {pages} {094423} (\bibinfo {year} {2008})}\BibitemShut {NoStop}%
\bibitem [{\citenamefont {Moessner}\ and\ \citenamefont
  {Chalker}(1998{\natexlab{b}})}]{Moessner1998lowtempproperties}%
  \BibitemOpen
  \bibfield  {author} {\bibinfo {author} {\bibfnamefont {R.}~\bibnamefont
  {Moessner}}\ and\ \bibinfo {author} {\bibfnamefont {J.~T.}\ \bibnamefont
  {Chalker}},\ }\bibfield  {title} {\bibinfo {title} {{Low-temperature
  properties of classical geometrically frustrated antiferromagnets}},\ }\href
  {https://doi.org/10.1103/PhysRevB.58.12049} {\bibfield  {journal} {\bibinfo
  {journal} {Phys. Rev. B}\ }\textbf {\bibinfo {volume} {58}},\ \bibinfo
  {pages} {12049} (\bibinfo {year} {1998}{\natexlab{b}})}\BibitemShut {NoStop}%
\bibitem [{\citenamefont {Sachdev}(1992)}]{Sachdev1992Kagome}%
  \BibitemOpen
  \bibfield  {author} {\bibinfo {author} {\bibfnamefont {S.}~\bibnamefont
  {Sachdev}},\ }\bibfield  {title} {\bibinfo {title} {{Kagom\'e- and
  triangular-lattice {H}eisenberg antiferromagnets: Ordering from quantum
  fluctuations and quantum-disordered ground states with unconfined bosonic
  spinons}},\ }\href {https://doi.org/10.1103/PhysRevB.45.12377} {\bibfield
  {journal} {\bibinfo  {journal} {Phys. Rev. B}\ }\textbf {\bibinfo {volume}
  {45}},\ \bibinfo {pages} {12377} (\bibinfo {year} {1992})}\BibitemShut
  {NoStop}%
\bibitem [{\citenamefont {Yan}\ \emph {et~al.}(2011)\citenamefont {Yan},
  \citenamefont {Huse},\ and\ \citenamefont {White}}]{Yan2011KagomeDMRG}%
  \BibitemOpen
  \bibfield  {author} {\bibinfo {author} {\bibfnamefont {S.}~\bibnamefont
  {Yan}}, \bibinfo {author} {\bibfnamefont {D.~A.}\ \bibnamefont {Huse}},\ and\
  \bibinfo {author} {\bibfnamefont {S.~R.}\ \bibnamefont {White}},\ }\bibfield
  {title} {\bibinfo {title} {{Spin-Liquid Ground State of the {$S=1/2$} Kagome
  {H}eisenberg Antiferromagnet}},\ }\href
  {https://doi.org/10.1126/science.1201080} {\bibfield  {journal} {\bibinfo
  {journal} {Science}\ }\textbf {\bibinfo {volume} {332}},\ \bibinfo {pages}
  {1173} (\bibinfo {year} {2011})}\BibitemShut {NoStop}%
\bibitem [{\citenamefont {Han}\ \emph {et~al.}(2012)\citenamefont {Han},
  \citenamefont {Helton}, \citenamefont {Chu}, \citenamefont {Nocera},
  \citenamefont {Rodriguez-Rivera}, \citenamefont {Broholm},\ and\
  \citenamefont {Lee}}]{han2012fractionalized}%
  \BibitemOpen
  \bibfield  {author} {\bibinfo {author} {\bibfnamefont {T.-H.}\ \bibnamefont
  {Han}}, \bibinfo {author} {\bibfnamefont {J.~S.}\ \bibnamefont {Helton}},
  \bibinfo {author} {\bibfnamefont {S.}~\bibnamefont {Chu}}, \bibinfo {author}
  {\bibfnamefont {D.~G.}\ \bibnamefont {Nocera}}, \bibinfo {author}
  {\bibfnamefont {J.~A.}\ \bibnamefont {Rodriguez-Rivera}}, \bibinfo {author}
  {\bibfnamefont {C.}~\bibnamefont {Broholm}},\ and\ \bibinfo {author}
  {\bibfnamefont {Y.~S.}\ \bibnamefont {Lee}},\ }\bibfield  {title} {\bibinfo
  {title} {{Fractionalized excitations in the spin-liquid state of a
  kagome-lattice antiferromagnet}},\ }\href
  {https://doi.org/https://doi.org/10.1038/nature11659} {\bibfield  {journal}
  {\bibinfo  {journal} {Nature}\ }\textbf {\bibinfo {volume} {492}},\ \bibinfo
  {pages} {406} (\bibinfo {year} {2012})}\BibitemShut {NoStop}%
\bibitem [{\citenamefont {Taillefumier}\ \emph {et~al.}(2017)\citenamefont
  {Taillefumier}, \citenamefont {Benton}, \citenamefont {Yan}, \citenamefont
  {Jaubert},\ and\ \citenamefont {Shannon}}]{Taillefumier2017CompetingSL}%
  \BibitemOpen
  \bibfield  {author} {\bibinfo {author} {\bibfnamefont {M.}~\bibnamefont
  {Taillefumier}}, \bibinfo {author} {\bibfnamefont {O.}~\bibnamefont
  {Benton}}, \bibinfo {author} {\bibfnamefont {H.}~\bibnamefont {Yan}},
  \bibinfo {author} {\bibfnamefont {L.~D.~C.}\ \bibnamefont {Jaubert}},\ and\
  \bibinfo {author} {\bibfnamefont {N.}~\bibnamefont {Shannon}},\ }\bibfield
  {title} {\bibinfo {title} {{Competing Spin Liquids and Hidden Spin-Nematic
  Order in Spin Ice with Frustrated Transverse Exchange}},\ }\href
  {https://doi.org/10.1103/PhysRevX.7.041057} {\bibfield  {journal} {\bibinfo
  {journal} {Phys. Rev. X}\ }\textbf {\bibinfo {volume} {7}},\ \bibinfo {pages}
  {041057} (\bibinfo {year} {2017})}\BibitemShut {NoStop}%
\bibitem [{\citenamefont {Benton}\ \emph {et~al.}(2018)\citenamefont {Benton},
  \citenamefont {Jaubert}, \citenamefont {Singh}, \citenamefont {Oitmaa},\ and\
  \citenamefont {Shannon}}]{Benton2018QuantumSL}%
  \BibitemOpen
  \bibfield  {author} {\bibinfo {author} {\bibfnamefont {O.}~\bibnamefont
  {Benton}}, \bibinfo {author} {\bibfnamefont {L.~D.~C.}\ \bibnamefont
  {Jaubert}}, \bibinfo {author} {\bibfnamefont {R.~R.~P.}\ \bibnamefont
  {Singh}}, \bibinfo {author} {\bibfnamefont {J.}~\bibnamefont {Oitmaa}},\ and\
  \bibinfo {author} {\bibfnamefont {N.}~\bibnamefont {Shannon}},\ }\bibfield
  {title} {\bibinfo {title} {{Quantum Spin Ice with Frustrated Transverse
  Exchange: From a $\ensuremath{\pi}$-Flux Phase to a Nematic Quantum Spin
  Liquid}},\ }\href {https://doi.org/10.1103/PhysRevLett.121.067201} {\bibfield
   {journal} {\bibinfo  {journal} {Phys. Rev. Lett.}\ }\textbf {\bibinfo
  {volume} {121}},\ \bibinfo {pages} {067201} (\bibinfo {year}
  {2018})}\BibitemShut {NoStop}%
\bibitem [{\citenamefont {Hagym\'asi}\ \emph {et~al.}(2021)\citenamefont
  {Hagym\'asi}, \citenamefont {Sch\"afer}, \citenamefont {Moessner},\ and\
  \citenamefont {Luitz}}]{Hagymasi2021possibleinv}%
  \BibitemOpen
  \bibfield  {author} {\bibinfo {author} {\bibfnamefont {I.}~\bibnamefont
  {Hagym\'asi}}, \bibinfo {author} {\bibfnamefont {R.}~\bibnamefont
  {Sch\"afer}}, \bibinfo {author} {\bibfnamefont {R.}~\bibnamefont
  {Moessner}},\ and\ \bibinfo {author} {\bibfnamefont {D.~J.}\ \bibnamefont
  {Luitz}},\ }\bibfield  {title} {\bibinfo {title} {{Possible Inversion
  Symmetry Breaking in the {$S=1/2$} Pyrochlore {H}eisenberg Magnet}},\ }\href
  {https://doi.org/10.1103/PhysRevLett.126.117204} {\bibfield  {journal}
  {\bibinfo  {journal} {Phys. Rev. Lett.}\ }\textbf {\bibinfo {volume} {126}},\
  \bibinfo {pages} {117204} (\bibinfo {year} {2021})}\BibitemShut {NoStop}%
\bibitem [{\citenamefont {Schäfer}\ \emph {et~al.}(2022)\citenamefont
  {Schäfer}, \citenamefont {Placke}, \citenamefont {Benton},\ and\
  \citenamefont {Moessner}}]{schafer2022abundancehardhex}%
  \BibitemOpen
  \bibfield  {author} {\bibinfo {author} {\bibfnamefont {R.}~\bibnamefont
  {Schäfer}}, \bibinfo {author} {\bibfnamefont {B.}~\bibnamefont {Placke}},
  \bibinfo {author} {\bibfnamefont {O.}~\bibnamefont {Benton}},\ and\ \bibinfo
  {author} {\bibfnamefont {R.}~\bibnamefont {Moessner}},\ }\href@noop {}
  {\bibinfo {title} {{Abundance of hard-hexagon crystals in the quantum
  pyrochlore antiferromagnet}}} (\bibinfo {year} {2022}),\ \Eprint
  {https://arxiv.org/abs/2210.07235} {arXiv:2210.07235 [cond-mat.str-el]}
  \BibitemShut {NoStop}%
\bibitem [{\citenamefont {Astrakhantsev}\ \emph {et~al.}(2021)\citenamefont
  {Astrakhantsev}, \citenamefont {Westerhout}, \citenamefont {Tiwari},
  \citenamefont {Choo}, \citenamefont {Chen}, \citenamefont {Fischer},
  \citenamefont {Carleo},\ and\ \citenamefont
  {Neupert}}]{Astrakhantsev2021brokensymm}%
  \BibitemOpen
  \bibfield  {author} {\bibinfo {author} {\bibfnamefont {N.}~\bibnamefont
  {Astrakhantsev}}, \bibinfo {author} {\bibfnamefont {T.}~\bibnamefont
  {Westerhout}}, \bibinfo {author} {\bibfnamefont {A.}~\bibnamefont {Tiwari}},
  \bibinfo {author} {\bibfnamefont {K.}~\bibnamefont {Choo}}, \bibinfo {author}
  {\bibfnamefont {A.}~\bibnamefont {Chen}}, \bibinfo {author} {\bibfnamefont
  {M.~H.}\ \bibnamefont {Fischer}}, \bibinfo {author} {\bibfnamefont
  {G.}~\bibnamefont {Carleo}},\ and\ \bibinfo {author} {\bibfnamefont
  {T.}~\bibnamefont {Neupert}},\ }\bibfield  {title} {\bibinfo {title}
  {{Broken-Symmetry Ground States of the {H}eisenberg Model on the Pyrochlore
  Lattice}},\ }\href {https://doi.org/10.1103/PhysRevX.11.041021} {\bibfield
  {journal} {\bibinfo  {journal} {Phys. Rev. X}\ }\textbf {\bibinfo {volume}
  {11}},\ \bibinfo {pages} {041021} (\bibinfo {year} {2021})}\BibitemShut
  {NoStop}%
\bibitem [{\citenamefont {Hering}\ \emph {et~al.}(2022)\citenamefont {Hering},
  \citenamefont {Noculak}, \citenamefont {Ferrari}, \citenamefont {Iqbal},\
  and\ \citenamefont {Reuther}}]{Hering2022}%
  \BibitemOpen
  \bibfield  {author} {\bibinfo {author} {\bibfnamefont {M.}~\bibnamefont
  {Hering}}, \bibinfo {author} {\bibfnamefont {V.}~\bibnamefont {Noculak}},
  \bibinfo {author} {\bibfnamefont {F.}~\bibnamefont {Ferrari}}, \bibinfo
  {author} {\bibfnamefont {Y.}~\bibnamefont {Iqbal}},\ and\ \bibinfo {author}
  {\bibfnamefont {J.}~\bibnamefont {Reuther}},\ }\bibfield  {title} {\bibinfo
  {title} {{Dimerization tendencies of the pyrochlore {H}eisenberg
  antiferromagnet: A functional renormalization group perspective}},\ }\href
  {https://doi.org/10.1103/PhysRevB.105.054426} {\bibfield  {journal} {\bibinfo
   {journal} {Phys. Rev. B}\ }\textbf {\bibinfo {volume} {105}},\ \bibinfo
  {pages} {054426} (\bibinfo {year} {2022})}\BibitemShut {NoStop}%
\bibitem [{\citenamefont {Nakamura}\ \emph
  {et~al.}(1997{\natexlab{a}})\citenamefont {Nakamura}, \citenamefont
  {Yoshimoto}, \citenamefont {Shiga}, \citenamefont {Nishi},\ and\
  \citenamefont {Kakurai}}]{Nakamura1997distwind}%
  \BibitemOpen
  \bibfield  {author} {\bibinfo {author} {\bibfnamefont {H.}~\bibnamefont
  {Nakamura}}, \bibinfo {author} {\bibfnamefont {K.}~\bibnamefont {Yoshimoto}},
  \bibinfo {author} {\bibfnamefont {M.}~\bibnamefont {Shiga}}, \bibinfo
  {author} {\bibfnamefont {M.}~\bibnamefont {Nishi}},\ and\ \bibinfo {author}
  {\bibfnamefont {K.}~\bibnamefont {Kakurai}},\ }\bibfield  {title} {\bibinfo
  {title} {{Strong antiferromagnetic spin fluctuations and the quantum
  spin-liquid state in geometrically frustrated
  $\ensuremath{\beta}\ensuremath{-}\mathrm{Mn}$, and the transition to a
  spin-glass state caused by non-magnetic impurity}},\ }\href
  {https://doi.org/10.1088/0953-8984/9/22/022} {\bibfield  {journal} {\bibinfo
  {journal} {Journal of Physics: Condensed Matter}\ }\textbf {\bibinfo {volume}
  {9}},\ \bibinfo {pages} {4701} (\bibinfo {year}
  {1997}{\natexlab{a}})}\BibitemShut {NoStop}%
\bibitem [{\citenamefont {Canals}\ and\ \citenamefont
  {Lacroix}(2000)}]{Canals200bMnmeanfield}%
  \BibitemOpen
  \bibfield  {author} {\bibinfo {author} {\bibfnamefont {B.}~\bibnamefont
  {Canals}}\ and\ \bibinfo {author} {\bibfnamefont {C.}~\bibnamefont
  {Lacroix}},\ }\bibfield  {title} {\bibinfo {title} {{Mean-field study of the
  disordered ground state in the $\ensuremath{\beta}\ensuremath{-}\mathrm{Mn}$
  lattice}},\ }\href {https://doi.org/10.1103/PhysRevB.61.11251} {\bibfield
  {journal} {\bibinfo  {journal} {Phys. Rev. B}\ }\textbf {\bibinfo {volume}
  {61}},\ \bibinfo {pages} {11251} (\bibinfo {year} {2000})}\BibitemShut
  {NoStop}%
\bibitem [{\citenamefont {Isakov}\ \emph {et~al.}(2008)\citenamefont {Isakov},
  \citenamefont {Hopkinson},\ and\ \citenamefont
  {Kee}}]{Isakov2008FateTrillium}%
  \BibitemOpen
  \bibfield  {author} {\bibinfo {author} {\bibfnamefont {S.~V.}\ \bibnamefont
  {Isakov}}, \bibinfo {author} {\bibfnamefont {J.~M.}\ \bibnamefont
  {Hopkinson}},\ and\ \bibinfo {author} {\bibfnamefont {H.-Y.}\ \bibnamefont
  {Kee}},\ }\bibfield  {title} {\bibinfo {title} {{Fate of partial order on
  trillium and distorted windmill lattices}},\ }\href
  {https://doi.org/10.1103/PhysRevB.78.014404} {\bibfield  {journal} {\bibinfo
  {journal} {Phys. Rev. B}\ }\textbf {\bibinfo {volume} {78}},\ \bibinfo
  {pages} {014404} (\bibinfo {year} {2008})}\BibitemShut {NoStop}%
\bibitem [{\citenamefont {Bergholtz}\ \emph {et~al.}(2010)\citenamefont
  {Bergholtz}, \citenamefont {L\"auchli},\ and\ \citenamefont
  {Moessner}}]{Bergholtz2010SymmBreakHyp}%
  \BibitemOpen
  \bibfield  {author} {\bibinfo {author} {\bibfnamefont {E.~J.}\ \bibnamefont
  {Bergholtz}}, \bibinfo {author} {\bibfnamefont {A.~M.}\ \bibnamefont
  {L\"auchli}},\ and\ \bibinfo {author} {\bibfnamefont {R.}~\bibnamefont
  {Moessner}},\ }\bibfield  {title} {\bibinfo {title} {{Symmetry Breaking on
  the Three-Dimensional Hyperkagome Lattice of
  ${\mathrm{Na}}_{4}{\mathrm{Ir}}_{3}{\mathrm{O}}_{8}$}},\ }\href
  {https://doi.org/10.1103/PhysRevLett.105.237202} {\bibfield  {journal}
  {\bibinfo  {journal} {Phys. Rev. Lett.}\ }\textbf {\bibinfo {volume} {105}},\
  \bibinfo {pages} {237202} (\bibinfo {year} {2010})}\BibitemShut {NoStop}%
\bibitem [{\citenamefont {Schiffer}\ \emph {et~al.}(1995)\citenamefont
  {Schiffer}, \citenamefont {Ramirez}, \citenamefont {Huse}, \citenamefont
  {Gammel}, \citenamefont {Yaron}, \citenamefont {Bishop},\ and\ \citenamefont
  {Valentino}}]{Schiffer1995frustrGa}%
  \BibitemOpen
  \bibfield  {author} {\bibinfo {author} {\bibfnamefont {P.}~\bibnamefont
  {Schiffer}}, \bibinfo {author} {\bibfnamefont {A.~P.}\ \bibnamefont
  {Ramirez}}, \bibinfo {author} {\bibfnamefont {D.~A.}\ \bibnamefont {Huse}},
  \bibinfo {author} {\bibfnamefont {P.~L.}\ \bibnamefont {Gammel}}, \bibinfo
  {author} {\bibfnamefont {U.}~\bibnamefont {Yaron}}, \bibinfo {author}
  {\bibfnamefont {D.~J.}\ \bibnamefont {Bishop}},\ and\ \bibinfo {author}
  {\bibfnamefont {A.~J.}\ \bibnamefont {Valentino}},\ }\bibfield  {title}
  {\bibinfo {title} {{Frustration Induced Spin Freezing in a Site-Ordered
  Magnet: Gadolinium Gallium Garnet}},\ }\href
  {https://doi.org/10.1103/PhysRevLett.74.2379} {\bibfield  {journal} {\bibinfo
   {journal} {Phys. Rev. Lett.}\ }\textbf {\bibinfo {volume} {74}},\ \bibinfo
  {pages} {2379} (\bibinfo {year} {1995})}\BibitemShut {NoStop}%
\bibitem [{\citenamefont {Nakamura}\ \emph
  {et~al.}(1997{\natexlab{b}})\citenamefont {Nakamura}, \citenamefont
  {Yoshimoto}, \citenamefont {Shiga}, \citenamefont {Nishi},\ and\
  \citenamefont {Kakurai}}]{Nakamura1997BetaMn}%
  \BibitemOpen
  \bibfield  {author} {\bibinfo {author} {\bibfnamefont {H.}~\bibnamefont
  {Nakamura}}, \bibinfo {author} {\bibfnamefont {K.}~\bibnamefont {Yoshimoto}},
  \bibinfo {author} {\bibfnamefont {M.}~\bibnamefont {Shiga}}, \bibinfo
  {author} {\bibfnamefont {M.}~\bibnamefont {Nishi}},\ and\ \bibinfo {author}
  {\bibfnamefont {K.}~\bibnamefont {Kakurai}},\ }\bibfield  {title} {\bibinfo
  {title} {{Strong antiferromagnetic spin fluctuations and the quantum
  spin-liquid state in geometrically frustrated
  $\ensuremath{\beta}\ensuremath{-}\mathrm{Mn}$, and the transition to a
  spin-glass state caused by non-magnetic impurity}},\ }\href
  {https://doi.org/10.1088/0953-8984/9/22/022} {\bibfield  {journal} {\bibinfo
  {journal} {Journal of Physics: Condensed Matter}\ }\textbf {\bibinfo {volume}
  {9}},\ \bibinfo {pages} {4701} (\bibinfo {year}
  {1997}{\natexlab{b}})}\BibitemShut {NoStop}%
\bibitem [{\citenamefont {Okamoto}\ \emph {et~al.}(2007)\citenamefont
  {Okamoto}, \citenamefont {Nohara}, \citenamefont {Aruga-Katori},\ and\
  \citenamefont {Takagi}}]{Okamoto2007NaIr}%
  \BibitemOpen
  \bibfield  {author} {\bibinfo {author} {\bibfnamefont {Y.}~\bibnamefont
  {Okamoto}}, \bibinfo {author} {\bibfnamefont {M.}~\bibnamefont {Nohara}},
  \bibinfo {author} {\bibfnamefont {H.}~\bibnamefont {Aruga-Katori}},\ and\
  \bibinfo {author} {\bibfnamefont {H.}~\bibnamefont {Takagi}},\ }\bibfield
  {title} {\bibinfo {title} {{Spin-Liquid State in the {$S=\frac{1}{2}$}
  Hyperkagome Antiferromagnet
  {${\mathrm{Na}}_{4}{\mathrm{Ir}}_{3}{\mathrm{O}}_{8}$}}},\ }\href
  {https://doi.org/10.1103/PhysRevLett.99.137207} {\bibfield  {journal}
  {\bibinfo  {journal} {Phys. Rev. Lett.}\ }\textbf {\bibinfo {volume} {99}},\
  \bibinfo {pages} {137207} (\bibinfo {year} {2007})}\BibitemShut {NoStop}%
\bibitem [{\citenamefont {Paddison}\ \emph {et~al.}(2013)\citenamefont
  {Paddison}, \citenamefont {Stewart}, \citenamefont {Manuel}, \citenamefont
  {Courtois}, \citenamefont {McIntyre}, \citenamefont {Rainford},\ and\
  \citenamefont {Goodwin}}]{Paddison2013emergfrustrbMn}%
  \BibitemOpen
  \bibfield  {author} {\bibinfo {author} {\bibfnamefont {J.~A.~M.}\
  \bibnamefont {Paddison}}, \bibinfo {author} {\bibfnamefont {J.~R.}\
  \bibnamefont {Stewart}}, \bibinfo {author} {\bibfnamefont {P.}~\bibnamefont
  {Manuel}}, \bibinfo {author} {\bibfnamefont {P.}~\bibnamefont {Courtois}},
  \bibinfo {author} {\bibfnamefont {G.~J.}\ \bibnamefont {McIntyre}}, \bibinfo
  {author} {\bibfnamefont {B.~D.}\ \bibnamefont {Rainford}},\ and\ \bibinfo
  {author} {\bibfnamefont {A.~L.}\ \bibnamefont {Goodwin}},\ }\bibfield
  {title} {\bibinfo {title} {{Emergent Frustration in {C}o-doped
  $\ensuremath{\beta}$-{M}n}},\ }\href
  {https://doi.org/10.1103/PhysRevLett.110.267207} {\bibfield  {journal}
  {\bibinfo  {journal} {Phys. Rev. Lett.}\ }\textbf {\bibinfo {volume} {110}},\
  \bibinfo {pages} {267207} (\bibinfo {year} {2013})}\BibitemShut {NoStop}%
\bibitem [{\citenamefont {Chillal}\ \emph {et~al.}(2020)\citenamefont
  {Chillal}, \citenamefont {Iqbal}, \citenamefont {Jeschke}, \citenamefont
  {Rodriguez-Rivera}, \citenamefont {Bewley}, \citenamefont {Manuel},
  \citenamefont {Khalyavin}, \citenamefont {Steffens}, \citenamefont {Thomale},
  \citenamefont {Islam}, \citenamefont {Reuther},\ and\ \citenamefont
  {Lake}}]{Chillal2020Evidence}%
  \BibitemOpen
  \bibfield  {author} {\bibinfo {author} {\bibfnamefont {S.}~\bibnamefont
  {Chillal}}, \bibinfo {author} {\bibfnamefont {Y.}~\bibnamefont {Iqbal}},
  \bibinfo {author} {\bibfnamefont {H.~O.}\ \bibnamefont {Jeschke}}, \bibinfo
  {author} {\bibfnamefont {J.~A.}\ \bibnamefont {Rodriguez-Rivera}}, \bibinfo
  {author} {\bibfnamefont {R.}~\bibnamefont {Bewley}}, \bibinfo {author}
  {\bibfnamefont {P.}~\bibnamefont {Manuel}}, \bibinfo {author} {\bibfnamefont
  {D.}~\bibnamefont {Khalyavin}}, \bibinfo {author} {\bibfnamefont
  {P.}~\bibnamefont {Steffens}}, \bibinfo {author} {\bibfnamefont
  {R.}~\bibnamefont {Thomale}}, \bibinfo {author} {\bibfnamefont {A.~T. M.~N.}\
  \bibnamefont {Islam}}, \bibinfo {author} {\bibfnamefont {J.}~\bibnamefont
  {Reuther}},\ and\ \bibinfo {author} {\bibfnamefont {B.}~\bibnamefont
  {Lake}},\ }\bibfield  {title} {\bibinfo {title} {{Evidence for a
  three-dimensional quantum spin liquid in
  {${\mathrm{PbCuTe}}_{2}{\mathrm{O}}_{6}$}}},\ }\href
  {https://doi.org/10.1038/s41467-020-15594-1} {\bibfield  {journal} {\bibinfo
  {journal} {Nature Communications}\ }\textbf {\bibinfo {volume} {11}},\
  \bibinfo {pages} {2348} (\bibinfo {year} {2020})}\BibitemShut {NoStop}%
\bibitem [{\citenamefont {Koteswararao}\ \emph {et~al.}(2014)\citenamefont
  {Koteswararao}, \citenamefont {Kumar}, \citenamefont {Khuntia}, \citenamefont
  {Bhowal}, \citenamefont {Panda}, \citenamefont {Rahman}, \citenamefont
  {Mahajan}, \citenamefont {Dasgupta}, \citenamefont {Baenitz}, \citenamefont
  {Kim},\ and\ \citenamefont {Chou}}]{Koteswararao2014Magprop}%
  \BibitemOpen
  \bibfield  {author} {\bibinfo {author} {\bibfnamefont {B.}~\bibnamefont
  {Koteswararao}}, \bibinfo {author} {\bibfnamefont {R.}~\bibnamefont {Kumar}},
  \bibinfo {author} {\bibfnamefont {P.}~\bibnamefont {Khuntia}}, \bibinfo
  {author} {\bibfnamefont {S.}~\bibnamefont {Bhowal}}, \bibinfo {author}
  {\bibfnamefont {S.~K.}\ \bibnamefont {Panda}}, \bibinfo {author}
  {\bibfnamefont {M.~R.}\ \bibnamefont {Rahman}}, \bibinfo {author}
  {\bibfnamefont {A.~V.}\ \bibnamefont {Mahajan}}, \bibinfo {author}
  {\bibfnamefont {I.}~\bibnamefont {Dasgupta}}, \bibinfo {author}
  {\bibfnamefont {M.}~\bibnamefont {Baenitz}}, \bibinfo {author} {\bibfnamefont
  {K.~H.}\ \bibnamefont {Kim}},\ and\ \bibinfo {author} {\bibfnamefont {F.~C.}\
  \bibnamefont {Chou}},\ }\bibfield  {title} {\bibinfo {title} {{Magnetic
  properties and heat capacity of the three-dimensional frustrated
  {$S=\frac{1}{2}$} antiferromagnet
  {${\mathrm{PbCuTe}}_{2}{\mathrm{O}}_{6}$}}},\ }\href
  {https://doi.org/10.1103/PhysRevB.90.035141} {\bibfield  {journal} {\bibinfo
  {journal} {Phys. Rev. B}\ }\textbf {\bibinfo {volume} {90}},\ \bibinfo
  {pages} {035141} (\bibinfo {year} {2014})}\BibitemShut {NoStop}%
\bibitem [{\citenamefont {Khuntia}\ \emph {et~al.}(2016)\citenamefont
  {Khuntia}, \citenamefont {Bert}, \citenamefont {Mendels}, \citenamefont
  {Koteswararao}, \citenamefont {Mahajan}, \citenamefont {Baenitz},
  \citenamefont {Chou}, \citenamefont {Baines}, \citenamefont {Amato},\ and\
  \citenamefont {Furukawa}}]{Khuntia2016muonspin}%
  \BibitemOpen
  \bibfield  {author} {\bibinfo {author} {\bibfnamefont {P.}~\bibnamefont
  {Khuntia}}, \bibinfo {author} {\bibfnamefont {F.}~\bibnamefont {Bert}},
  \bibinfo {author} {\bibfnamefont {P.}~\bibnamefont {Mendels}}, \bibinfo
  {author} {\bibfnamefont {B.}~\bibnamefont {Koteswararao}}, \bibinfo {author}
  {\bibfnamefont {A.~V.}\ \bibnamefont {Mahajan}}, \bibinfo {author}
  {\bibfnamefont {M.}~\bibnamefont {Baenitz}}, \bibinfo {author} {\bibfnamefont
  {F.~C.}\ \bibnamefont {Chou}}, \bibinfo {author} {\bibfnamefont
  {C.}~\bibnamefont {Baines}}, \bibinfo {author} {\bibfnamefont
  {A.}~\bibnamefont {Amato}},\ and\ \bibinfo {author} {\bibfnamefont
  {Y.}~\bibnamefont {Furukawa}},\ }\bibfield  {title} {\bibinfo {title} {{Spin
  Liquid State in the 3D Frustrated Antiferromagnet
  {${\mathrm{PbCuTe}}_{2}{\mathrm{O}}_{6}$}: NMR and Muon Spin Relaxation
  Studies}},\ }\href {https://doi.org/10.1103/PhysRevLett.116.107203}
  {\bibfield  {journal} {\bibinfo  {journal} {Phys. Rev. Lett.}\ }\textbf
  {\bibinfo {volume} {116}},\ \bibinfo {pages} {107203} (\bibinfo {year}
  {2016})}\BibitemShut {NoStop}%
\bibitem [{\citenamefont {Chern}\ and\ \citenamefont
  {Kim}(2021)}]{Chern2021PSGJ1J2}%
  \BibitemOpen
  \bibfield  {author} {\bibinfo {author} {\bibfnamefont {L.~E.}\ \bibnamefont
  {Chern}}\ and\ \bibinfo {author} {\bibfnamefont {Y.~B.}\ \bibnamefont
  {Kim}},\ }\bibfield  {title} {\bibinfo {title} {{Theoretical study of quantum
  spin liquids in {$S=\frac{1}{2}$} hyper-hyperkagome magnets: Classification,
  heat capacity, and dynamical spin structure factor}},\ }\href
  {https://doi.org/10.1103/PhysRevB.104.094413} {\bibfield  {journal} {\bibinfo
   {journal} {Phys. Rev. B}\ }\textbf {\bibinfo {volume} {104}},\ \bibinfo
  {pages} {094413} (\bibinfo {year} {2021})}\BibitemShut {NoStop}%
\bibitem [{\citenamefont {Jin}\ and\ \citenamefont
  {Zhou}(2020)}]{Jin2020Classicalquantum}%
  \BibitemOpen
  \bibfield  {author} {\bibinfo {author} {\bibfnamefont {H.-K.}\ \bibnamefont
  {Jin}}\ and\ \bibinfo {author} {\bibfnamefont {Y.}~\bibnamefont {Zhou}},\
  }\bibfield  {title} {\bibinfo {title} {{Classical and quantum order in
  hyperkagome antiferromagnets}},\ }\href
  {https://doi.org/10.1103/PhysRevB.101.054408} {\bibfield  {journal} {\bibinfo
   {journal} {Phys. Rev. B}\ }\textbf {\bibinfo {volume} {101}},\ \bibinfo
  {pages} {054408} (\bibinfo {year} {2020})}\BibitemShut {NoStop}%
\bibitem [{\citenamefont {Samarakoon}\ \emph {et~al.}(2017)\citenamefont
  {Samarakoon}, \citenamefont {Banerjee}, \citenamefont {Zhang}, \citenamefont
  {Kamiya}, \citenamefont {Nagler}, \citenamefont {Tennant}, \citenamefont
  {Lee},\ and\ \citenamefont {Batista}}]{Samarakoon2017ClassicalKitaev}%
  \BibitemOpen
  \bibfield  {author} {\bibinfo {author} {\bibfnamefont {A.~M.}\ \bibnamefont
  {Samarakoon}}, \bibinfo {author} {\bibfnamefont {A.}~\bibnamefont
  {Banerjee}}, \bibinfo {author} {\bibfnamefont {S.-S.}\ \bibnamefont {Zhang}},
  \bibinfo {author} {\bibfnamefont {Y.}~\bibnamefont {Kamiya}}, \bibinfo
  {author} {\bibfnamefont {S.~E.}\ \bibnamefont {Nagler}}, \bibinfo {author}
  {\bibfnamefont {D.~A.}\ \bibnamefont {Tennant}}, \bibinfo {author}
  {\bibfnamefont {S.-H.}\ \bibnamefont {Lee}},\ and\ \bibinfo {author}
  {\bibfnamefont {C.~D.}\ \bibnamefont {Batista}},\ }\bibfield  {title}
  {\bibinfo {title} {Comprehensive study of the dynamics of a classical kitaev
  spin liquid},\ }\href {https://doi.org/10.1103/PhysRevB.96.134408} {\bibfield
   {journal} {\bibinfo  {journal} {Phys. Rev. B}\ }\textbf {\bibinfo {volume}
  {96}},\ \bibinfo {pages} {134408} (\bibinfo {year} {2017})}\BibitemShut
  {NoStop}%
\bibitem [{\citenamefont {Samarakoon}\ \emph {et~al.}(2018)\citenamefont
  {Samarakoon}, \citenamefont {Wachtel}, \citenamefont {Yamaji}, \citenamefont
  {Tennant}, \citenamefont {Batista},\ and\ \citenamefont
  {Kim}}]{Samarakoon2018quantclass}%
  \BibitemOpen
  \bibfield  {author} {\bibinfo {author} {\bibfnamefont {A.~M.}\ \bibnamefont
  {Samarakoon}}, \bibinfo {author} {\bibfnamefont {G.}~\bibnamefont {Wachtel}},
  \bibinfo {author} {\bibfnamefont {Y.}~\bibnamefont {Yamaji}}, \bibinfo
  {author} {\bibfnamefont {D.~A.}\ \bibnamefont {Tennant}}, \bibinfo {author}
  {\bibfnamefont {C.~D.}\ \bibnamefont {Batista}},\ and\ \bibinfo {author}
  {\bibfnamefont {Y.~B.}\ \bibnamefont {Kim}},\ }\bibfield  {title} {\bibinfo
  {title} {Classical and quantum spin dynamics of the honeycomb
  $\mathrm{\ensuremath{\Gamma}}$ model},\ }\href
  {https://doi.org/10.1103/PhysRevB.98.045121} {\bibfield  {journal} {\bibinfo
  {journal} {Phys. Rev. B}\ }\textbf {\bibinfo {volume} {98}},\ \bibinfo
  {pages} {045121} (\bibinfo {year} {2018})}\BibitemShut {NoStop}%
\bibitem [{\citenamefont {Hosoi}\ \emph {et~al.}(2022)\citenamefont {Hosoi},
  \citenamefont {Zhang}, \citenamefont {Patri},\ and\ \citenamefont
  {Kim}}]{Hosoi2022Uncovering}%
  \BibitemOpen
  \bibfield  {author} {\bibinfo {author} {\bibfnamefont {M.}~\bibnamefont
  {Hosoi}}, \bibinfo {author} {\bibfnamefont {E.~Z.}\ \bibnamefont {Zhang}},
  \bibinfo {author} {\bibfnamefont {A.~S.}\ \bibnamefont {Patri}},\ and\
  \bibinfo {author} {\bibfnamefont {Y.~B.}\ \bibnamefont {Kim}},\ }\bibfield
  {title} {\bibinfo {title} {Uncovering footprints of dipolar-octupolar quantum
  spin ice from neutron scattering signatures},\ }\href
  {https://doi.org/10.1103/PhysRevLett.129.097202} {\bibfield  {journal}
  {\bibinfo  {journal} {Phys. Rev. Lett.}\ }\textbf {\bibinfo {volume} {129}},\
  \bibinfo {pages} {097202} (\bibinfo {year} {2022})}\BibitemShut {NoStop}%
\bibitem [{\citenamefont {Smith}\ \emph {et~al.}(2022)\citenamefont {Smith},
  \citenamefont {Benton}, \citenamefont {Yahne}, \citenamefont {Placke},
  \citenamefont {Sch\"afer}, \citenamefont {Gaudet}, \citenamefont {Dudemaine},
  \citenamefont {Fitterman}, \citenamefont {Beare}, \citenamefont {Wildes},
  \citenamefont {Bhattacharya}, \citenamefont {DeLazzer}, \citenamefont
  {Buhariwalla}, \citenamefont {Butch}, \citenamefont {Movshovich},
  \citenamefont {Garrett}, \citenamefont {Marjerrison}, \citenamefont {Clancy},
  \citenamefont {Kermarrec}, \citenamefont {Luke}, \citenamefont {Bianchi},
  \citenamefont {Ross},\ and\ \citenamefont {Gaulin}}]{Smith2022Case}%
  \BibitemOpen
  \bibfield  {author} {\bibinfo {author} {\bibfnamefont {E.~M.}\ \bibnamefont
  {Smith}}, \bibinfo {author} {\bibfnamefont {O.}~\bibnamefont {Benton}},
  \bibinfo {author} {\bibfnamefont {D.~R.}\ \bibnamefont {Yahne}}, \bibinfo
  {author} {\bibfnamefont {B.}~\bibnamefont {Placke}}, \bibinfo {author}
  {\bibfnamefont {R.}~\bibnamefont {Sch\"afer}}, \bibinfo {author}
  {\bibfnamefont {J.}~\bibnamefont {Gaudet}}, \bibinfo {author} {\bibfnamefont
  {J.}~\bibnamefont {Dudemaine}}, \bibinfo {author} {\bibfnamefont
  {A.}~\bibnamefont {Fitterman}}, \bibinfo {author} {\bibfnamefont
  {J.}~\bibnamefont {Beare}}, \bibinfo {author} {\bibfnamefont {A.~R.}\
  \bibnamefont {Wildes}}, \bibinfo {author} {\bibfnamefont {S.}~\bibnamefont
  {Bhattacharya}}, \bibinfo {author} {\bibfnamefont {T.}~\bibnamefont
  {DeLazzer}}, \bibinfo {author} {\bibfnamefont {C.~R.~C.}\ \bibnamefont
  {Buhariwalla}}, \bibinfo {author} {\bibfnamefont {N.~P.}\ \bibnamefont
  {Butch}}, \bibinfo {author} {\bibfnamefont {R.}~\bibnamefont {Movshovich}},
  \bibinfo {author} {\bibfnamefont {J.~D.}\ \bibnamefont {Garrett}}, \bibinfo
  {author} {\bibfnamefont {C.~A.}\ \bibnamefont {Marjerrison}}, \bibinfo
  {author} {\bibfnamefont {J.~P.}\ \bibnamefont {Clancy}}, \bibinfo {author}
  {\bibfnamefont {E.}~\bibnamefont {Kermarrec}}, \bibinfo {author}
  {\bibfnamefont {G.~M.}\ \bibnamefont {Luke}}, \bibinfo {author}
  {\bibfnamefont {A.~D.}\ \bibnamefont {Bianchi}}, \bibinfo {author}
  {\bibfnamefont {K.~A.}\ \bibnamefont {Ross}},\ and\ \bibinfo {author}
  {\bibfnamefont {B.~D.}\ \bibnamefont {Gaulin}},\ }\bibfield  {title}
  {\bibinfo {title} {Case for a ${\mathrm{u}(1)}_{\ensuremath{\pi}}$ quantum
  spin liquid ground state in the dipole-octupole pyrochlore
  ${\mathrm{ce}}_{2}{\mathrm{zr}}_{2}{\mathrm{o}}_{7}$},\ }\href
  {https://doi.org/10.1103/PhysRevX.12.021015} {\bibfield  {journal} {\bibinfo
  {journal} {Phys. Rev. X}\ }\textbf {\bibinfo {volume} {12}},\ \bibinfo
  {pages} {021015} (\bibinfo {year} {2022})}\BibitemShut {NoStop}%
\bibitem [{\citenamefont {Bhardwaj}\ \emph {et~al.}(2022)\citenamefont
  {Bhardwaj}, \citenamefont {Zhang}, \citenamefont {Yan}, \citenamefont
  {Moessner}, \citenamefont {Nevidomskyy},\ and\ \citenamefont
  {Changlani}}]{Bhardwaj2022-sf}%
  \BibitemOpen
  \bibfield  {author} {\bibinfo {author} {\bibfnamefont {A.}~\bibnamefont
  {Bhardwaj}}, \bibinfo {author} {\bibfnamefont {S.}~\bibnamefont {Zhang}},
  \bibinfo {author} {\bibfnamefont {H.}~\bibnamefont {Yan}}, \bibinfo {author}
  {\bibfnamefont {R.}~\bibnamefont {Moessner}}, \bibinfo {author}
  {\bibfnamefont {A.~H.}\ \bibnamefont {Nevidomskyy}},\ and\ \bibinfo {author}
  {\bibfnamefont {H.~J.}\ \bibnamefont {Changlani}},\ }\bibfield  {title}
  {\bibinfo {title} {Sleuthing out exotic quantum spin liquidity in the
  pyrochlore magnet ${\mathrm{ce}}_{2}{\mathrm{zr}}_{2}{\mathrm{o}}_{7}$},\
  }\href {https://doi.org/10.1038/s41535-022-00458-2} {\bibfield  {journal}
  {\bibinfo  {journal} {npj Quantum Materials}\ }\textbf {\bibinfo {volume}
  {7}},\ \bibinfo {pages} {51} (\bibinfo {year} {2022})}\BibitemShut {NoStop}%
\bibitem [{\citenamefont {Zhang}\ \emph {et~al.}(2019)\citenamefont {Zhang},
  \citenamefont {Changlani}, \citenamefont {Plumb}, \citenamefont
  {Tchernyshyov},\ and\ \citenamefont {Moessner}}]{Zhang2019Dynamicalstruct}%
  \BibitemOpen
  \bibfield  {author} {\bibinfo {author} {\bibfnamefont {S.}~\bibnamefont
  {Zhang}}, \bibinfo {author} {\bibfnamefont {H.~J.}\ \bibnamefont
  {Changlani}}, \bibinfo {author} {\bibfnamefont {K.~W.}\ \bibnamefont
  {Plumb}}, \bibinfo {author} {\bibfnamefont {O.}~\bibnamefont
  {Tchernyshyov}},\ and\ \bibinfo {author} {\bibfnamefont {R.}~\bibnamefont
  {Moessner}},\ }\bibfield  {title} {\bibinfo {title} {{Dynamical Structure
  Factor of the Three-Dimensional Quantum Spin Liquid Candidate
  {${\mathrm{NaCaNi}}_{2}{\mathrm{F}}_{7}$}}},\ }\href
  {https://doi.org/10.1103/PhysRevLett.122.167203} {\bibfield  {journal}
  {\bibinfo  {journal} {Phys. Rev. Lett.}\ }\textbf {\bibinfo {volume} {122}},\
  \bibinfo {pages} {167203} (\bibinfo {year} {2019})}\BibitemShut {NoStop}%
\bibitem [{\citenamefont {Pohle}\ \emph {et~al.}(2021)\citenamefont {Pohle},
  \citenamefont {Yan},\ and\ \citenamefont {Shannon}}]{Pohle2021ca10}%
  \BibitemOpen
  \bibfield  {author} {\bibinfo {author} {\bibfnamefont {R.}~\bibnamefont
  {Pohle}}, \bibinfo {author} {\bibfnamefont {H.}~\bibnamefont {Yan}},\ and\
  \bibinfo {author} {\bibfnamefont {N.}~\bibnamefont {Shannon}},\ }\bibfield
  {title} {\bibinfo {title} {{Theory of
  {${\mathrm{Ca}}_{10}{\mathrm{Cr}}_{7}{\mathrm{O}}_{28}$} as a bilayer
  breathing-kagome magnet: Classical thermodynamics and semiclassical
  dynamics}},\ }\href {https://doi.org/10.1103/PhysRevB.104.024426} {\bibfield
  {journal} {\bibinfo  {journal} {Phys. Rev. B}\ }\textbf {\bibinfo {volume}
  {104}},\ \bibinfo {pages} {024426} (\bibinfo {year} {2021})}\BibitemShut
  {NoStop}%
\bibitem [{\citenamefont {Bai}\ \emph {et~al.}(2019)\citenamefont {Bai},
  \citenamefont {Paddison}, \citenamefont {Kapit}, \citenamefont {Koohpayeh},
  \citenamefont {Wen}, \citenamefont {Dutton}, \citenamefont {Savici},
  \citenamefont {Kolesnikov}, \citenamefont {Granroth}, \citenamefont
  {Broholm}, \citenamefont {Chalker},\ and\ \citenamefont
  {Mourigal}}]{Bai2019}%
  \BibitemOpen
  \bibfield  {author} {\bibinfo {author} {\bibfnamefont {X.}~\bibnamefont
  {Bai}}, \bibinfo {author} {\bibfnamefont {J.~A.~M.}\ \bibnamefont
  {Paddison}}, \bibinfo {author} {\bibfnamefont {E.}~\bibnamefont {Kapit}},
  \bibinfo {author} {\bibfnamefont {S.~M.}\ \bibnamefont {Koohpayeh}}, \bibinfo
  {author} {\bibfnamefont {J.-J.}\ \bibnamefont {Wen}}, \bibinfo {author}
  {\bibfnamefont {S.~E.}\ \bibnamefont {Dutton}}, \bibinfo {author}
  {\bibfnamefont {A.~T.}\ \bibnamefont {Savici}}, \bibinfo {author}
  {\bibfnamefont {A.~I.}\ \bibnamefont {Kolesnikov}}, \bibinfo {author}
  {\bibfnamefont {G.~E.}\ \bibnamefont {Granroth}}, \bibinfo {author}
  {\bibfnamefont {C.~L.}\ \bibnamefont {Broholm}}, \bibinfo {author}
  {\bibfnamefont {J.~T.}\ \bibnamefont {Chalker}},\ and\ \bibinfo {author}
  {\bibfnamefont {M.}~\bibnamefont {Mourigal}},\ }\bibfield  {title} {\bibinfo
  {title} {Magnetic excitations of the classical spin liquid
  ${\mathrm{mgcr}}_{2}{\mathrm{o}}_{4}$},\ }\href
  {https://doi.org/10.1103/PhysRevLett.122.097201} {\bibfield  {journal}
  {\bibinfo  {journal} {Phys. Rev. Lett.}\ }\textbf {\bibinfo {volume} {122}},\
  \bibinfo {pages} {097201} (\bibinfo {year} {2019})}\BibitemShut {NoStop}%
\bibitem [{\citenamefont {Franke}\ \emph {et~al.}(2022)\citenamefont {Franke},
  \citenamefont {C\ifmmode \u{a}\else \u{a}\fi{}lug\ifmmode~\u{a}\else
  \u{a}\fi{}ru}, \citenamefont {Nunnenkamp},\ and\ \citenamefont
  {Knolle}}]{Franke2022}%
  \BibitemOpen
  \bibfield  {author} {\bibinfo {author} {\bibfnamefont {O.}~\bibnamefont
  {Franke}}, \bibinfo {author} {\bibfnamefont {D.}~\bibnamefont {C\ifmmode
  \u{a}\else \u{a}\fi{}lug\ifmmode~\u{a}\else \u{a}\fi{}ru}}, \bibinfo {author}
  {\bibfnamefont {A.}~\bibnamefont {Nunnenkamp}},\ and\ \bibinfo {author}
  {\bibfnamefont {J.}~\bibnamefont {Knolle}},\ }\bibfield  {title} {\bibinfo
  {title} {Thermal spin dynamics of kitaev magnets: Scattering continua and
  magnetic field induced phases within a stochastic semiclassical approach},\
  }\href {https://doi.org/10.1103/PhysRevB.106.174428} {\bibfield  {journal}
  {\bibinfo  {journal} {Phys. Rev. B}\ }\textbf {\bibinfo {volume} {106}},\
  \bibinfo {pages} {174428} (\bibinfo {year} {2022})}\BibitemShut {NoStop}%
\bibitem [{\citenamefont {Hopkinson}\ and\ \citenamefont
  {Kee}(2006)}]{Hopkinson2006Trillium}%
  \BibitemOpen
  \bibfield  {author} {\bibinfo {author} {\bibfnamefont {J.~M.}\ \bibnamefont
  {Hopkinson}}\ and\ \bibinfo {author} {\bibfnamefont {H.-Y.}\ \bibnamefont
  {Kee}},\ }\bibfield  {title} {\bibinfo {title} {{Geometric frustration
  inherent to the trillium lattice, a sublattice of the {B}20 structure}},\
  }\href {https://doi.org/10.1103/PhysRevB.74.224441} {\bibfield  {journal}
  {\bibinfo  {journal} {Phys. Rev. B}\ }\textbf {\bibinfo {volume} {74}},\
  \bibinfo {pages} {224441} (\bibinfo {year} {2006})}\BibitemShut {NoStop}%
\bibitem [{\citenamefont {Luttinger}\ and\ \citenamefont
  {Tisza}(1946)}]{Luttinger1946LT}%
  \BibitemOpen
  \bibfield  {author} {\bibinfo {author} {\bibfnamefont {J.~M.}\ \bibnamefont
  {Luttinger}}\ and\ \bibinfo {author} {\bibfnamefont {L.}~\bibnamefont
  {Tisza}},\ }\bibfield  {title} {\bibinfo {title} {{Theory of Dipole
  Interaction in Crystals}},\ }\href {https://doi.org/10.1103/PhysRev.70.954}
  {\bibfield  {journal} {\bibinfo  {journal} {Phys. Rev.}\ }\textbf {\bibinfo
  {volume} {70}},\ \bibinfo {pages} {954} (\bibinfo {year} {1946})}\BibitemShut
  {NoStop}%
\bibitem [{\citenamefont {Sklan}\ and\ \citenamefont
  {Henley}(2013{\natexlab{a}})}]{Sklan2013Noncop}%
  \BibitemOpen
  \bibfield  {author} {\bibinfo {author} {\bibfnamefont {S.~R.}\ \bibnamefont
  {Sklan}}\ and\ \bibinfo {author} {\bibfnamefont {C.~L.}\ \bibnamefont
  {Henley}},\ }\bibfield  {title} {\bibinfo {title} {{Nonplanar ground states
  of frustrated antiferromagnets on an octahedral lattice}},\ }\href
  {https://doi.org/10.1103/PhysRevB.88.024407} {\bibfield  {journal} {\bibinfo
  {journal} {Phys. Rev. B}\ }\textbf {\bibinfo {volume} {88}},\ \bibinfo
  {pages} {024407} (\bibinfo {year} {2013}{\natexlab{a}})}\BibitemShut
  {NoStop}%
\bibitem [{\citenamefont {Chen}\ and\ \citenamefont
  {Balents}(2008)}]{Chen2008SpinOrbit}%
  \BibitemOpen
  \bibfield  {author} {\bibinfo {author} {\bibfnamefont {G.}~\bibnamefont
  {Chen}}\ and\ \bibinfo {author} {\bibfnamefont {L.}~\bibnamefont {Balents}},\
  }\bibfield  {title} {\bibinfo {title} {{Spin-orbit effects in
  ${\text{Na}}_{4}{\text{Ir}}_{3}{\text{O}}_{8}$: A hyper-kagome lattice
  antiferromagnet}},\ }\href {https://doi.org/10.1103/PhysRevB.78.094403}
  {\bibfield  {journal} {\bibinfo  {journal} {Phys. Rev. B}\ }\textbf {\bibinfo
  {volume} {78}},\ \bibinfo {pages} {094403} (\bibinfo {year}
  {2008})}\BibitemShut {NoStop}%
\bibitem [{\citenamefont {\ifmmode \check{Z}\else
  \v{Z}\fi{}ivkovi\ifmmode~\acute{c}\else \'{c}\fi{}}\ \emph
  {et~al.}(2021)\citenamefont {\ifmmode \check{Z}\else
  \v{Z}\fi{}ivkovi\ifmmode~\acute{c}\else \'{c}\fi{}}, \citenamefont {Favre},
  \citenamefont {Salazar~Mejia}, \citenamefont {Jeschke}, \citenamefont
  {Magrez}, \citenamefont {Dabholkar}, \citenamefont {Noculak}, \citenamefont
  {Freitas}, \citenamefont {Jeong}, \citenamefont {Hegde}, \citenamefont
  {Testa}, \citenamefont {Babkevich}, \citenamefont {Su}, \citenamefont
  {Manuel}, \citenamefont {Luetkens}, \citenamefont {Baines}, \citenamefont
  {Baker}, \citenamefont {Wosnitza}, \citenamefont {Zaharko}, \citenamefont
  {Iqbal}, \citenamefont {Reuther},\ and\ \citenamefont
  {R\o{}nnow}}]{zivkovic21}%
  \BibitemOpen
  \bibfield  {author} {\bibinfo {author} {\bibfnamefont {I.}~\bibnamefont
  {\ifmmode \check{Z}\else \v{Z}\fi{}ivkovi\ifmmode~\acute{c}\else
  \'{c}\fi{}}}, \bibinfo {author} {\bibfnamefont {V.}~\bibnamefont {Favre}},
  \bibinfo {author} {\bibfnamefont {C.}~\bibnamefont {Salazar~Mejia}}, \bibinfo
  {author} {\bibfnamefont {H.~O.}\ \bibnamefont {Jeschke}}, \bibinfo {author}
  {\bibfnamefont {A.}~\bibnamefont {Magrez}}, \bibinfo {author} {\bibfnamefont
  {B.}~\bibnamefont {Dabholkar}}, \bibinfo {author} {\bibfnamefont
  {V.}~\bibnamefont {Noculak}}, \bibinfo {author} {\bibfnamefont {R.~S.}\
  \bibnamefont {Freitas}}, \bibinfo {author} {\bibfnamefont {M.}~\bibnamefont
  {Jeong}}, \bibinfo {author} {\bibfnamefont {N.~G.}\ \bibnamefont {Hegde}},
  \bibinfo {author} {\bibfnamefont {L.}~\bibnamefont {Testa}}, \bibinfo
  {author} {\bibfnamefont {P.}~\bibnamefont {Babkevich}}, \bibinfo {author}
  {\bibfnamefont {Y.}~\bibnamefont {Su}}, \bibinfo {author} {\bibfnamefont
  {P.}~\bibnamefont {Manuel}}, \bibinfo {author} {\bibfnamefont
  {H.}~\bibnamefont {Luetkens}}, \bibinfo {author} {\bibfnamefont
  {C.}~\bibnamefont {Baines}}, \bibinfo {author} {\bibfnamefont {P.~J.}\
  \bibnamefont {Baker}}, \bibinfo {author} {\bibfnamefont {J.}~\bibnamefont
  {Wosnitza}}, \bibinfo {author} {\bibfnamefont {O.}~\bibnamefont {Zaharko}},
  \bibinfo {author} {\bibfnamefont {Y.}~\bibnamefont {Iqbal}}, \bibinfo
  {author} {\bibfnamefont {J.}~\bibnamefont {Reuther}},\ and\ \bibinfo {author}
  {\bibfnamefont {H.~M.}\ \bibnamefont {R\o{}nnow}},\ }\bibfield  {title}
  {\bibinfo {title} {{Magnetic Field Induced Quantum Spin Liquid in the Two
  Coupled Trillium Lattices of
  {${\mathrm{K}}_{2}{\mathrm{Ni}}_{2}({\mathrm{SO}}_{4}{)}_{3}$}}},\ }\href
  {https://doi.org/10.1103/PhysRevLett.127.157204} {\bibfield  {journal}
  {\bibinfo  {journal} {Phys. Rev. Lett.}\ }\textbf {\bibinfo {volume} {127}},\
  \bibinfo {pages} {157204} (\bibinfo {year} {2021})}\BibitemShut {NoStop}%
\bibitem [{\citenamefont {Rehn}\ \emph {et~al.}(2017)\citenamefont {Rehn},
  \citenamefont {Sen},\ and\ \citenamefont {Moessner}}]{Rehn2017classicalZ2}%
  \BibitemOpen
  \bibfield  {author} {\bibinfo {author} {\bibfnamefont {J.}~\bibnamefont
  {Rehn}}, \bibinfo {author} {\bibfnamefont {A.}~\bibnamefont {Sen}},\ and\
  \bibinfo {author} {\bibfnamefont {R.}~\bibnamefont {Moessner}},\ }\bibfield
  {title} {\bibinfo {title} {{Fractionalized {${\mathbb{Z}}_{2}$} Classical
  {H}eisenberg Spin Liquids}},\ }\href
  {https://doi.org/10.1103/PhysRevLett.118.047201} {\bibfield  {journal}
  {\bibinfo  {journal} {Phys. Rev. Lett.}\ }\textbf {\bibinfo {volume} {118}},\
  \bibinfo {pages} {047201} (\bibinfo {year} {2017})}\BibitemShut {NoStop}%
\bibitem [{\citenamefont {Yan}\ \emph {et~al.}(2023{\natexlab{a}})\citenamefont
  {Yan}, \citenamefont {Benton}, \citenamefont {Moessner},\ and\ \citenamefont
  {Nevidomskyy}}]{Han2305.00155}%
  \BibitemOpen
  \bibfield  {author} {\bibinfo {author} {\bibfnamefont {H.}~\bibnamefont
  {Yan}}, \bibinfo {author} {\bibfnamefont {O.}~\bibnamefont {Benton}},
  \bibinfo {author} {\bibfnamefont {R.}~\bibnamefont {Moessner}},\ and\
  \bibinfo {author} {\bibfnamefont {A.~H.}\ \bibnamefont {Nevidomskyy}},\
  }\href@noop {} {\bibinfo {title} {{Classification of Classical Spin Liquids:
  Typology and Resulting Landscape}}} (\bibinfo {year} {2023}{\natexlab{a}}),\
  \Eprint {https://arxiv.org/abs/arXiv:2305.00155} {arXiv:2305.00155}
  \BibitemShut {NoStop}%
\bibitem [{\citenamefont {Yan}\ \emph {et~al.}(2023{\natexlab{b}})\citenamefont
  {Yan}, \citenamefont {Benton}, \citenamefont {Nevidomskyy},\ and\
  \citenamefont {Moessner}}]{Han2305.19189}%
  \BibitemOpen
  \bibfield  {author} {\bibinfo {author} {\bibfnamefont {H.}~\bibnamefont
  {Yan}}, \bibinfo {author} {\bibfnamefont {O.}~\bibnamefont {Benton}},
  \bibinfo {author} {\bibfnamefont {A.~H.}\ \bibnamefont {Nevidomskyy}},\ and\
  \bibinfo {author} {\bibfnamefont {R.}~\bibnamefont {Moessner}},\ }\href@noop
  {} {\bibinfo {title} {{Classification of Classical Spin Liquids: Detailed
  Formalism and Suite of Examples}}} (\bibinfo {year} {2023}{\natexlab{b}}),\
  \Eprint {https://arxiv.org/abs/arXiv:2305.19189} {arXiv:2305.19189}
  \BibitemShut {NoStop}%
\bibitem [{\citenamefont {Davier}\ \emph {et~al.}(2023)\citenamefont {Davier},
  \citenamefont {Albarrac{\'i}n}, \citenamefont {Rosales},\ and\ \citenamefont
  {Pujol}}]{Davier2304.10906}%
  \BibitemOpen
  \bibfield  {author} {\bibinfo {author} {\bibfnamefont {N.}~\bibnamefont
  {Davier}}, \bibinfo {author} {\bibfnamefont {F.~G.}\ \bibnamefont
  {Albarrac{\'i}n}}, \bibinfo {author} {\bibfnamefont {D.}~\bibnamefont
  {Rosales}},\ and\ \bibinfo {author} {\bibfnamefont {P.}~\bibnamefont
  {Pujol}},\ }\href@noop {} {\bibinfo {title} {{A combined approach to analyze
  and classify families of classical spin liquids}}} (\bibinfo {year} {2023}),\
  \Eprint {https://arxiv.org/abs/arXiv:2304.10906} {arXiv:2304.10906}
  \BibitemShut {NoStop}%
\bibitem [{\citenamefont {Masuda}\ \emph {et~al.}(2012)\citenamefont {Masuda},
  \citenamefont {Okubo},\ and\ \citenamefont {Kawamura}}]{Masuda2012Hyst}%
  \BibitemOpen
  \bibfield  {author} {\bibinfo {author} {\bibfnamefont {H.}~\bibnamefont
  {Masuda}}, \bibinfo {author} {\bibfnamefont {T.}~\bibnamefont {Okubo}},\ and\
  \bibinfo {author} {\bibfnamefont {H.}~\bibnamefont {Kawamura}},\ }\bibfield
  {title} {\bibinfo {title} {Finite-temperature transition of the
  antiferromagnetic {H}eisenberg model on a distorted kagome lattice},\ }\href
  {https://doi.org/10.1103/PhysRevLett.109.057201} {\bibfield  {journal}
  {\bibinfo  {journal} {Phys. Rev. Lett.}\ }\textbf {\bibinfo {volume} {109}},\
  \bibinfo {pages} {057201} (\bibinfo {year} {2012})}\BibitemShut {NoStop}%
\bibitem [{\citenamefont {Zheng}\ and\ \citenamefont
  {Zhang}(1998)}]{Zheng1998ThermalHysteresis}%
  \BibitemOpen
  \bibfield  {author} {\bibinfo {author} {\bibfnamefont {G.~P.}\ \bibnamefont
  {Zheng}}\ and\ \bibinfo {author} {\bibfnamefont {J.~X.}\ \bibnamefont
  {Zhang}},\ }\bibfield  {title} {\bibinfo {title} {{Thermal hysteresis scaling
  for first-order phase transitions}},\ }\href
  {https://doi.org/10.1088/0953-8984/10/2/006} {\bibfield  {journal} {\bibinfo
  {journal} {Journal of Physics: Condensed Matter}\ }\textbf {\bibinfo {volume}
  {10}},\ \bibinfo {pages} {275} (\bibinfo {year} {1998})}\BibitemShut
  {NoStop}%
\bibitem [{\citenamefont {Okubo}\ \emph {et~al.}(2011)\citenamefont {Okubo},
  \citenamefont {Nguyen},\ and\ \citenamefont
  {Kawamura}}]{Okubo2011multiple-q}%
  \BibitemOpen
  \bibfield  {author} {\bibinfo {author} {\bibfnamefont {T.}~\bibnamefont
  {Okubo}}, \bibinfo {author} {\bibfnamefont {T.~H.}\ \bibnamefont {Nguyen}},\
  and\ \bibinfo {author} {\bibfnamefont {H.}~\bibnamefont {Kawamura}},\
  }\bibfield  {title} {\bibinfo {title} {{Cubic and noncubic multiple-$q$
  states in the {H}eisenberg antiferromagnet on the pyrochlore lattice}},\
  }\href {https://doi.org/10.1103/PhysRevB.84.144432} {\bibfield  {journal}
  {\bibinfo  {journal} {Phys. Rev. B}\ }\textbf {\bibinfo {volume} {84}},\
  \bibinfo {pages} {144432} (\bibinfo {year} {2011})}\BibitemShut {NoStop}%
\bibitem [{\citenamefont {Shimokawa}\ \emph {et~al.}(2019)\citenamefont
  {Shimokawa}, \citenamefont {Okubo},\ and\ \citenamefont
  {Kawamura}}]{Shimokawa2019multiqhex}%
  \BibitemOpen
  \bibfield  {author} {\bibinfo {author} {\bibfnamefont {T.}~\bibnamefont
  {Shimokawa}}, \bibinfo {author} {\bibfnamefont {T.}~\bibnamefont {Okubo}},\
  and\ \bibinfo {author} {\bibfnamefont {H.}~\bibnamefont {Kawamura}},\
  }\bibfield  {title} {\bibinfo {title} {{Multiple-$q$ states of the
  {${J}_{1}\ensuremath{-}{J}_{2}$} classical honeycomb-lattice {H}eisenberg
  antiferromagnet under a magnetic field}},\ }\href
  {https://doi.org/10.1103/PhysRevB.100.224404} {\bibfield  {journal} {\bibinfo
   {journal} {Phys. Rev. B}\ }\textbf {\bibinfo {volume} {100}},\ \bibinfo
  {pages} {224404} (\bibinfo {year} {2019})}\BibitemShut {NoStop}%
\bibitem [{\citenamefont {Brown}(2006)}]{brown2006magneticform}%
  \BibitemOpen
  \bibfield  {author} {\bibinfo {author} {\bibfnamefont {P.}~\bibnamefont
  {Brown}},\ }\href@noop {} {\emph {\bibinfo {title} {{Magnetic form
  factors}}}}\ (\bibinfo  {publisher} {Wiley Online Library},\ \bibinfo {year}
  {2006})\BibitemShut {NoStop}%
\bibitem [{\citenamefont {Pohle}\ \emph {et~al.}(2017)\citenamefont {Pohle},
  \citenamefont {Yan},\ and\ \citenamefont {Shannon}}]{pohle2017spinCa}%
  \BibitemOpen
  \bibfield  {author} {\bibinfo {author} {\bibfnamefont {R.}~\bibnamefont
  {Pohle}}, \bibinfo {author} {\bibfnamefont {H.}~\bibnamefont {Yan}},\ and\
  \bibinfo {author} {\bibfnamefont {N.}~\bibnamefont {Shannon}},\ }\href@noop
  {} {\bibinfo {title} {{How many spin liquids are there in
  {Ca$_{10}$Cr$_7$O$_{28}$}?}}} (\bibinfo {year} {2017}),\ \Eprint
  {https://arxiv.org/abs/1711.03778} {arXiv:1711.03778 [cond-mat.str-el]}
  \BibitemShut {NoStop}%
\bibitem [{\citenamefont {Balla}\ \emph {et~al.}(2020)\citenamefont {Balla},
  \citenamefont {Iqbal},\ and\ \citenamefont {Penc}}]{Balla2020phasediagfcc}%
  \BibitemOpen
  \bibfield  {author} {\bibinfo {author} {\bibfnamefont {P.}~\bibnamefont
  {Balla}}, \bibinfo {author} {\bibfnamefont {Y.}~\bibnamefont {Iqbal}},\ and\
  \bibinfo {author} {\bibfnamefont {K.}~\bibnamefont {Penc}},\ }\bibfield
  {title} {\bibinfo {title} {{Degenerate manifolds, helimagnets, and
  multi-{$Q$} chiral phases in the classical {H}eisenberg antiferromagnet on
  the face-centered-cubic lattice}},\ }\href
  {https://doi.org/10.1103/PhysRevResearch.2.043278} {\bibfield  {journal}
  {\bibinfo  {journal} {Phys. Rev. Research}\ }\textbf {\bibinfo {volume}
  {2}},\ \bibinfo {pages} {043278} (\bibinfo {year} {2020})}\BibitemShut
  {NoStop}%
\bibitem [{\citenamefont {Mambrini}\ \emph {et~al.}(2006)\citenamefont
  {Mambrini}, \citenamefont {L\"auchli}, \citenamefont {Poilblanc},\ and\
  \citenamefont {Mila}}]{Mambrini2006plaq}%
  \BibitemOpen
  \bibfield  {author} {\bibinfo {author} {\bibfnamefont {M.}~\bibnamefont
  {Mambrini}}, \bibinfo {author} {\bibfnamefont {A.}~\bibnamefont {L\"auchli}},
  \bibinfo {author} {\bibfnamefont {D.}~\bibnamefont {Poilblanc}},\ and\
  \bibinfo {author} {\bibfnamefont {F.}~\bibnamefont {Mila}},\ }\bibfield
  {title} {\bibinfo {title} {{Plaquette valence-bond crystal in the frustrated
  {H}eisenberg quantum antiferromagnet on the square lattice}},\ }\href
  {https://doi.org/10.1103/PhysRevB.74.144422} {\bibfield  {journal} {\bibinfo
  {journal} {Phys. Rev. B}\ }\textbf {\bibinfo {volume} {74}},\ \bibinfo
  {pages} {144422} (\bibinfo {year} {2006})}\BibitemShut {NoStop}%
\bibitem [{\citenamefont {Yamamoto}\ \emph {et~al.}(2014)\citenamefont
  {Yamamoto}, \citenamefont {Marmorini},\ and\ \citenamefont
  {Danshita}}]{Yamamoto2014quantumphasediagTri}%
  \BibitemOpen
  \bibfield  {author} {\bibinfo {author} {\bibfnamefont {D.}~\bibnamefont
  {Yamamoto}}, \bibinfo {author} {\bibfnamefont {G.}~\bibnamefont
  {Marmorini}},\ and\ \bibinfo {author} {\bibfnamefont {I.}~\bibnamefont
  {Danshita}},\ }\bibfield  {title} {\bibinfo {title} {{Quantum Phase Diagram
  of the Triangular-Lattice {$XXZ$} Model in a Magnetic Field}},\ }\href
  {https://doi.org/10.1103/PhysRevLett.112.127203} {\bibfield  {journal}
  {\bibinfo  {journal} {Phys. Rev. Lett.}\ }\textbf {\bibinfo {volume} {112}},\
  \bibinfo {pages} {127203} (\bibinfo {year} {2014})}\BibitemShut {NoStop}%
\bibitem [{\citenamefont {Iqbal}\ \emph {et~al.}(2019)\citenamefont {Iqbal},
  \citenamefont {M\"uller}, \citenamefont {Ghosh}, \citenamefont {Gingras},
  \citenamefont {Jeschke}, \citenamefont {Rachel}, \citenamefont {Reuther},\
  and\ \citenamefont {Thomale}}]{Iqbal2019quantclass}%
  \BibitemOpen
  \bibfield  {author} {\bibinfo {author} {\bibfnamefont {Y.}~\bibnamefont
  {Iqbal}}, \bibinfo {author} {\bibfnamefont {T.}~\bibnamefont {M\"uller}},
  \bibinfo {author} {\bibfnamefont {P.}~\bibnamefont {Ghosh}}, \bibinfo
  {author} {\bibfnamefont {M.~J.~P.}\ \bibnamefont {Gingras}}, \bibinfo
  {author} {\bibfnamefont {H.~O.}\ \bibnamefont {Jeschke}}, \bibinfo {author}
  {\bibfnamefont {S.}~\bibnamefont {Rachel}}, \bibinfo {author} {\bibfnamefont
  {J.}~\bibnamefont {Reuther}},\ and\ \bibinfo {author} {\bibfnamefont
  {R.}~\bibnamefont {Thomale}},\ }\bibfield  {title} {\bibinfo {title}
  {{Quantum and Classical Phases of the Pyrochlore {H}eisenberg Model with
  Competing Interactions}},\ }\href {https://doi.org/10.1103/PhysRevX.9.011005}
  {\bibfield  {journal} {\bibinfo  {journal} {Phys. Rev. X}\ }\textbf {\bibinfo
  {volume} {9}},\ \bibinfo {pages} {011005} (\bibinfo {year}
  {2019})}\BibitemShut {NoStop}%
\bibitem [{\citenamefont {Hanna}\ \emph {et~al.}(2021)\citenamefont {Hanna},
  \citenamefont {Islam}, \citenamefont {Feyerherm}, \citenamefont
  {Siemensmeyer}, \citenamefont {Karmakar}, \citenamefont {Chillal},\ and\
  \citenamefont {Lake}}]{Hanna2021Crystalgrowth}%
  \BibitemOpen
  \bibfield  {author} {\bibinfo {author} {\bibfnamefont {A.~R.~N.}\
  \bibnamefont {Hanna}}, \bibinfo {author} {\bibfnamefont {A.~T. M.~N.}\
  \bibnamefont {Islam}}, \bibinfo {author} {\bibfnamefont {R.}~\bibnamefont
  {Feyerherm}}, \bibinfo {author} {\bibfnamefont {K.}~\bibnamefont
  {Siemensmeyer}}, \bibinfo {author} {\bibfnamefont {K.}~\bibnamefont
  {Karmakar}}, \bibinfo {author} {\bibfnamefont {S.}~\bibnamefont {Chillal}},\
  and\ \bibinfo {author} {\bibfnamefont {B.}~\bibnamefont {Lake}},\ }\bibfield
  {title} {\bibinfo {title} {{Crystal growth, characterization, and phase
  transition of {${\mathrm{PbCuTe}}_{2}{\mathrm{O}}_{6}$}}},\ }\href
  {https://doi.org/10.1103/PhysRevMaterials.5.113401} {\bibfield  {journal}
  {\bibinfo  {journal} {Phys. Rev. Materials}\ }\textbf {\bibinfo {volume}
  {5}},\ \bibinfo {pages} {113401} (\bibinfo {year} {2021})}\BibitemShut
  {NoStop}%
\bibitem [{\citenamefont {Thurn}\ \emph {et~al.}(2021)\citenamefont {Thurn},
  \citenamefont {Eibisch}, \citenamefont {Ata}, \citenamefont {Winkler},
  \citenamefont {Lunkenheimer}, \citenamefont {K{\'e}zsm{\'a}rki},
  \citenamefont {Tutsch}, \citenamefont {Saito}, \citenamefont {Hartmann},
  \citenamefont {Zimmermann}, \citenamefont {Hanna}, \citenamefont {Islam},
  \citenamefont {Chillal}, \citenamefont {Lake}, \citenamefont {Wolf},\ and\
  \citenamefont {Lang}}]{Thurn2021ferroel}%
  \BibitemOpen
  \bibfield  {author} {\bibinfo {author} {\bibfnamefont {C.}~\bibnamefont
  {Thurn}}, \bibinfo {author} {\bibfnamefont {P.}~\bibnamefont {Eibisch}},
  \bibinfo {author} {\bibfnamefont {A.}~\bibnamefont {Ata}}, \bibinfo {author}
  {\bibfnamefont {M.}~\bibnamefont {Winkler}}, \bibinfo {author} {\bibfnamefont
  {P.}~\bibnamefont {Lunkenheimer}}, \bibinfo {author} {\bibfnamefont
  {I.}~\bibnamefont {K{\'e}zsm{\'a}rki}}, \bibinfo {author} {\bibfnamefont
  {U.}~\bibnamefont {Tutsch}}, \bibinfo {author} {\bibfnamefont
  {Y.}~\bibnamefont {Saito}}, \bibinfo {author} {\bibfnamefont
  {S.}~\bibnamefont {Hartmann}}, \bibinfo {author} {\bibfnamefont
  {J.}~\bibnamefont {Zimmermann}}, \bibinfo {author} {\bibfnamefont {A.~R.~N.}\
  \bibnamefont {Hanna}}, \bibinfo {author} {\bibfnamefont {A.~T. M.~N.}\
  \bibnamefont {Islam}}, \bibinfo {author} {\bibfnamefont {S.}~\bibnamefont
  {Chillal}}, \bibinfo {author} {\bibfnamefont {B.}~\bibnamefont {Lake}},
  \bibinfo {author} {\bibfnamefont {B.}~\bibnamefont {Wolf}},\ and\ \bibinfo
  {author} {\bibfnamefont {M.}~\bibnamefont {Lang}},\ }\bibfield  {title}
  {\bibinfo {title} {{Spin liquid and ferroelectricity close to a quantum
  critical point in {${\mathrm{PbCuTe}}_{2}{\mathrm{O}}_{6}$}}},\ }\href
  {https://doi.org/10.1038/s41535-021-00395-6} {\bibfield  {journal} {\bibinfo
  {journal} {npj Quantum Materials}\ }\textbf {\bibinfo {volume} {6}},\
  \bibinfo {pages} {95} (\bibinfo {year} {2021})}\BibitemShut {NoStop}%
\bibitem [{\citenamefont {Wang}\ \emph {et~al.}(2020)\citenamefont {Wang},
  \citenamefont {Cai}, \citenamefont {Chen}, \citenamefont {Prokof'ev},\ and\
  \citenamefont {Svistunov}}]{Wang2020quantclasscorr}%
  \BibitemOpen
  \bibfield  {author} {\bibinfo {author} {\bibfnamefont {T.}~\bibnamefont
  {Wang}}, \bibinfo {author} {\bibfnamefont {X.}~\bibnamefont {Cai}}, \bibinfo
  {author} {\bibfnamefont {K.}~\bibnamefont {Chen}}, \bibinfo {author}
  {\bibfnamefont {N.~V.}\ \bibnamefont {Prokof'ev}},\ and\ \bibinfo {author}
  {\bibfnamefont {B.~V.}\ \bibnamefont {Svistunov}},\ }\bibfield  {title}
  {\bibinfo {title} {{Quantum-to-classical correspondence in two-dimensional
  {H}eisenberg models}},\ }\href {https://doi.org/10.1103/PhysRevB.101.035132}
  {\bibfield  {journal} {\bibinfo  {journal} {Phys. Rev. B}\ }\textbf {\bibinfo
  {volume} {101}},\ \bibinfo {pages} {035132} (\bibinfo {year}
  {2020})}\BibitemShut {NoStop}%
\bibitem [{\citenamefont {Sklan}\ and\ \citenamefont
  {Henley}(2013{\natexlab{b}})}]{Sklan2013noncoplanaroctahedra}%
  \BibitemOpen
  \bibfield  {author} {\bibinfo {author} {\bibfnamefont {S.~R.}\ \bibnamefont
  {Sklan}}\ and\ \bibinfo {author} {\bibfnamefont {C.~L.}\ \bibnamefont
  {Henley}},\ }\bibfield  {title} {\bibinfo {title} {{Nonplanar ground states
  of frustrated antiferromagnets on an octahedral lattice}},\ }\href
  {https://doi.org/10.1103/PhysRevB.88.024407} {\bibfield  {journal} {\bibinfo
  {journal} {Phys. Rev. B}\ }\textbf {\bibinfo {volume} {88}},\ \bibinfo
  {pages} {024407} (\bibinfo {year} {2013}{\natexlab{b}})}\BibitemShut
  {NoStop}%
\bibitem [{\citenamefont {Miyatake}\ \emph {et~al.}(1986)\citenamefont
  {Miyatake}, \citenamefont {Yamamoto}, \citenamefont {Kim}, \citenamefont
  {Toyonaga},\ and\ \citenamefont {Nagai}}]{Miyatake1986heatbath}%
  \BibitemOpen
  \bibfield  {author} {\bibinfo {author} {\bibfnamefont {Y.}~\bibnamefont
  {Miyatake}}, \bibinfo {author} {\bibfnamefont {M.}~\bibnamefont {Yamamoto}},
  \bibinfo {author} {\bibfnamefont {J.~J.}\ \bibnamefont {Kim}}, \bibinfo
  {author} {\bibfnamefont {M.}~\bibnamefont {Toyonaga}},\ and\ \bibinfo
  {author} {\bibfnamefont {O.}~\bibnamefont {Nagai}},\ }\bibfield  {title}
  {\bibinfo {title} {{On the implementation of the {\textquotesingle}heat
  bath{\textquotesingle} algorithms for {M}onte {C}arlo simulations of
  classical {H}eisenberg spin systems}},\ }\href
  {https://doi.org/10.1088/0022-3719/19/14/020} {\bibfield  {journal} {\bibinfo
   {journal} {Journal of Physics C: Solid State Physics}\ }\textbf {\bibinfo
  {volume} {19}},\ \bibinfo {pages} {2539} (\bibinfo {year}
  {1986})}\BibitemShut {NoStop}%
\bibitem [{\citenamefont {Creutz}(1987)}]{Creutz1987overrelax}%
  \BibitemOpen
  \bibfield  {author} {\bibinfo {author} {\bibfnamefont {M.}~\bibnamefont
  {Creutz}},\ }\bibfield  {title} {\bibinfo {title} {{Overrelaxation and
  {M}onte {C}arlo simulation}},\ }\href
  {https://doi.org/10.1103/PhysRevD.36.515} {\bibfield  {journal} {\bibinfo
  {journal} {Phys. Rev. D}\ }\textbf {\bibinfo {volume} {36}},\ \bibinfo
  {pages} {515} (\bibinfo {year} {1987})}\BibitemShut {NoStop}%
\end{thebibliography}%

\end{document}